\documentclass[aps,prd,showpacs,notitlepage,nofootinbib,preprintnumbers,amsmath,amssymb]{revtex4-1}

\usepackage[dvipsnames]{xcolor}
\usepackage{graphics,graphicx}
\usepackage{dcolumn}
\usepackage{bm}
\usepackage{epsfig}
\usepackage{hyperref} 
\usepackage{mathbbol}
\usepackage{epstopdf}
\usepackage{simplewick}
\usepackage[utf8]{inputenc} 
\usepackage{slashed}
\usepackage{stackrel}
\usepackage{mathrsfs}
\usepackage{xcolor}
	
\usepackage{tikz}
\usetikzlibrary{decorations.pathmorphing, decorations.markings, shapes.geometric, shapes.misc, calc, math}
\tikzstyle{gluon}=[decorate, decoration={coil,aspect=0.8, amplitude=1.5pt,  segment length=3pt}]

\usepackage{stackrel}
\usepackage[most]{tcolorbox}

\def\eq#1{{Eq.~(\ref{#1})}}
\def\fig#1{{Fig.~\ref{#1}}}
\newcommand{\ben}{\begin{eqnarray*}}
\newcommand{\een}{\end{eqnarray*}}
\newcommand{\un}[1]{\underline{#1}}

\newcommand{\pd}{\partial}

\newcommand{\tr}{\mbox{tr}}
\newcommand{\thalf}{\tfrac{1}{2}}
\newcommand{\llangle}{\Big\langle \!\! \Big\langle}
\newcommand{\rrangle}{\Big\rangle \!\! \Big\rangle}

\newcommand{\as}{\alpha_s}

\newcommand{\dhd}{{\textstyle d}
\lower.03ex\hbox{\kern-0.38em$^{\scriptstyle-}$}\kern-0.05em{}}
\newcommand{\dbar}{{\textstyle \delta}
\lower.03ex\hbox{\kern-0.38em$^{\scriptstyle-}$}\kern-0.05em{}}

\newcommand{\ul}[1]{\underline{#1}}

\makeatletter
\DeclareRobustCommand{\cev}[1]{%
  {\mathpalette\do@cev{#1}}%
}
\newcommand{\do@cev}[2]{%
  \vbox{\offinterlineskip
    \sbox\z@{$\m@th#1 x$}%
    \ialign{##\cr
      \hidewidth\reflectbox{$\m@th#1\vec{}\mkern4mu$}\hidewidth\cr
      \noalign{\kern-\ht\z@}
      $\m@th#1#2$\cr
    }%
  }%
}
\makeatother

\begin{document}

\title{Helicity Evolution at Small $x$: \\ Quark to Gluon and Gluon to Quark Transition Operators}

\author{Jeremy Borden} 
         \email[Email: ]{borden.75@buckeyemail.osu.edu}
         \affiliation{Department of Physics, The Ohio State
           University, Columbus, OH 43210, USA}

\author{Yuri V. Kovchegov} 
         \email[Email: ]{kovchegov.1@osu.edu}
         \affiliation{Department of Physics, The Ohio State
           University, Columbus, OH 43210, USA}

\author{Ming Li} 
         \email[Email: ]{li.13449@osu.edu}
         \affiliation{Department of Physics, The Ohio State
           University, Columbus, OH 43210, USA}

\begin{abstract}
We include the quark to gluon and gluon to quark shock-wave transition operators into the small Bjorken-$x$ evolution equations for helicity in the flavor-singlet channel derived earlier in \cite{Kovchegov:2015pbl, Kovchegov:2018znm, Cougoulic:2022gbk}. While such transitions do not affect the large-$N_c$ version of the evolution equations for helicity, the large-$N_c \& N_f$ equations are affected. ($N_c$ and $N_f$ are the numbers of quark colors and flavors, respectively.) We derive the corresponding corrected large-$N_c \& N_f$ equations for the polarized dipole amplitudes contributing to the flavor-singlet quark and gluon helicity distributions in the double-logarithmic approximation (DLA), resumming powers of $\as \, \ln^2 (1/x)$ with $\as$ the strong coupling constant. We solve these equations iteratively and extract the polarized splitting functions up to four loops. We show that our splitting functions agree with the fixed-order perturbative calculations up to and including the existing three-loops results \cite{Altarelli:1977zs,Dokshitzer:1977sg,Mertig:1995ny,Moch:2014sna}. Similar to the large-$N_c$ helicity evolution in the shock-wave approach \cite{Borden:2023ugd}, our large-$N_c \& N_f$ small-$x$ splitting functions agree with those obtained in the infrared evolution equations framework from \cite{Bartels:1996wc,Blumlein:1996hb} up to three loops, but appear to slightly disagree at four loops. 
\end{abstract}


\maketitle

\tableofcontents


\section{Introduction}
\label{sec:intro}

In the past decade significant advances have been achieved toward constructing small-$x$ evolution equations for helicity parton distribution functions (hPDFs) and related to them $g_1$ structure function of a nucleon \cite{Kovchegov:2015pbl, Hatta:2016aoc, Kovchegov:2016zex, Kovchegov:2016weo, Kovchegov:2017jxc, Kovchegov:2017lsr, Kovchegov:2018znm, Kovchegov:2019rrz, Boussarie:2019icw, Cougoulic:2019aja, Kovchegov:2020hgb, Cougoulic:2020tbc, Chirilli:2021lif, Kovchegov:2021lvz, Cougoulic:2022gbk, Borden:2023ugd, Adamiak:2023okq}. Most of this work was carried out using the $s$-channel evolution/shock wave formalism \cite{Mueller:1994rr,Mueller:1994jq,Mueller:1995gb,Balitsky:1995ub,Balitsky:1998ya,Kovchegov:1999yj,Kovchegov:1999ua,Jalilian-Marian:1997dw,Jalilian-Marian:1997gr,Weigert:2000gi,Iancu:2001ad,Iancu:2000hn,Ferreiro:2001qy}, which had been developed earlier to address the problem of gluon saturation in the proton and nuclear wave functions. The aim of developing small-$x$ evolution for helicity distributions and observables is to help solve the proton spin puzzle \cite{EuropeanMuon:1987isl, Jaffe:1989jz, Ji:1996ek, Boer:2011fh, Aidala:2012mv, Accardi:2012qut, Leader:2013jra, Aschenauer:2013woa, Aschenauer:2015eha,  Proceedings:2020eah, Ji:2020ena, AbdulKhalek:2021gbh} by theoretically and phenomenologically constraining the amount of the proton spin carried by small-$x$ partons. Phenomenological implementation of the formalism from \cite{Kovchegov:2015pbl, Hatta:2016aoc, Kovchegov:2016zex, Kovchegov:2016weo, Kovchegov:2017jxc, Kovchegov:2017lsr, Kovchegov:2018znm, Kovchegov:2019rrz, Cougoulic:2019aja, Kovchegov:2020hgb, Cougoulic:2020tbc, Kovchegov:2021lvz, Cougoulic:2022gbk, Borden:2023ugd, Adamiak:2023okq} to describe the polarized deep inelastic scattering (DIS) and semi-inclusive DIS (SIDIS) small-$x$ data has been recently carried out in \cite{Adamiak:2021ppq, Adamiak:2023yhz}.

Derivation of the helicity evolution equations at small $x$ \cite{Kovchegov:2015pbl, Kovchegov:2016zex, Kovchegov:2018znm, Cougoulic:2022gbk} utilizes the sub-eikonal corrections to the eikonal shock wave formalism computed in 
\cite{Altinoluk:2014oxa,Balitsky:2015qba,Balitsky:2016dgz, Kovchegov:2017lsr, Kovchegov:2018znm, Chirilli:2018kkw, Jalilian-Marian:2018iui, Jalilian-Marian:2019kaf, Altinoluk:2020oyd, Boussarie:2020vzf, Boussarie:2020fpb, Kovchegov:2021iyc, Altinoluk:2021lvu, Kovchegov:2022kyy, Altinoluk:2022jkk, Altinoluk:2023qfr,Altinoluk:2023dww, Li:2023tlw}. The first step is to simplify the distribution functions, such as the flavor-singlet quark and gluon hPDFs $\Delta \Sigma (x, Q^2)$ and $\Delta G (x, Q^2)$, at small $x$ by rewriting them in terms of the dipole scattering amplitudes on the target proton. These dipole amplitudes are different from the case of the eikonal spin-independent scattering of \cite{Mueller:1994rr,Mueller:1994jq,Mueller:1995gb,Balitsky:1995ub,Balitsky:1998ya,Kovchegov:1999yj,Kovchegov:1999ua,Jalilian-Marian:1997dw,Jalilian-Marian:1997gr,Weigert:2000gi,Iancu:2001ad,Iancu:2000hn,Ferreiro:2001qy}, where the dipole amplitudes are correlators of pairs of infinite light-cone Wilson lines. Since Wilson lines do not couple to the target proton helicity, for helicity observables one needs to improve the precision of the calculation by employing the sub-eikonal corrections to the Wilson lines: these corrections are suppressed by one power of energy compared to the eikonal terms. The sub-eikonal corrections come in the form of insertions of one or two local sub-eikonal operators into the infinite light-cone Wilson line \cite{Altinoluk:2014oxa,Balitsky:2015qba,Balitsky:2016dgz, Kovchegov:2017lsr, Kovchegov:2018znm, Chirilli:2018kkw, Jalilian-Marian:2018iui, Jalilian-Marian:2019kaf, Altinoluk:2020oyd, Kovchegov:2021iyc, Altinoluk:2021lvu, Kovchegov:2022kyy, Altinoluk:2022jkk, Altinoluk:2023qfr,Altinoluk:2023dww, Li:2023tlw}. We will call the resulting objects `polarized Wilson lines': they are detailed in Sec.~\ref{sec:polWlines} below. When a polarized Wilson line is combined in a color trace with the regular infinite light-cone Wilson line (with one of them conjugated) and the resulting operator is evaluated in the polarized nucleon state, one obtains the so-called `polarized dipole scattering amplitude'. Since there exist different types of polarized Wilson lines, there exist different types of polarized dipole amplitudes \cite{Cougoulic:2022gbk}, in contrast to the single unpolarized dipole amplitude in the eikonal scattering case (for a given color representation of the partons). The hPDFs $\Delta \Sigma$ and $\Delta G$, along with the $g_1$ structure function, can all be expressed at small $x$ in terms of the polarized dipole amplitudes  \cite{Kovchegov:2015pbl, Kovchegov:2016zex, Kovchegov:2018znm, Cougoulic:2022gbk}. 

In the shock wave formalism, the small-$x$ evolution equations for helicity are written for the polarized Wilson lines correlators with the regular light-cone Wilson lines \cite{Kovchegov:2015pbl, Kovchegov:2016zex, Kovchegov:2018znm, Chirilli:2021lif,  Cougoulic:2022gbk}. At the leading order, the equations re-sum powers of the double-logarithmic parameters $\as \, \ln^2 (1/x)$ and $\as \, \ln (1/x) \, \ln (Q^2 / \Lambda^2)$, with $\as$ the strong coupling constant, $Q$ the hard momentum scale, and $\Lambda$ the infrared (IR) cutoff. This is the double logarithmic approximation (DLA). The equations are not closed in general (cf. \cite{Balitsky:1995ub}). They close in the large-$N_c$ limit \cite{tHooft:1977xjm} and, unlike the unpolarized case, in the large-$N_c \& N_f$ limit \cite{Veneziano:1976wm} as well. The large-$N_c$ equations, whose current version was derived in \cite{Cougoulic:2022gbk}, were solved numerically in the same reference, and analytically in \cite{Borden:2023ugd}. The gluon--gluon polarized anomalous dimension for the polarized Dokshitzer-Gribov-Lipatov-Altarelli-Parisi (DGLAP) evolution equations \cite{Gribov:1972ri,Altarelli:1977zs,Dokshitzer:1977sg} obtained in \cite{Borden:2023ugd} was shown to agree with the small-$x$ large-$N_c$ limit of the existing perturbative calculations
\cite{Altarelli:1977zs,Dokshitzer:1977sg,Zijlstra:1993sh,Mertig:1995ny,Moch:1999eb,vanNeerven:2000uj,Vermaseren:2005qc,Moch:2014sna,Blumlein:2021ryt,Blumlein:2021lmf,Davies:2022ofz,Blumlein:2022gpp} to the three known loops. The same three-loop agreement with the known gluon--gluon splitting function was observed by iteratively solving the large-$N_c$ helicity evolution equations in \cite{Cougoulic:2022gbk}. The gluon-gluon polarized anomalous dimension from \cite{Borden:2023ugd} appears to differ from that extracted out of the work by Bartels, Ermolaev and Ryskin (BER) \cite{Bartels:1996wc}. The latter reference used the infrared evolution equations (IREE) approach \cite{Gorshkov:1966ht,Kirschner:1983di,Kirschner:1994rq,Kirschner:1994vc,Griffiths:1999dj} to extract small-$x$ expressions for quark and gluon hPDFs. The gluon-gluon polarized anomalous dimensions of \cite{Cougoulic:2022gbk} and \cite{Bartels:1996wc}, when expanded as power series in $\as$, agree to the three known loops, and disagree beyond that.

The large-$N_c \& N_f$ helicity evolution equations from \cite{Cougoulic:2022gbk} were solved numerically in \cite{Adamiak:2023okq}. In the latter reference an iterative solution of these large-$N_c \& N_f$ equations was constructed, resulting in the polarized splitting functions, which, in the terms containing $N_f$, appeared to deviate from the existing known finite-order $\overline{\text{MS}}$ results \cite{Altarelli:1977zs,Dokshitzer:1977sg,Mertig:1995ny,Moch:2014sna} at the next-to-leading order (NLO). The disagreement could not be attributed to a scheme dependence \cite{Adamiak:2023okq}. In addition, an analytic solution of the large-$N_c \& N_f$ helicity evolution equations from \cite{Cougoulic:2022gbk} was constructed by a subset of the authors using the same general approach as that used in \cite{Borden:2023ugd}, also exhibiting a deviation in $N_f$-dependent terms from the NLO fixed-order calculations \cite{Altarelli:1977zs,Dokshitzer:1977sg,Mertig:1995ny,Moch:2014sna}. A possible explanation of this disagreement is that a contribution is missing in the large-$N_c \& N_f$ helicity evolution equations from \cite{Cougoulic:2022gbk}: this is the possibility we will explore in this work, ultimately identifying the missing contribution and correcting the evolution equations by including it. 

The helicity evolution derived in \cite{Kovchegov:2015pbl, Kovchegov:2016zex, Kovchegov:2018znm, Cougoulic:2022gbk} only involved the polarized Wilson lines where the incoming quark becomes an outgoing quark after interacting with the shock wave, or an incoming gluon becomes an outgoing gluon. The {\sl transition} operators, where the incoming (anti-)quark interacts with the shock wave and becomes an outgoing gluon, and, vice versa, an incoming gluon becomes the outgoing (anti-)quark (see \fig{fig:newoperators} below) were considered in \cite{Kovchegov:2015pbl, Kovchegov:2016zex}, but were not included in the calculation since they were perceived to cancel for the flavor-singlet quantities \cite{Kovchegov:2015pbl, Kovchegov:2016zex}, not contributing to the flavor-singlet helicity evolution equations in the DLA analysis performed there.  The contribution of such transition operators was extensively studied by Chirilli in \cite{Chirilli:2021lif}: they enter the DLA helicity evolution equations derived in that reference. Note that the flavor-singlet versus non-singlet decomposition is done differently in \cite{Chirilli:2021lif} from \cite{Kovchegov:2015pbl, Kovchegov:2016zex, Cougoulic:2022gbk}. The importance of the transition operators for the flavor non-singlet helicity evolution was also recognized in \cite{Kovchegov:2016zex}, but the contribution of these operators was (correctly) neglected in that reference as sub-leading in the large-$N_c$ limit employed there, leading to the same small-$x$ asymptotics as that obtained by BER in \cite{Bartels:1995iu} for the flavor non-singlet case (if one takes the large-$N_c$ limit of the latter).

In this work we show that the $q/{\bar q} \to G$ and $G \to q/{\bar q}$ shock-wave transition operators do indeed generate DLA terms and contribute to helicity evolution equations in the flavor-singlet channel. We explicitly derive and include the correction due to the transition operators into the large-$N_c \& N_f$ helicity evolution equations derived in \cite{Cougoulic:2022gbk}. The large-$N_c$ helicity evolution equations from \cite{Cougoulic:2022gbk} remain unchanged, since they include the gluon contributions only. We verify the corrected large-$N_c \& N_f$ helicity evolution equations we obtain below by solving them iteratively: this results in the polarized DGLAP splitting functions agreeing with BER \cite{Bartels:1996wc} to three loops. Our splitting functions also agree with the (small-$x$ and large-$N_c \& N_f$ limit of the) fixed-order $\overline{\text{MS}}$ polarized splitting functions \cite{Altarelli:1977zs,Dokshitzer:1977sg,Mertig:1995ny,Moch:2014sna} at one and two loops. The agreement at three loops is achieved after a minor scheme transformation: the same is true for BER splitting functions \cite{Moch:2014sna}. The observed agreement applies to all $N_c$- and $N_f$-dependent terms in the splitting functions. 
We also predict the 4-loop polarized DGLAP splitting functions at small-$x$ and large-$N_c \& N_f$.

Our paper is structured as follows. In Sec.~\ref{sec:polWlines} we list the polarized Wilson lines and polarized dipole amplitudes employed in \cite{Kovchegov:2015pbl, Kovchegov:2018znm, Cougoulic:2022gbk}. We derive the corrected helicity evolution equations in Sec.~\ref{sec:derivation} using the light-cone operator treatment (LCOT) method \cite{Kovchegov:2017lsr,Kovchegov:2018znm,Cougoulic:2022gbk,Kovchegov:2021iyc}. We first derive the $q/{\bar q} \to G$ and $G \to q/{\bar q}$ shock-wave transition operators in Sec.~\ref{sec:operators}. The correction to the evolution of the adjoint type-1 dipole amplitude (in the notation of \cite{Cougoulic:2022gbk}) is derived in Sec.~\ref{sec:Gadj_evol}. This correction contains an expectation value of a new operator, which we denote ${\widetilde Q}$. This operator is proportional to the quark helicity transverse momentum-dependent (TMD) PDF and, hence, cannot be called a (polarized) dipole amplitude. The evolution of ${\widetilde Q}$ is derived in   Sec.~\ref{sec:Qtilde_evol}, and it is found to be related to the quark helicity TMD and the flavor-singlet PDF $\Delta \Sigma$. For completeness, we also find the contribution of the transition operators to the evolution of the type-1 fundamental dipole amplitude $Q$ in Sec.~\ref{sec:Q_evol}: such contribution is suppressed in the large-$N_c \& N_f$ limit. The corrected large-$N_c \& N_f$ helicity evolution equations are summarized in Sec.~\ref{sec:evolution}. (These corrected equations are re-derived in Appendix~\ref{A} using the light-cone perturbation theory (LCPT) \cite{Lepage:1980fj, Brodsky:1997de}.) We cross-check these equations against known polarized DGLAP anomalous dimensions \cite{Altarelli:1977zs,Dokshitzer:1977sg,Mertig:1995ny,Moch:2014sna} by performing an iterative solution in Sec.~\ref{sec:iterative} and extracting the anomalous dimension perturbatively, finding a complete agreement at the first three loops with the $\overline{\text{MS}}$ polarized splitting functions \cite{Altarelli:1977zs,Dokshitzer:1977sg,Mertig:1995ny,Moch:2014sna} (modulo a scheme transformation at 3 loops) mentioned above and predicting the 4-loop polarized DGLAP splitting functions at low $x$. We conclude in Sec.~\ref{sec:conclusions}.


\section{Polarized Wilson lines: a recap}
\label{sec:polWlines}

In this Section we briefly review the main results of the earlier works that we will build on below. For a quark scattering on a quark and gluon background field, the sub-eikonal $S$-matrix is \cite{Altinoluk:2014oxa,Balitsky:2015qba,Balitsky:2016dgz, Kovchegov:2017lsr, Kovchegov:2018znm, Chirilli:2018kkw, Jalilian-Marian:2018iui, Jalilian-Marian:2019kaf, Altinoluk:2020oyd, Kovchegov:2021iyc, Altinoluk:2021lvu, Kovchegov:2022kyy, Altinoluk:2022jkk, Altinoluk:2023qfr,Altinoluk:2023dww, Li:2023tlw}
\begin{align}\label{Vxy_sub-eikonal}
V_{\un{x}, \un{y}; \sigma', \sigma} \bigg|_{\textrm{sub-eikonal}}  \equiv \sigma \, \delta_{\sigma, \sigma'} \, \left[ V_{\un x}^{\textrm{G} [1]} + V_{\un x}^{\textrm{q} [1]} \right] \, \delta^2 ({\un x} - {\un y}) + \delta_{\sigma, \sigma'} \, \left[ V_{{\ul x}, {\un y}}^{\textrm{G} [2]} + V_{{\ul x}}^{\textrm{q} [2]} \, \delta^2 ({\un x} - {\un y}) \right],
\end{align}
where the fundamental polarized Wilson lines of the first and second type (denoted by [1] and [2] in the superscript, respectively) are
\begin{subequations}\label{VqG}
\begin{align}
& V_{\un x}^{\textrm{G} [1]}  = \frac{i \, g \, p_1^+}{s} \int\limits_{-\infty}^{\infty} d{x}^- V_{\un{x}} [ \infty, x^-] \, F^{12} (x^-, {\un x}) \, \, V_{\un{x}} [ x^-, -\infty]  , \label{VG1} \\
& V_{\un x}^{\textrm{q} [1]}  = \frac{g^2 p_1^+}{2 \, s} \int\limits_{-\infty}^{\infty} \!\! d{x}_1^- \! \int\limits_{x_1^-}^\infty d x_2^- V_{\un{x}} [ \infty, x_2^-] \, t^b \, \psi_{\beta} (x_2^-,\un{x}) \, U_{\un{x}}^{ba} [x_2^-, x_1^-] \, \left[ \gamma^+ \gamma^5 \right]_{\alpha \beta} \, \bar{\psi}_\alpha (x_1^-,\un{x}) \, t^a \, V_{\un{x}} [ x_1^-, -\infty] , \label{Vq1} \\
& V_{{\ul x}, {\un y}}^{\textrm{G} [2]}  = - \frac{i \, p_1^+}{s} \int\limits_{-\infty}^{\infty} d{z}^- d^2 z \ V_{\un{x}} [ \infty, z^-] \, \delta^2 (\un{x} - \un{z}) \, \cev{D}^i (z^-, {\un z}) \, D^i  (z^-, {\un z}) \, V_{\un{y}} [ z^-, -\infty] \, \delta^2 (\un{y} - \un{z}) , \label{VxyG2} \\
& V_{{\ul x}}^{\textrm{q} [2]} = - \frac{g^2 p_1^+}{2 \, s} \int\limits_{-\infty}^{\infty} \!\! d{x}_1^- \! \int\limits_{x_1^-}^\infty d x_2^- V_{\un{x}} [ \infty, x_2^-] \, t^b \, \psi_{\beta} (x_2^-,\un{x}) \, U_{\un{x}}^{ba} [x_2^-, x_1^-] \, \left[ \gamma^+ \right]_{\alpha \beta} \, \bar{\psi}_\alpha (x_1^-,\un{x}) \, t^a \, V_{\un{x}} [ x_1^-, -\infty] \label{Vq2}.
\end{align}
\end{subequations}
We employ the light-cone coordinates defined as $x^\pm = (x^0 \pm x^3)/\sqrt{2}$, the transverse position vectors are ${\un x} = (x^1, x^2)$, the Latin index $i = 1,2$ labels transverse components, and $g$ is the QCD coupling. Our target nucleon is moving with large momentum $p_1^+$ in the light-cone $x^+$-direction, while $s$ is the center-of-mass energy squared of the projectile--target system. The incoming quark has polarization $\sigma$, while the outgoing quark has polarization $\sigma'$. The background gluon field $A^\mu$ brings in two sub-eikonal operators: the transverse--transverse component of the field strength tensor $F^{12}$ and the $\cev{D}^i \, D^i$ operator, with the right- and left-acting covariant derivatives defined by $D^i = \pd^i - i g A^i$ and $\cev{D}^i = \cev{\pd}^i + i g A^i$, respectively. The background quark fields $\psi$ and $\bar \psi$ bring in the vector ($\gamma^+$) and the axial vector ($\gamma^+ \, \gamma^5$) current operators, both of which are also sub-eikonal: the vector current operator $V_{{\ul x}}^{\textrm{q} [2]}$ does not contribute to the helicity evolution constructed in \cite{Kovchegov:2015pbl,  Kovchegov:2016zex, Kovchegov:2017lsr, Kovchegov:2018znm, Cougoulic:2022gbk}. We have also employed the fundamental ($V$) and adjoint ($U$) light-cone Wilson lines,
\begin{align}\label{Vline}
V_{\un{x}} [x^-_f,x^-_i] = \mathcal{P} \exp \left[ ig \int\limits_{x^-_i}^{x^-_f} d{x}^- A^+ (0^+, x^-, \un{x}) \right], \ \ \ 
U_{\un{x}} [x^-_f,x^-_i] = \mathcal{P} \exp \left[ ig \int\limits_{x^-_i}^{x^-_f} d{x}^- {\cal A}^+ (0^+, x^-, \un{x}) \right].
\end{align}

For the gluon scattering on the same background fields with the initial helicity $\lambda$ and the final helicity $\lambda'$ we have the following sub-eikonal $S$-matrix,
\begin{align}\label{Uxy_sub-eikonal}
(U_{{\ul x}, {\un y}; \lambda', \lambda})^{b a} \bigg|_{\textrm{sub-eikonal}}  \equiv \lambda \, \delta_{\lambda, \lambda'} \, \left( U_{\un x}^{\textrm{G} [1]} + U_{\un x}^{\textrm{q} [1]} \right)^{b a} \, \delta^2 ({\un x} - {\un y}) + \delta_{\lambda, \lambda'} \, \left(U_{{\ul x}, {\un y}}^{\textrm{G} [2]} + U_{{\ul x}}^{\textrm{q} [2]} \, \delta^2 ({\un x} - {\un y}) \right)^{b a} 
\end{align}
with the polarized Wilson lines
\begin{subequations}\label{UqG}
\begin{align}
& (U_{\un x}^{\textrm{G} [1]})^{ba} = \frac{2 \, i \, g \, p_1^+}{s} \int\limits_{-\infty}^{\infty} d{x}^- (U_{\un{x}} [ \infty, x^-])^{bb'} \, ({\cal F}^{12})^{b'a'} (x^-, {\un x}) \, (U_{\un{x}} [ x^-, -\infty])^{a'a}  , \label{UG1} \\
& (U_{\un x}^{\textrm{q} [1]})^{ba} = \frac{g^2 p_1^+}{2 \, s} \!\! \int\limits_{-\infty}^{\infty} \!\! d{x}_1^- \! \int\limits_{x_1^-}^\infty d x_2^- (U_{\un{x}} [ \infty, x_2^-])^{bb'} \bar{\psi} (x_2^-,\un{x}) \, t^{b'} V_{\un{x}} [x_2^-,x_1^-] \, \gamma^+ \gamma^5 \, t^{a'} \psi (x_1^-,\un{x})  (U_{\un{x}} [ x_1^-, -\infty])^{a'a} + \mbox{c.c.}  ,  \label{Uq1} \\
& (U_{{\ul x}, {\un y}}^{\textrm{G} [2]})^{ba}  = - \frac{i \, p_1^+}{s} \int\limits_{-\infty}^{\infty} d{z}^- d^2 z \ (U_{\un{x}} [ \infty, z^-])^{bb'} \, \delta^2 (\un{x} - \un{z}) \,\cev{\underline{\mathscr{D}}}^{b'c} (z^-, {\un z}) \, \cdot \, \underline{\mathscr{D}}^{ca'}  (z^-, {\un z}) \, (U_{\un{y}} [ z^-, -\infty])^{a'a} \, \delta^2 (\un{y} - \un{z}) , \label{UG2}  \\
& (U_{{\ul x}}^{\textrm{q} [2]} )^{ba} = - \frac{g^2 p_1^+}{2 \, s} \int\limits_{-\infty}^{\infty} \!\! d{x}_1^- \! \int\limits_{x_1^-}^\infty d x_2^- (U_{\un{x}} [ \infty, x_2^-])^{bb'} \, \bar{\psi} (x_2^-,\un{x}) \, t^{b'} \, V_{\un{x}} [x_2^-,x_1^-] \, \gamma^+ \, t^{a'} \, \psi (x_1^-,\un{x}) \,  (U_{\un{x}} [ x_1^-, -\infty])^{a'a} - \mbox{c.c.} , \label{Uq2}
\end{align}
\end{subequations}
where ${\cal F}^{12}$ and $\cev{\underline{\mathscr{D}}} \, \cdot \, \underline{\mathscr{D}}$ are now in the adjoint representation. Again, $U_{{\ul x}}^{\textrm{q} [2]}$ does not contribute to helicity evolution in \cite{Kovchegov:2015pbl,  Kovchegov:2016zex, Kovchegov:2017lsr, Kovchegov:2018znm, Cougoulic:2022gbk}.

The polarized dipole amplitudes needed to construct the large-$N_c \& N_f$ helicity evolution equations are
\begin{subequations}\label{QGi}
\begin{align}
Q_{10}(s) &= \frac{1}{2 \, N_c} \, \mbox{Re} \, \llangle \mbox{T} \, \tr \left[ V_{\un 0} \,  V_{\un 1}^{\textrm{pol} [1] \,\dagger} \right] + \mbox{T} \,  \tr \left[ V_{\un 1}^{\textrm{pol} [1]} \, V_{\un 0}^\dagger \right]   \rrangle (s)  , \label{Qdef} \\
G^i_{10}(s) &= \frac{1}{2 \, N_c} \, \mbox{Re} \, \llangle \mbox{T} \, \tr \left[ V_{\un 0} \,  V_{\un 1}^{i \, \textrm{G} [2] \, \dagger} \right] + \mbox{T} \,  \tr \left[ V_{\un 1}^{i \, \textrm{G} [2] } \, V_{\un 0}^\dagger \right]   \rrangle (s)  , \label{Gi_repeat}
\end{align}
\end{subequations}
with 
\begin{align}
    V_{\un x}^{\textrm{pol} [1]} = V_{\un x}^{\textrm{G} [1]} + V_{\un x}^{\textrm{q} [1]}
\end{align}
and with the new polarized Wilson line
\begin{align}\label{Vi}
V_{\un{z}}^{i \, \textrm{G} [2]} \equiv \frac{p_1^+}{2 s} \, \int\limits_{-\infty}^{\infty} d {z}^- \, V_{\un{z}} [ \infty, z^-] \, \left[ {D}^i (z^-, \un{z}) - \cev{D}^i (z^-, \un{z}) \right] \, V_{\un{z}} [ z^-, -\infty] 
\end{align}
related to the one in \eq{VxyG2} (see \cite{Cougoulic:2022gbk}). Here T stands for time-ordering. The double angle brackets are defined as the usual saturation physics averaging in the (now, polarized) target wave function, denoted by single angle brackets, with the result multiplied by a power of the center-of-mass energy squared $s$ \cite{Kovchegov:2015pbl},
\begin{align}\label{double_angle_brackets}
\llangle \ldots \rrangle \equiv s \,  \Big\langle \ldots \Big\rangle .
\end{align}
We also use the abbreviated notation for infinite Wilson lines $V_{\un x} = V_{\un x} [\infty, - \infty]$ and $V_{\un 1} = V_{{\un x}_1}$, $V_{\un 0} = V_{{\un x}_0}$, such that, for instance, $Q_{10} (s)$ depends on the transverse positions ${\un x}_1$, ${\un x}_0$ in addition to its dependence on energy $s$.

Since the $g_1$ structure function and hPDFs $\Delta\Sigma$ and $\Delta G$ at small $x$ depend on the polarized dipole amplitudes integrated over all impact parameters, we can integrate the amplitudes over the impact parameters while keeping the dipole transverse separation ${\un x}_{10} \equiv {\un x}_1 - {\un x}_0$ fixed, defining 
\begin{align}\label{Q_int}
Q (x^2_{10} , s) \equiv \int d^2 \left( \frac{x_0 + x_1}{2} \right) \, Q_{10} (s)
\end{align}
for the dipole amplitude $Q$. (We will use the notation where ${\un x}_{ij} = {\un x}_i - {\un x}_j$ and $x_{ij} = |{\un x}_{ij}|$.)  Integrating \eq{Gi_repeat} over the impact parameters we write the most general transverse tensor decomposition for the result of such integration as \cite{Kovchegov:2017lsr, Cougoulic:2022gbk}
\begin{align}
\int d^2 \left( \frac{x_{1} + x_0}{2} \right) \, G^{i}_{10} (s) = (x_{10})_\bot^i \, G_1
  (x_{10}^2, s) + \epsilon^{ij} \, (x_{10})_\bot^j \, G_2
  (x_{10}^2, s) ,
\end{align}
obtaining two new dipole amplitudes $G_1$ and $G_2$: only $G_2$ contributes to helicity evolution \cite{Cougoulic:2022gbk}.

The adjoint dipole amplitude of the first ($Q$) kind is defined by analogy to \eq{Qdef} as
\begin{align}\label{G_adj_def}
G^\textrm{adj}_{10} (s) \equiv \frac{1}{2 (N_c^2 -1)} \, \mbox{Re} \, \llangle \mbox{T} \, \mbox{Tr} \left[ U_{\ul 0} \, U_{{\un 1}}^{\textrm{pol} [1] \, \dagger} \right] + \mbox{T} \, \mbox{Tr} \left[ U_{{\un 1}}^{\textrm{pol} [1]} \, U_{\ul 0}^\dagger \right] \rrangle (s) 
\end{align}
with
\begin{align}\label{Upol1_def}
    U_{\un x}^{\textrm{pol} [1]} = U_{\un x}^{\textrm{G} [1]} + U_{\un x}^{\textrm{q} [1]}.
\end{align}
In the large-$N_c \& N_f$ limit this dipole amplitude reduces to \cite{Kovchegov:2018znm, Cougoulic:2022gbk}
\begin{align}\label{Gadj_Gtilde}
G^{\textrm{adj}}_{10} (s) = 4 \, S_{10} (s) \, {\widetilde G}_{10} (s)
\end{align}
where 
\begin{align}
S_{10} (s) = \frac{1}{N_c} \, \left\langle \mbox{T} \, \tr \left[ V_{\un 0} \,  V_{\un 1}^{\dagger} \right] \right\rangle (s)
\end{align}
is the eikonal (unpolarized) dipole $S$-matrix and
\begin{align}\label{G_dfn_NcNf}
& {\widetilde G}_{10} (s) = \frac{1}{2 N_c} \, \mbox{Re} \, \llangle \mbox{T} \, \mbox{tr} \left[ V_{\ul 0} \, W_{{\un 1}}^{\textrm{pol} [1] \,\dagger} \right] + \mbox{T} \, \mbox{tr} \left[ W_{{\un 1}}^{\textrm{pol} [1] } \, V_{\ul 0}^\dagger \right] \rrangle (s)
\end{align}
with 
\begin{align}\label{Wpol_dfn}
W_{{\un x}}^{\textrm{pol} [1] } &= V_{{\un x}}^{\textrm{G} [1]} - \frac{g^2 p_1^+}{4s} \, \int\limits_{-\infty}^{\infty}dx_1^- \int\limits_{x_1^-}^{\infty}dx_2^- \; V_{{\un x}}[\infty,x_2^-] \, \psi_{\alpha}(x_2^-,{\un x}) \, \left(\frac{1}{2}\gamma^+\gamma_5\right)_{\beta\alpha}\bar{\psi}_{\beta}(x_1^-,{\un x}) \, V_{{\un x}}[x_1^-,-\infty] \, .
\end{align}
(Note that the relative sign is different on the right-hand side of \eq{Wpol_dfn} as compared to Eq.~(137) of \cite{Cougoulic:2022gbk}: the difference is due to fermionic field ordering. The typo in Eq.~(137) of \cite{Cougoulic:2022gbk} does not propagate.) 

The large-$N_c \& N_f$ helicity evolution equations are written for the dipole amplitudes $Q$, $G_2$ and ${\widetilde G}$, along with their ``neighbor" dipole amplitudes $\overline \Gamma$, $\Gamma_2$, and ${\widetilde \Gamma}$, introduced to properly account for the light-cone lifetime ordering  \cite{Kovchegov:2015pbl, Kovchegov:2018znm, Cougoulic:2022gbk}. 

For completeness, we list the expressions for the flavor-singlet hPDFs obtained in \cite{Kovchegov:2015pbl, Kovchegov:2016zex, Kovchegov:2017lsr, Kovchegov:2018znm, Cougoulic:2022gbk} in the DLA limit:
\begin{subequations}
    \begin{align}
    & \Delta \Sigma (x, Q^2) = - \frac{N_c \, N_f}{2 \pi^3} \:  \int\limits_{\Lambda^2/s}^1 \frac{d z}{z} \,  \int\limits_{\frac{1}{zs}}^{\min \left\{ \frac{1}{z Q^2} , \frac{1}{\Lambda^2} \right\}} \frac{d x^{2}_{10}}{x_{10}^2}  \, \left[  Q (x^2_{10} , zs) + 2 \, G_2 (x^2_{10} , zs) \right], \label{DeltaSigma} \\
    & \label{JM_DeltaG}
\Delta G (x, Q^2) = \frac{2 N_c}{\as \pi^2} \ G_2 \left(  x_{10}^2 = \frac{1}{Q^2} ,  s = \frac{Q^2}{x} \right).
\end{align}
\end{subequations}
Here $z$ is the minus momentum fraction carried by the anti-quark (or quark) line in the definition of the quark (anti-quark) PDF. Below we will find a correction to the integration limits of \eq{DeltaSigma}.

The $g_1$ structure function is \cite{Kovchegov:2015pbl, Cougoulic:2022gbk}
\begin{align}\label{g1_DLA}
g_1 (x, Q^2)  = - \sum_f \frac{N_c \, Z^2_f}{4 \pi^3} \int\limits_{\Lambda^2/s}^1 \frac{dz}{z} \,  \int\limits^{\min \left\{ \frac{1}{z Q^2} , \frac{1}{\Lambda^2} \right\}}_\frac{1}{zs} \frac{d x^2_{10}}{x_{10}^2} \, \left[ Q (x_{10}^2, zs) + 2 \, G_2 (x_{10}^2, zs) \right], 
\end{align}
with $z$ the minus momentum fraction of the virtual photon's momentum carried by the (longitudinally) softer line between the quark and the anti-quark the virtual photon splits into.


\section{Derivation of the Quark to Gluon and Gluon to Quark Contribution to Helicity Evolution}
\label{sec:derivation}

In this Section we derive corrections to the large-$N_c \& N_f$ helicity evolution equations due to the $q/{\bar q} \to G$ and $G \to q/{\bar q}$ shock-wave transition operators using LCOT \cite{Kovchegov:2017lsr,Kovchegov:2018znm,Cougoulic:2022gbk,Kovchegov:2021iyc}. In Appendix~\ref{A} we re-derive the same corrections to the helicity evolution equations using LCPT \cite{Lepage:1980fj, Brodsky:1997de}.

\subsection{Quark to Gluon and Gluon to Quark Transition Operators}
\label{sec:operators}

\begin{figure}[h]
    \centering
    \includegraphics[width=\textwidth]{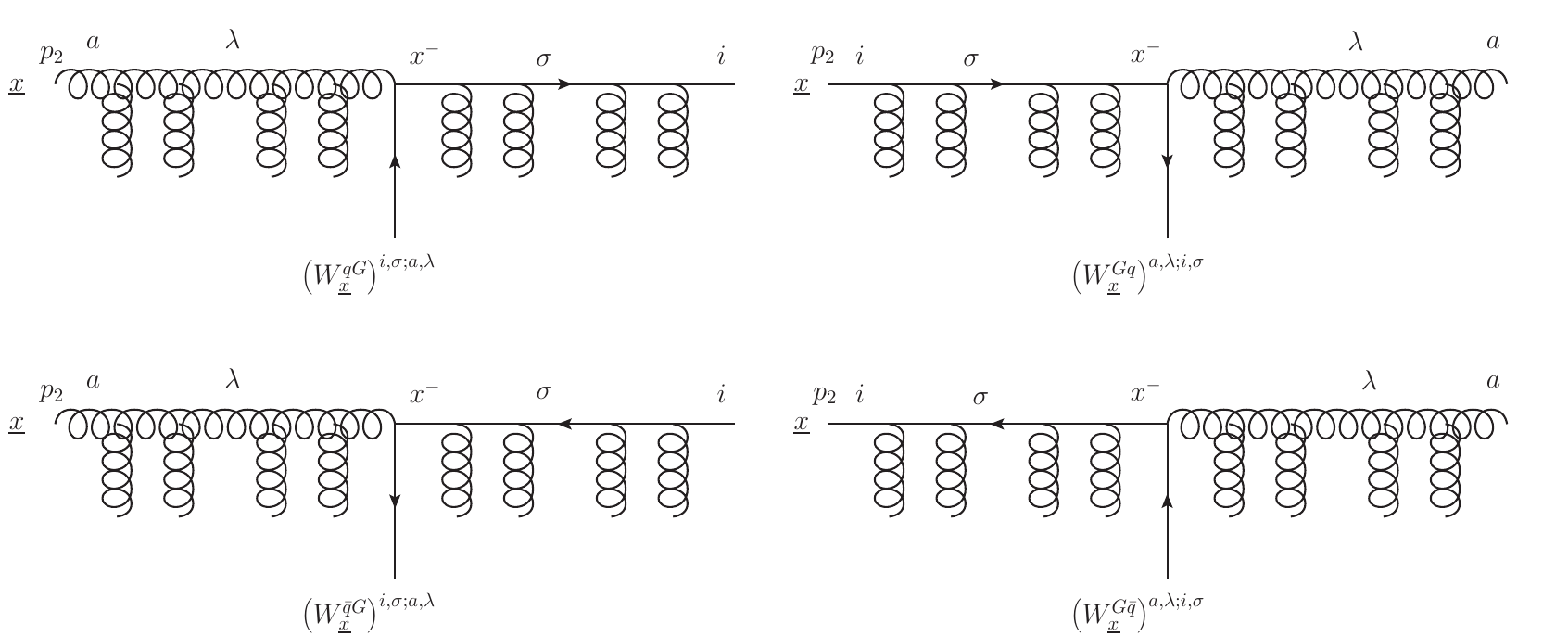}
    \caption{A diagrammatic representation of the $q/{\bar q} \to G$ and $G \to q/{\bar q}$ transition operators. The incoming particle's momentum $p_2$ has a large light-cone minus component. The polarizations and colors of the incoming and outgoing partons are labeled explicitly.}
    \label{fig:newoperators}
\end{figure}

We begin the calculation by constructing the (anti-)quark to gluon and gluon to (anti-)quark transition operators. They are shown diagrammatically in \fig{fig:newoperators}. The background gluon and quark fields in \fig{fig:newoperators} are represented by vertical corkscrew and straight lines, respectively. The non-eikonal interaction is due to the quark background field: the interactions with the gluon background field in \fig{fig:newoperators} are, therefore, assumed to be eikonal, and will be described by light-cone Wilson lines in the appropriate (adjoint or fundamental) color representations.  

A straightforward calculation along the lines of those done in \cite{Kovchegov:2017lsr, Kovchegov:2018znm, Kovchegov:2021iyc} yields for the transition operators defined in \fig{fig:newoperators}
\begin{subequations}\label{Ws}
\begin{align}
& W_{\un x}^{Gq} [\infty, - \infty] = \frac{- i g}{2 \, \sqrt{\sqrt{2} \, p_2^-}} \, \delta_{\lambda, \sigma} \, \int\limits_{-\infty}^\infty d x^- \, U_{\un x}^{ab} [\infty, x^-] \, {\bar \psi} (x^-, {\un x}) \, t^b \, \left[ \rho(+) + \rho(-) - \sigma \, \left( \rho(+) - \rho(-) \right) \right] \, V_{\un x} [x^-, - \infty], \\
& W_{\un x}^{G{\bar q}} [\infty, - \infty] = \frac{i g}{2 \, \sqrt{\sqrt{2} \, p_2^-}} \, \delta_{\lambda, \sigma} \, \int\limits_{-\infty}^\infty d x^- \, U_{\un x}^{ab} [\infty, x^-] \, V_{\un x} [-\infty, x^-]  \, \left[ {\bar \rho}(+) + {\bar \rho}(-) + \sigma \, \left( {\bar \rho}(+) - {\bar \rho}(-) \right) \right] \, t^b \, {\psi} (x^-, {\un x}), \\
& W_{\un x}^{qG} [\infty, - \infty] =  \frac{- i g}{2 \, \sqrt{\sqrt{2} \, p_2^-}} \, \delta_{\lambda, \sigma} \, \int\limits_{-\infty}^\infty d x^- \, V_{\un x} [\infty, x^-]  \, \left[ {\bar \rho}(+) + {\bar \rho}(-) - \sigma \, \left( {\bar \rho}(+) - {\bar \rho}(-) \right) \right] \, t^b \, {\psi} (x^-, {\un x}) \, U_{\un x}^{ba} [x^-, -\infty] , \\
& W_{\un x}^{{\bar q}G} [\infty, - \infty] = \frac{i g}{2 \, \sqrt{\sqrt{2} \, p_2^-}} \, \delta_{\lambda, \sigma} \, \int\limits_{-\infty}^\infty d x^- \, {\bar \psi} (x^-, {\un x}) \, t^b \, \left[ \rho(+) + \rho(-) + \sigma \, \left( \rho(+) - \rho(-) \right) \right] \, V_{\un x} [x^-, \infty]  \, U_{\un x}^{ba} [x^-, -\infty] .
\end{align}
\end{subequations}
We are suppressing the polarization and color indices, shown in \fig{fig:newoperators}. The superscripts of the $W$-operators in Eq.~\eqref{Ws} denote the outgoing and incoming particles: for instance, $W_{\un x}^{Gq}$ denotes the transition operator for the incoming quark and outgoing gluon, as shown in the upper right panel of \fig{fig:newoperators}. Here we are working in $A^- =0$ gauge for the gluons; an incoming or outgoing gluon with momentum $k$ is described by the polarization four-vector $\epsilon_\lambda^\mu = ({\un \epsilon}_\lambda \cdot {\un k}/k^-, 0, {\un \epsilon}_\lambda)$ with ${\un \epsilon}_\lambda = -(1/\sqrt{2}) (\lambda, i)$ \cite{Lepage:1980fj}. The external quark and anti-quark lines are described by the ``anti"-Brodsky--Lepage spinors \cite{Lepage:1980fj, Kovchegov:2018znm} 
\begin{align}\label{anti-BLspinors}
u_\sigma (p) = \frac{1}{\sqrt{\sqrt{2} \, p^-}} \, [\sqrt{2} \, p^- + m \, \gamma^0 +  \gamma^0 \, {\un \gamma} \cdot {\un p} ] \,  \rho (\sigma), \ \ \ v_\sigma (p) = \frac{1}{\sqrt{\sqrt{2} \, p^-}} \, [\sqrt{2} \, p^- - m \, \gamma^0 +  \gamma^0 \, {\un \gamma} \cdot {\un p} ] \,  \rho (-\sigma),
\end{align}
with $p^\mu = \left( \frac{{\un p}^2+ m^2}{2 p^-}, p^-, {\un p} \right)$ and
\begin{align}
  \rho (+1) \, = \, \frac{1}{\sqrt{2}} \, \left(
  \begin{array}{c}
      1 \\ 0 \\ -1 \\ 0
  \end{array}
\right), \ \ \ \rho (-1) \, = \, \frac{1}{\sqrt{2}} \, \left(
  \begin{array}{c}
        0 \\ 1 \\ 0 \\ 1
  \end{array}
\right) .
\end{align}

Relations between the transition operators are
\begin{align}
\left( W_{\un x}^{Gq} [\infty, - \infty] \right)^\dagger = W_{\un x}^{qG}  [-\infty, \infty], \ \ \ \left( W_{\un x}^{G{\bar q}} [\infty, - \infty] \right)^\dagger = W_{\un x}^{{\bar q}G}  [-\infty, \infty] .
\end{align}

Defining
\begin{subequations}\label{W12}
\begin{align}
& W_{\un x}^{a; i \, [1]} [b^-, a^-] = \frac{i g}{2 \, \sqrt{\sqrt{2} \, p^-}} \, \int\limits_{a^-}^{b^-} d x^- \, U_{\un x}^{ab} [b^-, x^-] \, {\bar \psi}^{i'} (x^-, {\un x}) \, t^b \, \left[ \rho(+) - \rho(-) \right] \, \left( V_{\un x} [x^-, a^-] \right)^{i'i}, \\
& W_{\un x}^{a; i \, [2]} [b^-, a^-] = - \frac{i g}{2 \, \sqrt{\sqrt{2} \, p^-}} \, \int\limits_{a^-}^{b^-} d x^- \, U_{\un x}^{ab} [b^- , x^-] \, {\bar \psi}^{i'} (x^-, {\un x}) \, t^b \, \left[ \rho(+) + \rho(-) \right] \, \left( V_{\un x} [x^-, a^-] \right)^{i'i},
\end{align}
\end{subequations}
which are rows in quark color space,
we rewrite Eqs.~\eqref{Ws} as
\begin{subequations}\label{Ws2}
\begin{align}
& \left( W_{\un x}^{Gq}[\infty, - \infty] \right)^{a,\lambda ; i, \sigma} = \sigma \, \delta_{\lambda, \sigma} \, W_{\un x}^{a; i \, [1]} [\infty, - \infty] + \delta_{\lambda, \sigma} \, W_{\un x}^{a; i \, [2]} [\infty, - \infty] , \\
& \left( W_{\un x}^{G{\bar q}} [\infty, - \infty] \right)^{a,\lambda ; i, \sigma} = - \sigma \, \delta_{\lambda, \sigma} \, \left( W_{\un x}^{a; i \, [1]} [\infty, - \infty] \right)^\dagger + \delta_{\lambda, \sigma} \, \left( W_{\un x}^{a; i \, [2]} [\infty, - \infty] \right)^\dagger , \\
& \left( W_{\un x}^{qG} [\infty, - \infty] \right)^{i, \sigma ; a,\lambda} = \sigma \, \delta_{\lambda, \sigma} \, \left( W_{\un x}^{a; i \, [1]} [-\infty, \infty] \right)^\dagger + \delta_{\lambda, \sigma} \, \left( W_{\un x}^{a; i \, [2]} [-\infty, \infty] \right)^\dagger , \\
& \left( W_{\un x}^{{\bar q}G} [\infty, - \infty] \right)^{i, \sigma ; a,\lambda} =  - \sigma \, \delta_{\lambda, \sigma} \, W_{\un x}^{a; i \, [1]} [-\infty, \infty] + \delta_{\lambda, \sigma} \, W_{\un x}^{a; i \, [2]} [-\infty, \infty] .
\end{align}
\end{subequations}

Having constructed the shock-wave transition operators, we are now ready to calculate their contributions to the flavor-singlet helicity evolution.


\subsection{Contribution to the evolution of the adjoint dipole amplitude of the first type}
\label{sec:Gadj_evol}


\subsubsection{Diagrams calculation} 
\label{sec:ABCD}

\begin{figure}[ht]
\centering
\includegraphics[width= 0.95 \textwidth]{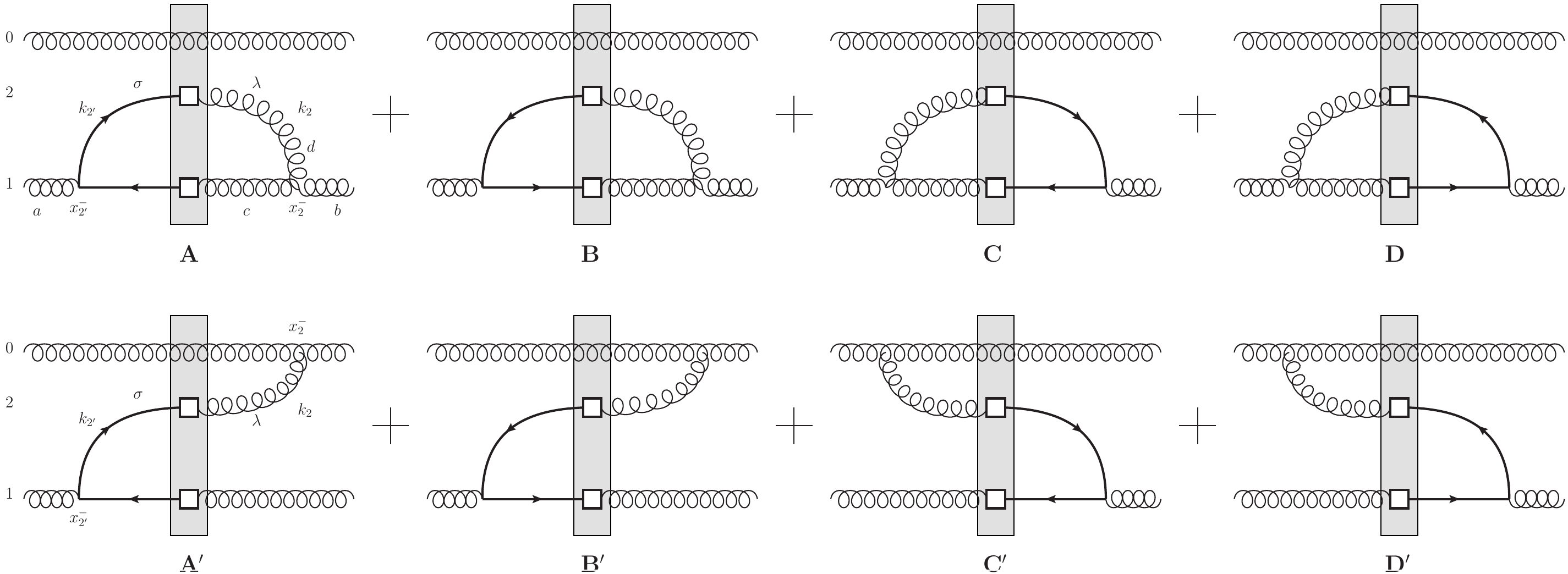}
\caption{The diagrams for the evolution of $G^\textrm{adj}_{10}$ due to the $q/{\bar q} \to G$ and $G \to q/{\bar q}$ shock-wave transition operators calculated here. The shaded rectangle denotes the shock wave, while the white square denotes the non-eikonal interaction with the shock wave mediated by the transition operators from \fig{fig:newoperators}.}
\label{FIG:Gadj_evol}
\end{figure}

The shock-wave transition operators \eqref{Ws2} may contribute to the flavor-singlet helicity evolution of the adjoint type-1 polarized dipole amplitude $G^\textrm{adj}_{10}$ defined in \eq{G_adj_def}. The diagrams in the shock wave formalism are shown in \fig{FIG:Gadj_evol}, with the shock wave represented by a vertical shaded rectangle. These diagrams arise due to the contribution of the polarized Wilson line operator $\left( U_{\un x}^{\textrm{q} [1]} \right)^{ba}$ to $G^\textrm{adj}_{10}$. The diagrams in \fig{FIG:Gadj_evol} include two transition operators from Eqs.~\eqref{Ws2}: they were not included in the analysis of \cite{Kovchegov:2015pbl, Kovchegov:2018znm, Cougoulic:2022gbk}.

To calculate diagrams A and A' from \fig{FIG:Gadj_evol} we will need the following background-field propagator, which can be constructed along the lines of \cite{Kovchegov:2017lsr, Kovchegov:2018znm, Kovchegov:2021iyc, Cougoulic:2022gbk} (quarks are assumed to be massless):
\begin{align}\label{psibar_a+}
\int\limits_{-\infty}^0 dx_{2'}^- \, & 
\int\limits_0^\infty dx_2^- \, 
\contraction[2ex]
{}
{{\bar \psi}}{_\alpha^i
(x_{2'}^- , \ul{x}_1) \:}
{a}
\: 
{\bar \psi}_\alpha^i (x_{2'}^- , \ul{x}_1) \:
a^{+ \, d} (x_2^- , \ul{x}_0) =
\frac{1}{4 \pi^3} \, \int\limits_0^{p_2^-} d k^- \,   \int d^2 x_2 \,  \frac{x_{20}^m}{x_{20}^2} \\ 
& \times \, \left[ \left( \epsilon_+^m \, {\bar \rho} (+)_\beta - \epsilon_-^m \, {\bar \rho} (-)_\beta \right) \, W_{\un 2}^{d;i \, [1]} [\infty, - \infty] + \left( \epsilon_+^m \, {\bar \rho} (+)_\beta + \epsilon_-^m \, {\bar \rho} (-)_\beta \right) \, W_{\un 2}^{d;i \, [2]} [\infty, - \infty] \right]  \notag \\
& \times \, \left[ i \, \sqrt{\sqrt{2} k^-} \, \ln \left( \frac{1}{x_{21} \Lambda} \right) + \frac{1}{\sqrt{\sqrt{2} k^-}} \,  \frac{x_{21}^l}{x_{21}^2} \, \gamma^0 \, \gamma^l \right]_{\beta\alpha} . \notag
\end{align}
The notations used in \eq{psibar_a+} are defined in \fig{FIG:Gadj_evol}, with the typical (large) minus momentum of the original (parent) dipole being $p_2^-$. The first term in the last square brackets of \eq{psibar_a+} does not contribute to the evolution at hand. The second term gives, after some algebra,
\begin{align}\label{AA'}
A + A' & = \frac{1}{N_c^2 -1} \, \frac{\as}{2 \pi^2} \, \sum_f \, \int\limits_0^{p_2^-} \frac{d k^-}{k^-} \, \int d^2 x_2 \, \llangle \left( U^\dagger_{\un 0} \right)^{ab} \, f^{dbc} \, \left[ i \, W_{\un 2}^{d \, [1]} [\infty, - \infty] \, t^a \, \left( W_{\un 1}^{c \, [2]} [\infty, - \infty] \right)^\dagger \frac{{\un x}_{20}}{x_{20}^2}  \cdot \frac{{\un x}_{21}}{x_{21}^2}  \right. \\ 
& + W_{\un 2}^{d \, [1]} [\infty, - \infty] \, t^a \, \left( W_{\un 1}^{c \, [1]} [\infty, - \infty] \right)^\dagger \frac{{\un x}_{20}}{x_{20}^2}  \times \frac{{\un x}_{21}}{x_{21}^2}  + i \, W_{\un 2}^{d \, [2]} [\infty, - \infty] \, t^a \, \left( W_{\un 1}^{c \, [1]} [\infty, - \infty] \right)^\dagger \frac{{\un x}_{20}}{x_{20}^2}  \cdot \frac{{\un x}_{21}}{x_{21}^2} \notag \\ 
&  \left.  + W_{\un 2}^{d \, [2]} [\infty, - \infty] \, t^a \, \left( W_{\un 1}^{c \, [2]} [\infty, - \infty] \right)^\dagger \frac{{\un x}_{20}}{x_{20}^2}  \times \frac{{\un x}_{21}}{x_{21}^2}  \right] \rrangle - ({\un x}_{20} \to {\un x}_{21}) , \notag
\end{align} 
where we have suppressed the quark color indices on the $W$'s.

Here, the double angle brackets are defined as 
\begin{align}\label{double_def}
\llangle W_{\un 2}^{d \, [k]} \, \ldots \, \left( W_{\un 1}^{c \, [l]} \right)^\dagger \rrangle \equiv \sqrt{2 k^- p_1^+} \, \sqrt{2 p_2^- p_1^+} \, \Big\langle W_{\un 2}^{d \, [k]} \, \ldots \, \left( W_{\un 1}^{c \, [l]} \right)^\dagger \Big\rangle 
\end{align} 
with $k, l = 1,2$, 
which is a generalization of \eq{double_angle_brackets} to the case of two non-eikonal interactions considered in \fig{FIG:Gadj_evol}.

Next we move on to diagrams B and B'. We now need another propagator, which one can also readily construct,
\begin{align}\label{psi_a+}
\int\limits_{-\infty}^0 dx_{2'}^- \, & 
\int\limits_0^\infty dx_2^- \, 
\contraction[2ex]
{}
{\psi}{_\alpha^i
(x_{2'}^- , \ul{x}_1) \:}
{a}
\: 
\psi_\alpha^i (x_{2'}^- , \ul{x}_1) \:
a^{+ \, d} (x_2^- , \ul{x}_0) =
\frac{1}{4 \pi^3} \, \int\limits_0^{p_2^-} d k^- \,   \int d^2 x_2 \,  \frac{x_{20}^m}{x_{20}^2} \\ 
& \times \, \left[ i \, \sqrt{\sqrt{2} k^-} \, \ln \left( \frac{1}{x_{21} \Lambda} \right) + \frac{1}{\sqrt{\sqrt{2} k^-}} \,  \frac{x_{21}^l}{x_{21}^2} \, \gamma^l \, \gamma^0 \right]_{\alpha\beta}  \notag \\
& \times \, \left[ \left( \epsilon_+^m \, {\rho} (-)_\beta - \epsilon_-^m \, {\rho} (+)_\beta \right) \, \left( W_{\un 2}^{d; i \, [1]} [\infty, - \infty] \right)^\dagger - \left( \epsilon_+^m \, {\rho} (-)_\beta + \epsilon_-^m \, {\rho} (+)_\beta \right) \, \left( W_{\un 2}^{d; i \, [2]} [\infty, - \infty] \right)^\dagger \right].  \notag 
\end{align}
Again the first term in the (now) first square brackets does not contribute to evolution. The second term yields
\begin{align}\label{BB'}
B + B' & = \frac{1}{N_c^2 -1} \, \frac{\as}{2 \pi^2} \, \sum_f \, \int\limits_0^{p_2^-} \frac{d k^-}{k^-} \, \int d^2 x_2 \, \llangle \left( U^\dagger_{\un 0} \right)^{ab} \, f^{dbc} \, \left[ - i \, W_{\un 1}^{c \, [2]} [\infty, - \infty] \, t^a \, \left( W_{\un 2}^{d \, [1]} [\infty, - \infty] \right)^\dagger \frac{{\un x}_{20}}{x_{20}^2}  \cdot \frac{{\un x}_{21}}{x_{21}^2}  \right. \notag \\ 
&  + W_{\un 1}^{c \, [1]} [\infty, - \infty] \, t^a \, \left( W_{\un 2}^{d \, [1]} [\infty, - \infty] \right)^\dagger \frac{{\un x}_{20}}{x_{20}^2}  \times \frac{{\un x}_{21}}{x_{21}^2}  - i \, W_{\un 1}^{c \, [1]} [\infty, - \infty] \, t^a \, \left( W_{\un 2}^{d \, [2]} [\infty, - \infty] \right)^\dagger \frac{{\un x}_{20}}{x_{20}^2}  \cdot \frac{{\un x}_{21}}{x_{21}^2}  \\
& \left. + W_{\un 1}^{c \, [2]} [\infty, - \infty] \, t^a \, \left( W_{\un 2}^{d \, [2]} [\infty, - \infty] \right)^\dagger \frac{{\un x}_{20}}{x_{20}^2}  \times \frac{{\un x}_{21}}{x_{21}^2}  \right] \rrangle - ({\un x}_{20} \to {\un x}_{21}) . \notag
\end{align} 
We observe that
\begin{align}
B + B' = (A + A')^* .
\end{align}

Considering diagrams C and C' next, we need the following propagator:
\begin{align}\label{psi_a+_2}
\int\limits_{-\infty}^0 dx_{2'}^- \, & 
\int\limits_0^\infty dx_2^- \, 
\contraction[2ex]
{}
{\psi}{_\alpha^i
(x_{2}^- , \ul{x}_1) \:}
{a}
\: 
\psi_\alpha^i (x_{2}^- , \ul{x}_1) \:
a^{+ \, d} (x_{2'}^- , \ul{x}_0) =
- \frac{1}{4 \pi^3} \, \int\limits_0^{p_2^-} d k^- \,   \int d^2 x_2 \,  \frac{x_{20}^m}{x_{20}^2} \\ 
& \times \, \left[ i \, \sqrt{\sqrt{2} k^-} \, \ln \left( \frac{1}{x_{21} \Lambda} \right) + \frac{1}{\sqrt{\sqrt{2} k^-}} \,  \frac{x_{21}^l}{x_{21}^2} \, \gamma^0 \, \gamma^l \right]_{\alpha\beta}  \notag \\
& \times \, \left[ \left( \epsilon_+^{m \, *} \, {\rho} (+)_\beta - \epsilon_-^{m \, *} \, {\rho} (-)_\beta \right) \, \left( W_{\un 2}^{d; i \, [1]} [- \infty, \infty] \right)^\dagger + \left( \epsilon_+^{m \, *} \, {\rho} (+)_\beta + \epsilon_-^{m \, *} \, {\rho} (-)_\beta \right) \, \left( W_{\un 2}^{d; i \, [2]} [- \infty, \infty] \right)^\dagger \right].  \notag 
\end{align}
Only the second term in the first square bracket contributes. We obtain
\begin{align}\label{CC'}
C + C' & = \frac{1}{N_c^2 -1} \, \frac{\as}{2 \pi^2} \, \sum_f \, \int\limits_0^{p_2^-} \frac{d k^-}{k^-} \, \int d^2 x_2 \, \llangle \left( U^\dagger_{\un 0} \right)^{ab} \, f^{dac} \, \left[ - i \, W_{\un 1}^{c \, [2]} [-\infty, \infty] \, t^b \, \left( W_{\un 2}^{d \, [1]} [- \infty, \infty] \right)^\dagger \frac{{\un x}_{20}}{x_{20}^2}  \cdot \frac{{\un x}_{21}}{x_{21}^2}  \right. \notag \\ 
&  + W_{\un 1}^{c \, [1]} [- \infty, \infty] \, t^b \, \left( W_{\un 2}^{d \, [1]} [- \infty, \infty] \right)^\dagger \frac{{\un x}_{20}}{x_{20}^2}  \times \frac{{\un x}_{21}}{x_{21}^2}  - i \, W_{\un 1}^{c \, [1]} [- \infty, \infty] \, t^b \, \left( W_{\un 2}^{d \, [2]} [- \infty, \infty] \right)^\dagger \frac{{\un x}_{20}}{x_{20}^2}  \cdot \frac{{\un x}_{21}}{x_{21}^2}  \\
& \left. + W_{\un 1}^{c \, [2]} [- \infty, \infty] \, t^b \, \left( W_{\un 2}^{d \, [2]} [- \infty, \infty] \right)^\dagger \frac{{\un x}_{20}}{x_{20}^2}  \times \frac{{\un x}_{21}}{x_{21}^2}  \right] \rrangle - ({\un x}_{20} \to {\un x}_{21}) . \notag
\end{align}

One can now guess that 
\begin{align}
D + D' = (C + C')^* ,
\end{align}
such that
\begin{align}\label{DD'}
D + D' & = \frac{1}{N_c^2 -1} \, \frac{\as}{2 \pi^2} \, \sum_f \, \int\limits_0^{p_2^-} \frac{d k^-}{k^-} \, \int d^2 x_2 \, \llangle \left( U^\dagger_{\un 0} \right)^{ab} \, f^{dac} \, \left[ i \, W_{\un 2}^{d \, [1]} [- \infty, \infty] \, t^b \, \left( W_{\un 1}^{c \, [2]} [-\infty, \infty] \right)^\dagger \frac{{\un x}_{20}}{x_{20}^2}  \cdot \frac{{\un x}_{21}}{x_{21}^2}  \right. \notag \\ 
&  + W_{\un 2}^{d \, [1]} [- \infty, \infty] \, t^b \, \left( W_{\un 1}^{c \, [1]} [- \infty, \infty] \right)^\dagger \frac{{\un x}_{20}}{x_{20}^2}  \times \frac{{\un x}_{21}}{x_{21}^2}  +  i \,  W_{\un 2}^{d \, [2]} [- \infty, \infty] \, t^b \, \left( W_{\un 1}^{c \, [1]} [- \infty, \infty] \right)^\dagger \frac{{\un x}_{20}}{x_{20}^2}  \cdot \frac{{\un x}_{21}}{x_{21}^2}  \\
& \left. + W_{\un 2}^{d \, [2]} [- \infty, \infty] \, t^b \, \left( W_{\un 1}^{c \, [2]} [- \infty, \infty]  \right)^\dagger \frac{{\un x}_{20}}{x_{20}^2}  \times \frac{{\un x}_{21}}{x_{21}^2}  \right] \rrangle - ({\un x}_{20} \to {\un x}_{21}) . \notag
\end{align} 
This is what one indeed obtains by a direct calculation using the propagator
\begin{align}\label{psibar_a+_2}
\int\limits_{-\infty}^0 dx_{2'}^- \, & 
\int\limits_0^\infty dx_2^- \, 
\contraction[2ex]
{}
{{\bar \psi}}{_\alpha^i
(x_{2}^- , \ul{x}_1) \:}
{a}
\: 
{\bar \psi}_\alpha^i (x_{2}^- , \ul{x}_1) \:
a^{+ \, d} (x_{2'}^- , \ul{x}_0) =
\frac{1}{4 \pi^3} \, \int\limits_0^{p_2^-} d k^- \,   \int d^2 x_2 \,  \frac{x_{20}^m}{x_{20}^2} \\ 
& \times \, \left[ \left( - \epsilon_+^{m \, *} \, {\bar \rho} (-)_\beta + \epsilon_-^{m \, *} \, {\bar \rho} (+)_\beta \right) \, W_{\un 2}^{d; i \, [1]} [- \infty, \infty] + \left( \epsilon_+^{m \, *} \, {\bar \rho} (-)_\beta + \epsilon_-^{m \, *} \, {\bar \rho} (+)_\beta \right) \, W_{\un 2}^{d; i \, [2]} [- \infty, \infty] \right]  \notag \\
& \times \, \left[ i \, \sqrt{\sqrt{2} k^-} \, \ln \left( \frac{1}{x_{21} \Lambda} \right) + \frac{1}{\sqrt{\sqrt{2} k^-}} \,  \frac{x_{21}^l}{x_{21}^2} \, \gamma^l \, \gamma^0 \right]_{\beta\alpha} . \notag
\end{align}

The sum of the diagrams from \fig{FIG:Gadj_evol} is obtained by adding the contributions found in Eqs.~\eqref{AA'}, \eqref{BB'}, \eqref{CC'}, and \eqref{DD'}.


\subsubsection{Simplifications in the DLA and at large $N_c \& N_f$}

Simplifying Eqs.~\eqref{AA'}, \eqref{BB'}, \eqref{CC'}, and \eqref{DD'} in the DLA, we neglect the terms with the cross products as not double-logarithmic, and write
\begin{align}\label{total}
& A+A'+B+B'+C+C'+D+D' \bigg|_{DLA} = - \frac{1}{N_c^2 -1} \, \frac{\as}{2 \pi} \, \sum_f \, \int\limits_0^{p_2^-} \frac{d k^-}{k^-} \, \int\limits^{x_{10}^2} \frac{d x_{21}^2}{x_{21}^2} \\ 
& \times \, \llangle \left( U^\dagger_{\un 0} \right)^{ab} \, f^{dbc} \, \left[ i \, W_{\un 2}^{d \, [1]} [\infty, - \infty] \, t^a \, \left( W_{\un 1}^{c \, [2]} [\infty, - \infty] \right)^\dagger + i \, W_{\un 2}^{d \, [2]} [\infty, - \infty] \, t^a \, \left( W_{\un 1}^{c \, [1]} [\infty, - \infty] \right)^\dagger  + \mbox{c.c.} \right] \notag \\ 
& + \left( U^\dagger_{\un 0} \right)^{ab} \, f^{dac} \, \left[ i \, W_{\un 2}^{d \, [1]} [- \infty, \infty] \, t^b \, \left( W_{\un 1}^{c \, [2]} [-\infty, \infty] \right)^\dagger +  i \,  W_{\un 2}^{d \, [2]} [- \infty, \infty] \, t^b \, \left( W_{\un 1}^{c \, [1]} [- \infty, \infty] \right)^\dagger + \mbox{c.c.}  \right] \rrangle . \notag 
\end{align}
Here c.c. stands for complex conjugate.

Using the definitions \eqref{W12} one can readily show that the expression in the first square brackets of \eq{total} is
\begin{align}\label{sq1}
&  i \, W_{\un 2}^{d \, [1]} [\infty, - \infty] \, t^a \, \left( W_{\un 1}^{c \, [2]} [\infty, - \infty] \right)^\dagger + i \, W_{\un 2}^{d \, [2]} [\infty, - \infty] \, t^a \, \left( W_{\un 1}^{c \, [1]} [\infty, - \infty] \right)^\dagger  + \mbox{c.c.}  = \\ 
 & - i \frac{g^2}{4 \sqrt{k^- \, p_2^-}} \, \int\limits_{-\infty}^\infty dy^- \, \int\limits_{-\infty}^\infty dz^- \, U_{\un 2}^{dd'} [\infty, y^-] \, U_{\un 1}^{cc'} [\infty, z^-] \, {\bar \psi} (y^-, {\un x}_2 ) \, t^{d'} \, \gamma^+ \, \gamma^5 \, V_{\un 2} [y^-, - \infty] \, t^a \, V_{\un 1} [-\infty, z^-] \, t^{c'} \, \psi (z^-, {\un x}_1) + \mbox{c.c.} . \notag 
\end{align}
The expression in the second square brackets of \eq{total} is obtained from the above by simply interchanging the $\infty \leftrightarrow - \infty$ limits: 
\begin{align}\label{sq2}
&  i \, W_{\un 2}^{d \, [1]} [- \infty, \infty] \, t^b \, \left( W_{\un 1}^{c \, [2]} [-\infty, \infty] \right)^\dagger +  i \,  W_{\un 2}^{d \, [2]} [- \infty, \infty] \, t^b \, \left( W_{\un 1}^{c \, [1]} [- \infty, \infty] \right)^\dagger  + \mbox{c.c.}  = \\ 
 & - i \frac{g^2}{4 \sqrt{k^- \, p_2^-}} \, \int\limits_{-\infty}^\infty dy^- \, \int\limits_{-\infty}^\infty dz^- \, U_{\un 2}^{dd'} [-\infty, y^-] \, U_{\un 1}^{cc'} [-\infty, z^-] \, {\bar \psi} (y^-, {\un x}_2 ) \, t^{d'} \, \gamma^+ \, \gamma^5 \, V_{\un 2} [y^-, \infty] \, t^b \, V_{\un 1} [\infty, z^-] \, t^{c'} \, \psi (z^-, {\un x}_1) + \mbox{c.c.} . \notag 
\end{align}

Employing \eq{sq1} along with
\begin{align}\label{Uab}
U^{ab} = 2 \, \tr [t^a \, V \, t^b \, V^\dagger],
\end{align}
applying the Fierz identity several times while neglecting the $N_c$-suppressed terms (either explicitly, or by neglecting the terms with fewer color traces, since each color trace brings in a factor of $N_c$), we arrive at
\begin{align}\label{sq12}
& \llangle \left( U^\dagger_{\un 0} \right)^{ab} \, f^{dbc} \, \left[ i \, W_{\un 2}^{d \, [1]} [\infty, - \infty] \, t^a \, \left( W_{\un 1}^{c \, [2]} [\infty, - \infty] \right)^\dagger + i \, W_{\un 2}^{d \, [2]} [\infty, - \infty] \, t^a \, \left( W_{\un 1}^{c \, [1]} [\infty, - \infty] \right)^\dagger  + \mbox{c.c.} \right] \rrangle \\
& =  \int\limits_{-\infty}^\infty dy^- \int\limits_{-\infty}^\infty dz^- \,  \left\langle \tr \left[ V_{\un 2} \, V^\dagger_{\un 0} \right] \right\rangle \, \left\langle \tr \left[V_{\un 0} \, V^\dagger_{\un 1} \right] \right\rangle \,  \llangle \frac{g^2}{16 \sqrt{k^- \, p_2^-}} \, {\bar \psi} (y^-, {\un x}_2 ) \, \gamma^+ \, \gamma^5 \, V_{\un 2} [y^-, \infty] \,  V_{\un 1} [\infty, z^-] \, \psi (z^-, {\un x}_1) \rrangle + \mbox{c.c.} . \notag
\end{align}
Similarly, for the expression \eqref{sq2}, we get
\begin{align}\label{sq22}
& \llangle \left( U^\dagger_{\un 0} \right)^{ab} \, f^{dac} \, \left[ i \, W_{\un 2}^{d \, [1]} [- \infty, \infty] \, t^b \, \left( W_{\un 1}^{c \, [2]} [-\infty, \infty] \right)^\dagger +  i \,  W_{\un 2}^{d \, [2]} [- \infty, \infty] \, t^b \, \left( W_{\un 1}^{c \, [1]} [- \infty, \infty] \right)^\dagger + \mbox{c.c.}  \right] \rrangle \\
& =  \int\limits_{-\infty}^\infty dy^- \int\limits_{-\infty}^\infty dz^- \,  \left\langle \tr \left[ V^\dagger_{\un 2} \, V_{\un 0} \right] \right\rangle \, \left\langle \tr \left[V^\dagger_{\un 0} \, V_{\un 1} \right] \right\rangle \,  \llangle \frac{g^2}{16 \sqrt{k^- \, p_2^-}} \, {\bar \psi} (y^-, {\un x}_2 ) \, \gamma^+ \, \gamma^5 \, V_{\un 2} [y^-, - \infty] \,  V_{\un 1} [- \infty, z^-] \, \psi (z^-, {\un x}_1) \rrangle + \mbox{c.c.} . \notag
\end{align}

Employing Eqs.~\eqref{sq12} and \eqref{sq22} in \eq{total} we obtain (at large $N_c \& N_f$)
\begin{align}\label{total2}
& A+A'+B+B'+C+C'+D+D' \bigg|_{DLA} = - \frac{\as}{2 \pi} \, \sum_f \, \int\limits_0^{p_2^-} \frac{d k^-}{k^-} \, \int\limits^{x_{10}^2} \frac{d x_{21}^2}{x_{21}^2} \, S_{20} (2 k^- p_1^+)  \, S_{10} (2 k^- p_1^+) \\ 
& \times \,   \llangle \frac{g^2}{16 \sqrt{k^- \, p_2^-}} \, \int\limits_{-\infty}^\infty dy^- \int\limits_{-\infty}^\infty dz^- \, \bigg[ {\bar \psi} (y^-, {\un x}_2 ) \, \gamma^+ \, \gamma^5 \, V_{\un 2} [y^-, \infty] \,  V_{\un 1} [\infty, z^-] \, \psi (z^-, {\un x}_1) \notag \\ 
& + {\bar \psi} (y^-, {\un x}_2 ) \, \gamma^+ \, \gamma^5 \, V_{\un 2} [y^-, - \infty] \,  V_{\un 1} [- \infty, z^-] \, \psi (z^-, {\un x}_1) + \mbox{c.c.} \bigg] \rrangle  . \notag 
\end{align}

By analogy to Eqs.~\eqref{G_dfn_NcNf} and \eqref{Wpol_dfn} we define a new object (which one cannot call a dipole scattering amplitude, since it is more like a quark helicity TMD, with one term containing a future-pointing (SIDIS) staple, and another term containing a past-pointing (Drell-Yan (DY)) staple),\footnote{Note that our normalization here does not have the $1/N_c$ factor of Eqs.~\eqref{QGi} or the $1/(N_c^2 -1)$ factor of \eq{G_dfn_NcNf}. In that respect, our ${\widetilde Q}_{12}$ is normalized similar to the dipole amplitude $Q$ from \eq{Qdef}: the latter, in the quark sector, has a sum over quark fields colors, and not the average. At the same time, ${\widetilde Q}_{12}$ is different from the amplitude $\widetilde G$ in \eq{G_dfn_NcNf}, in which the quark colors (in the quark sector) are averaged over.}
\begin{align}\label{Qtilde}
{\widetilde Q}_{12} (s) \equiv & \  \llangle \frac{g^2}{16 \sqrt{k^- \, p_2^-}} \, \int\limits_{-\infty}^\infty dy^- \int\limits_{-\infty}^\infty dz^- \, \bigg[ {\bar \psi} (y^-, {\un x}_2 ) \, \left( \frac{1}{2} \,  \gamma^+ \, \gamma^5 \right) \, V_{\un 2} [y^-, \infty] \,  V_{\un 1} [\infty, z^-] \, \psi (z^-, {\un x}_1)  \\ 
& + {\bar \psi} (y^-, {\un x}_2 ) \, \left( \frac{1}{2} \,  \gamma^+ \, \gamma^5 \right) \, V_{\un 2} [y^-, - \infty] \,  V_{\un 1} [- \infty, z^-] \, \psi (z^-, {\un x}_1) + \mbox{c.c.} \bigg] \rrangle (s) . \notag
\end{align}
This gives
\begin{align}\label{total3}
A+A'+B+B'+C+C'+D+D' \bigg|_{DLA} = - \frac{\as}{\pi} \, \sum_f \, \int\limits_0^{p_2^-} \frac{d k^-}{k^-} \, \int\limits^{x_{10}^2} \frac{d x_{21}^2}{x_{21}^2} \, S_{20} (2 k^- p_1^+)  \, S_{10} (2 k^- p_1^+) \, {\widetilde Q}_{12}  (2 k^- p_1^+) . 
\end{align}

In DLA we put $S=1$, since the unpolarized evolution is single-logarithmic. In addition, for simplicity, we replace $\sum_f \to N_f$, assuming that all flavors contribute equally (which, indeed, is not the case in phenomenology \cite{Adamiak:2023yhz}). Writing $k^- = z' p_2^-$ and generalizing the calculation to the case of the parent dipole not being the original projectile, and rather being produced by previous steps of the evolution and described by the minus momentum fraction $z > z'$ we obtain, after imposing the light-cone lifetime constraints \cite{Kovchegov:2015pbl, Kovchegov:2018znm, Cougoulic:2019aja},
\begin{align}\label{total4}
A+A'+B+B'+C+C'+D+D' \bigg|_{DLA} = - \frac{\as \, N_f}{\pi} \, \int\limits_\frac{\Lambda^2}{s}^{z} \frac{d z'}{z'} \, \int\limits^{x_{10}^2}_{1/(z' s)} \frac{d x_{21}^2}{x_{21}^2} \, {\widetilde Q}_{12}  (z' \, s) . 
\end{align}

Note that \eq{total4} gives the contribution of the transition operators to the evolution of $G^\textrm{adj}_{10} (zs)$. However, the large-$N_c \& N_f$ helicity evolution equations from \cite{Cougoulic:2022gbk} contain ${\widetilde G}_{10} (zs)$  instead of $G^\textrm{adj}_{10} (zs)$. Employing \eq{Gadj_Gtilde} we write
\begin{align}\label{Gadj_Gtilde2}
G^\textrm{adj}_{10} (zs) = 4 \, S_{10} (zs)  \, {\widetilde G}_{10} (zs) \approx 4 \, {\widetilde G}_{10} (zs),
\end{align}
with the last step valid in DLA, where $S_{10} =1$. We see that while \eq{total4} is the contribution to the evolution of $G^\textrm{adj}_{10} (zs)$, the contribution to the evolution of ${\widetilde G}_{10} (zs)$ in DLA is 4 times smaller.
Combining Eqs.~\eqref{Gadj_Gtilde2} and \eqref{total4} we arrive at the following contribution of the transition operators to the evolution of the dipole amplitude ${\widetilde G}_{10} (zs)$,
\begin{align}\label{total5}
{\widetilde G}_{10} (zs) \supset - \frac{\as \, N_f}{4 \pi} \, \int\limits_\frac{\Lambda^2}{s}^{z} \frac{d z'}{z'} \, \int\limits^{x_{10}^2}_{1/(z' s)} \frac{d x_{21}^2}{x_{21}^2} \ {\widetilde Q}_{12}  (z' \, s) .
\end{align}
We conclude that the transition operators do contribute to the evolution of ${\widetilde G}_{10} (zs)$. The evolution equation for ${\widetilde G}_{10} (zs)$ found in \cite{Cougoulic:2022gbk} needs to be augmented by including the new additive term from \eq{total5} into its right-hand side. The evolution equation for the corresponding neighbor dipole amplitude $\widetilde \Gamma$ needs to be modified in a similar way (with a slightly different upper integration limit on the dipole sizes). 

To complete the emerging new set of large-$N_c \& N_f$ helicity evolution equations we now need to derive an evolution equation for $\widetilde Q$. Since $\widetilde Q$ enters the evolution of $\widetilde G$ in \eq{total5} in the $x_{21} \ll x_{10}$ regime, it always describes the evolution of the smaller of the two daughter ``dipoles" ($x_{21} \ll x_{20} \approx x_{10}$), and the lifetime of the subsequent evolution in ${\widetilde Q}_{12}  (z' \, s)$ is bounded by $\sim z' \, x_{21}^2$ from above, a limit dependent on the transverse size of the $x_{21}$ ``dipole" and independent of $x_{10}$. We thus conclude that, unlike for $Q, G_2$ and ${\widetilde G}$, no ``neighbor dipole amplitude" is needed for $\widetilde Q$. We will only need to construct the evolution for the $\widetilde Q$ proper: this is what we will do next.


\subsection{Evolution for ${\widetilde Q}$}
\label{sec:Qtilde_evol}


\subsubsection{The DLA evolution equation for ${\widetilde Q}$}\label{subsec:Qtilde_evol_I}

\begin{figure}[ht]
\centering
\includegraphics[width= 0.95 \textwidth]{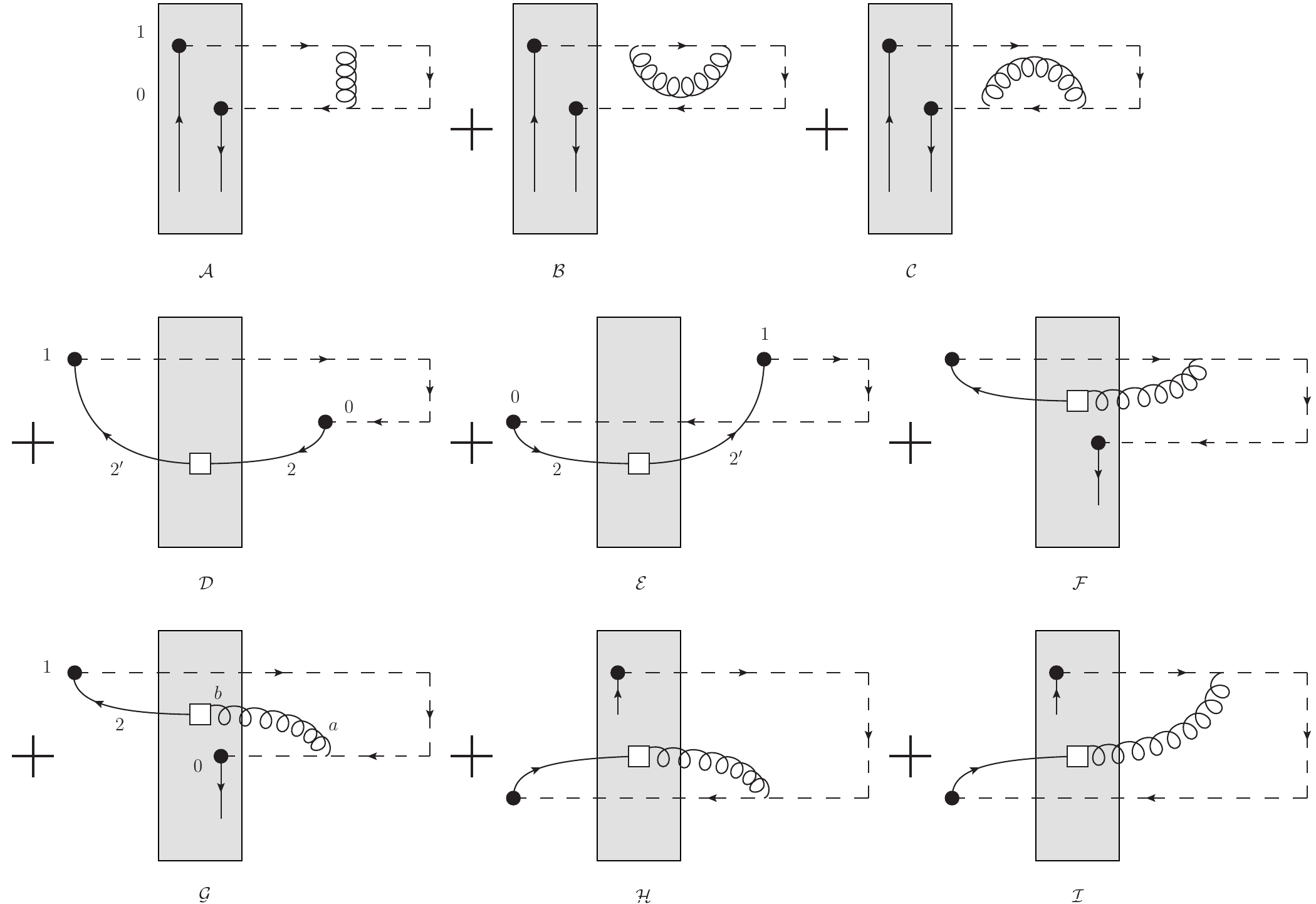}
\caption{Diagrams contributing to the DLA evolution of ${\widetilde Q}_{10} (zs)$. The shaded rectangle is, again, the shock wave. The white square denotes the non-eikonal interactions, either due to the transition operators \eqref{Ws} or due to the sub-eikonal operators introduced in Sec.~\ref{sec:polWlines}. Vertical straight lines denote the quark and anti-quark fields of the shock wave. The dashed line denotes the (future-pointing) Wilson-line staple.}
\label{FIG:Qtilde_evol}
\end{figure}

Diagrams contributing to the evolution of ${\widetilde Q}_{10} (zs)$ are shown in \fig{FIG:Qtilde_evol}, with only the future-pointing Wilson line staple shown: the diagrams for the past-pointing staple are constructed and calculated by analogy. Since ${\widetilde Q}_{10} (zs)$ operator is similar to the quark helicity TMD (SIDIS plus DY), the diagrams in \fig{FIG:Qtilde_evol} closely parallel those in Fig.~2 of \cite{Kovchegov:2018znm}. The analysis carried out in \cite{Kovchegov:2018znm} suggests that only diagrams $\cal D$ and $\cal E$ in our \fig{FIG:Qtilde_evol} should contribute in the DLA: this is what we will indeed confirm below. 

Diagrams $\cal A$, $\cal B$ and $\cal C$ from \fig{FIG:Qtilde_evol} are just the eikonal corrections for a ``half-dipole" amplitude \cite{Kovchegov:2015zha}. Their contribution is known from the unpolarized small-$x$ evolution \cite{Mueller:1994rr,Mueller:1994jq,Mueller:1995gb,Balitsky:1995ub,Balitsky:1998ya,Kovchegov:1999yj,Kovchegov:1999ua,Jalilian-Marian:1997dw,Jalilian-Marian:1997gr,Weigert:2000gi,Iancu:2001ad,Iancu:2000hn,Ferreiro:2001qy}:
\begin{align}\label{ABC}
{\cal A} + {\cal B} + {\cal C} = - \frac{\as \, N_c}{4 \pi^2} \, \int\limits_\frac{\Lambda^2}{s}^{z} \frac{d z'}{z'} \, \int d^2 x_2 \, \frac{x_{10}^2}{x_{21}^2 \, x_{20}^2} \, {\widetilde Q}_{10} (z' s) \approx - \frac{\as \, N_c}{4 \pi} \, \int\limits_\frac{\Lambda^2}{s}^{z} \frac{d z'}{z'} \, \left[ \int\limits^{x_{10}^2}_{1/(z' s)} \frac{d x_{21}^2}{x_{21}^2} + \int\limits^{x_{10}^2}_{1/(z' s)} \frac{d x_{20}^2}{x_{20}^2} \right] \, {\widetilde Q}_{10} (z' s) .
\end{align}
Here we include the past-pointing (DY) Wilson line staple's contribution as well to assemble the entire $\widetilde Q$ object. The last step in \eq{ABC} is the simplification in the DLA limit. 

Next, let us calculate diagrams $\cal F$ and $\cal G$ in \fig{FIG:Qtilde_evol}. We will need the propagator \eqref{psi_a+} again. After some algebra we end up with
\begin{align}\label{FG}
& {\cal F} + {\cal G} = - 2 P^+ \, \frac{g^4}{256 \, \pi^3} \int\limits_0^{p_2^-} \frac{d k^-}{k^-} \, \int\limits_{-\infty}^\infty dy^- \int\limits_{-\infty}^\infty dz^- \, \int d^2 x_2 \, \bigg\langle {\bar \psi} (y^-, {\un x}_0) \, V_{\un 0} [y^-, \infty] \, t^a \, V_{\un 1} \, U_{\un 2}^{ab} [\infty, z^-] \, V_{\un 2} [-\infty , z^-]  \\
& \times \, \left[ \frac{{\un x}_{20}}{x_{20}^2}  \cdot \frac{{\un x}_{21}}{x_{21}^2}  \, \gamma^+ \, \gamma^5 - i \, \frac{{\un x}_{20}}{x_{20}^2}  \times \frac{{\un x}_{21}}{x_{21}^2} \, \gamma^+  \right] \, t^b \, \psi (z^-, {\un x}_2) \bigg\rangle - ({\un x}_{20} \to {\un x}_{21}) + \mbox{c.c.} . \notag
\end{align}
Dropping the cross-product term as non-DLA and simplifying the dot-product term in DLA we obtain
\begin{align}\label{FG2}
& {\cal F} + {\cal G} = 2 P^+ \, \frac{g^4}{256 \, \pi^2} \int\limits_0^{p_2^-} \frac{d k^-}{k^-} \, \int\limits_{-\infty}^\infty dy^- \int\limits_{-\infty}^\infty dz^- \, \int\limits^{x_{10}^2} \frac{d x^2_{21}}{x^2_{21}} \, \bigg\langle {\bar \psi} (y^-, {\un x}_0) \, V_{\un 0} [y^-, \infty] \, t^a \, V_{\un 1} \, U_{\un 2}^{ab} [\infty, z^-] \, V_{\un 2} [-\infty , z^-]  \\
& \times \,  \gamma^+ \, \gamma^5 \, t^b \, \psi (z^-, {\un x}_2) \bigg\rangle + \mbox{c.c.}  . \notag
\end{align}
Taking the large-$N_c$ limit with the help of \eq{Uab} and the Fierz identity yields
\begin{align}\label{FG3}
& {\cal F} + {\cal G} = \frac{\as \, N_c}{4 \, \pi} \int\limits_\frac{\Lambda^2}{s}^{z} \frac{d z'}{z'} \,  \int\limits^{x_{10}^2}_{1/(z' s)} \frac{d x^2_{21}}{x^2_{21}} \, S_{21} (z' s)  \, {\widetilde Q}_{20} (z' s) \approx \frac{\as \, N_c}{4 \, \pi} \int\limits_\frac{\Lambda^2}{s}^{z} \frac{d z'}{z'} \,  \int\limits^{x_{10}^2}_{1/(z' s)} \frac{d x^2_{21}}{x^2_{21}} \, {\widetilde Q}_{20} (z' s) ,
\end{align}
where in the last step we have put $S = 1$, which is valid in DLA. Again we have included the past-pointing Wilson line staple's contribution by simply interchanging $\infty \leftrightarrow -\infty$ in the above.

Adding \eq{FG3} to the first term on the right of \eq{ABC} we get 
\begin{align}\label{Q-Q}
\frac{\as \, N_c}{4 \, \pi} \int\limits_\frac{\Lambda^2}{s}^{z} \frac{d z'}{z'} \,  \int\limits^{x_{10}^2}_{1/(z' s)} \frac{d x^2_{21}}{x^2_{21}} \, \left[ {\widetilde Q}_{20} (z' s) - {\widetilde Q}_{10} (z' s) \right] \approx 0.
\end{align}
This result, while in general non-zero, is not DLA. Therefore, it can be neglected in our DLA calculation.

It is now straightforward to conclude that diagrams $\cal H$ and $\cal I$ from \fig{FIG:Qtilde_evol} cancel the second term on the right of \eq{ABC} with the same DLA accuracy. 

Therefore, only the diagrams $\cal D$ and $\cal E$ from \fig{FIG:Qtilde_evol}, along with their past-pointing Wilson line staple counterparts, contribute to the evolution of the dipole amplitude ${\widetilde Q}$, corresponding to the diagram B and its complex conjugate in Fig.~2 of \cite{Kovchegov:2018znm}. We have confirmed the analysis carried out in \cite{Kovchegov:2018znm}.

Lastly, we need to calculate the diagrams $\cal D$ and $\cal E$. For the diagram $\cal E$ we will employ the propagator
\begin{align}\label{q_propagator1}
& \int\limits_{-\infty}^0 dx_{2}^- \, 
\int\limits_0^\infty dx_{2'}^- \, 
\contraction
{}
{\bar\psi^i_\alpha}
{(x_2^- , {\un x}_0) \:}
{psi^j_\beta}
\bar\psi^i_\alpha (x_2^- , {\un x}_0) \: 
\psi^j_\beta (x_{2'}^- , {\un x}_1) 
= \frac{1}{\pi} \sum_{\sigma}  \int\limits_0^{p_2^-} d k^- \, k^-  \int d^2 x_2 \, d^2 x_{2'} \, \left[ \int
 \frac{d^2 k_{2'}}{(2\pi)^2} \, 
e^{-i \ul{k}_{2'} \cdot \ul{x}_{2'1}} \,  \frac{1}{{\un k}_{2'}^2 }  \, \left( u_{\sigma} (k_{2'}) \right)_\beta \right] 
\\ & \times
\left( \sigma \, V_{\un 2}^{\textrm{pol} [1]} \, \delta^2 ({\un x}_{22'}) + V_{{\ul 2}', {\un 2}}^{\textrm{pol} [2]} \right)^{ji} \, \left[ \int \frac{d^2 k_2}{(2\pi)^2} \, e^{i \ul{k}_2 \cdot \ul{x}_{20} }  \, \frac{1}{{\un k}_2^2} \, \left( {\bar u}_\sigma (k_2) \right)_\alpha \right] , \notag
\end{align}
which includes a minus sign arising due to Wick contractions. 

For the diagram $\cal D$ we need the following propagator,
\begin{align}\label{q_propagator2}
& \int\limits_{-\infty}^0 dx_{2'}^- \, 
\int\limits_0^\infty dx_2^- \, 
\contraction
{}
{\bar\psi^i_\alpha}
{(x_2^- , {\un x}_0) \:}
{psi^j_\beta}
\bar\psi^i_\alpha (x_2^- , {\un x}_0) \: 
\psi^j_\beta (x_{2'}^- , {\un x}_1) 
= \frac{1}{\pi} \sum_{\sigma}  \int\limits_0^{p_2^-} d k^- \, k^-  \int d^2 x_2 \, d^2 x_{2'} \, \left[ \int
 \frac{d^2 k_{2'}}{(2\pi)^2} \, 
e^{i \ul{k}_{2'} \cdot \ul{x}_{2'1}} \,  \frac{1}{{\un k}_{2'}^2 }  \, \left( v_{\sigma} (k_{2'}) \right)_\beta \right] 
\\ & \times
\left( - \sigma \, V_{\un 2}^{\textrm{pol} [1] \, \dagger} \, \delta^2 ({\un x}_{22'}) + V_{{\ul 2}, {\un 2}'}^{\textrm{pol} [2] \, \dagger} \right)^{ji} \, \left[ \int \frac{d^2 k_2}{(2\pi)^2} \, e^{- i \ul{k}_2 \cdot \ul{x}_{20} }  \, \frac{1}{{\un k}_2^2} \, \left( {\bar v}_\sigma (k_2) \right)_\alpha \right] . \notag
\end{align}
Note the overall sign difference when comparing our \eq{q_propagator2} to Eq.~(92) in \cite{Cougoulic:2022gbk}: this is due to the Wick contraction included in our \eq{q_propagator2}. There is also a sign difference in front of the $\sigma$ term in \eq{q_propagator2} compared to Eq.~(92) in \cite{Cougoulic:2022gbk}. The typo does not propagate in Sec.~IV of \cite{Cougoulic:2022gbk}.

Using these propagators, we arrive at
\begin{align}\label{DE_app1}
{\cal D} + {\cal E} = \frac{\as}{4 \, \pi^2} \, \int\limits_\frac{\Lambda^2}{s}^{z} \frac{d z'}{z'} \, \left\{ - \int d^2 x_2 \, \frac{{\un x}_{20}}{x_{20}^2}  \cdot \frac{{\un x}_{21}}{x_{21}^2} \llangle \tr \left[ V_{\un 1} \, V_{\un 2}^{\textrm{pol} [1] \, \dagger} \right] \rrangle + i \int d^2 x_2 \, d^2 x_{2'} \, \frac{{\un x}_{2'1}}{x_{2'1}^2} \times \frac{{\un x}_{20}}{x_{20}^2}  \llangle \tr \left[ V_{\un 1} \, V_{{\ul 2}, {\un 2}'}^{\textrm{pol} [2] \, \dagger} \right] \rrangle  + \mbox{c.c.} \right\} 
\end{align}
where we have inserted a factor of $2$ to account for the past-pointing Wilson lines. Anticipating the impact-parameter integration, we perform a calculation parallel to that on page 12 of \cite{Cougoulic:2022gbk}, obtaining
\begin{align}
{\cal D} + {\cal E} = & \  - \frac{\as}{4 \, \pi^2} \, \int\limits_\frac{\Lambda^2}{s}^{z} \frac{d z'}{z'} \, \int d^2 x_2 \,  \left\{ \frac{{\un x}_{20}}{x_{20}^2}  \cdot \frac{{\un x}_{21}}{x_{21}^2} \llangle \tr \left[ V_{\un 1} \, V_{\un 2}^{\textrm{pol} [1] \, \dagger} \right] \rrangle \right. \\
& - \left.  \left[ - \epsilon^{ik} \, \frac{x_{20}^k + x_{21}^k}{x_{21}^2 \, x_{20}^2} + 2 \frac{{\un x}_{21} \times {\un x}_{20}}{x_{21}^2 \, x_{20}^2}  \left( \frac{x_{21}^i}{x_{21}^2} -  \frac{x_{20}^i}{x_{20}^2} \right) \right]  \llangle \tr \left[ V_{\un 1} \, V_{{\ul 2}}^{i \, \textrm{G} [2] \, \dagger} \right] \rrangle  + \mbox{c.c.} \right\} , \notag
\end{align}
or, equivalently, 
\begin{align}\label{DE}
{\cal D} + {\cal E} = & \  - \frac{\as \, N_c}{2 \, \pi^2} \, \int\limits_\frac{\Lambda^2}{s}^{z} \frac{d z'}{z'} \, \int d^2 x_2 \,  \left\{ \frac{{\un x}_{20}}{x_{20}^2}  \cdot \frac{{\un x}_{21}}{x_{21}^2} \, Q_{21} (z' s) \right. \\
& - \left.  \left[ - \epsilon^{ik} \, \frac{x_{20}^k + x_{21}^k}{x_{21}^2 \, x_{20}^2} + 2 \frac{{\un x}_{21} \times {\un x}_{20}}{x_{21}^2 \, x_{20}^2}  \left( \frac{x_{21}^i}{x_{21}^2} -  \frac{x_{20}^i}{x_{20}^2} \right) \right]  \, G_{21}^i (z' s) \right\} . \notag
\end{align}

Integrating \eq{DE} over the impact parameters and extracting the DLA contribution we get
\begin{align}\label{DE_app2}
\int d^2 \left( \frac{x_0 + x_1}{2} \right) \, [{\cal D} + {\cal E} ] = - \frac{\as \, N_c}{2 \, \pi} \, \int\limits_\frac{\Lambda^2}{s}^{z} \frac{d z'}{z'} \, \int\limits_{\max \left[ x_{10}^2, \frac{1}{z' s} \right]}^{\min \left[ \frac{z}{z'} x_{10}^2, \frac{1}{\Lambda^2} \right]} \frac{d x^2_{21}}{x^2_{21}} \, \left[ Q (x_{21}^2, z' s) + 2 \, G_2 (x_{21}^2, z' s) \right] ,
\end{align}
with the $x_{21}^2$-integration limits resulting from the structure of the kernel in \eq{DE}, lifetime limits, and from $1/\Lambda$ being the upper cutoff on the dipole sizes (an IR cutoff).

The evolution equation for the impact-parameter integrated $\widetilde Q$, 
\begin{align}
{\widetilde Q} (x^2_{10}, z s) \equiv \int d^2 \left( \frac{x_0 + x_1}{2} \right) \, {\widetilde Q}_{10} (z s) ,
\end{align}
is
\begin{align}\label{eq:Qevol}
{\widetilde Q} (x^2_{10}, z s) =  {\widetilde Q}^{(0)} (x^2_{10}, z s) - \frac{\as \, N_c}{2 \, \pi} \, \int\limits_\frac{\Lambda^2}{s}^{z} \frac{d z'}{z'} \, \int\limits_{\max \left[ x_{10}^2, \frac{1}{z' s} \right]}^{\min \left[ \frac{z}{z'} x_{10}^2, \frac{1}{\Lambda^2} \right]} \frac{d x^2_{21}}{x^2_{21}} \, \left[ Q (x_{21}^2, z' s) + 2 \, G_2 (x_{21}^2, z' s) \right] .
\end{align}
Here ${\widetilde Q}^{(0)}$ denotes the inhomogeneous term in the integral equation, resulting from the initial condition for the evolution of $\widetilde Q$. 

We have completed augmenting the large-$N_c \& N_f$ evolution equations from \cite{Cougoulic:2022gbk} by the contributions of the transition operators \eqref{Ws}: in addition to the correction to the evolution of $\widetilde G$ and $\widetilde \Gamma$ resulting from the additional term in \eq{total5}, we also have the evolution equation \eqref{eq:Qevol} for the new operator $\widetilde Q$. Before we summarize the new corrected large-$N_c \& N_f$ evolution equations, let us make a few observations.


\subsubsection{A relation between ${\widetilde Q}$ and the quark helicity TMD and PDF}

\label{sec:Qtilde_relations}

As we have mentioned above, the operator definition of $\widetilde Q$ in \eq{Qtilde} resembles that for the quark helicity TMD (SIDIS plus DY). Let us try to better quantify this similarity. 

We start with Eq.~(8) of \cite{Kovchegov:2018znm} for the SIDIS quark helicity TMD, in which we un-insert the complete set of states $X$:
\begin{align}\label{TMD11}
g_{1L}^{S, \, \textrm{SIDIS}} (x, k_T^2) = 2 \, \frac{2 p^+}{(2\pi)^3} \:  \int d^{2} \zeta \, d \zeta^- \, d^{2} \xi \, d \xi^-
\, e^{- i {\un k} \cdot ({\un \zeta} - {\un \xi})} 
\left\langle \bar\psi (\xi) \, \left( \thalf \gamma^+ \gamma^5 \right) \, V_{\ul \xi} [\xi^-, \infty] \,
 V_{\ul \zeta} [\infty , \zeta^-] \, \psi (\zeta) \right\rangle .
\end{align}
We have also inserted an overall factor of 2 to account for the anti-quark contribution and approximated $k \cdot (\zeta - \xi) \approx - {\un k} \cdot ({\un \zeta} - {\un \xi})$ in the exponent, which is valid at small $x$. Fourier-transforming \eq{TMD11} we arrive at
\begin{align}\label{TMD12}
\int d^2 k_\perp \, e^{i {\un k} \cdot {\un x}_{10}} \, g_{1L}^{S, \, \textrm{SIDIS}} (x, k_T^2) = \frac{2 p^+}{\pi} \,  \int d \zeta^- \, d \xi^- \, d^{2} \xi \,  \left\langle \bar\psi (\xi) \, \left( \thalf \gamma^+ \gamma^5 \right) \, V_{\ul \xi} [\xi^-, \infty] \, V_{{\ul \xi} + {\un x}_{10}} [\infty , \zeta^-] \, \psi (\xi + x_{10}) \right\rangle .
\end{align}
Similarly, for the DY quark helicity TMD, we write
\begin{align}\label{TMD_DY}
\int d^2 k_\perp \, e^{i {\un k} \cdot {\un x}_{10}} \, g_{1L}^{S, \, \textrm{DY}} (x, k_T^2) = \frac{2 p^+}{\pi} \,  \int d \zeta^- \, d \xi^- \, d^{2} \xi \,  \left\langle \bar\psi (\xi) \, \left( \thalf \gamma^+ \gamma^5 \right) \, V_{\ul \xi} [\xi^-, -\infty] \, V_{{\ul \xi} + {\un x}_{10}} [-\infty , \zeta^-] \, \psi (\xi + x_{10}) \right\rangle .
\end{align}

Comparing Eqs.~\eqref{TMD12} and \eqref{TMD_DY} to the definition of ${\widetilde Q}$ in \eq{Qtilde}, we readily see that 
\begin{align}\label{gQ_relation}
\int d^2 k_\perp \, e^{i {\un k} \cdot {\un x}_{10}} \, \left[ g_{1L}^{S, \, \textrm{SIDIS}} (x, k_T^2) + g_{1L}^{S, \, \textrm{DY}} (x, k_T^2) \right] = \frac{1}{\pi} \, \frac{16}{g^2} \, \frac{1}{2} \, \int d^2 \left( \frac{x_0 + x_1}{2} \right) \, {\widetilde Q}_{10} (s = Q^2/x) .
\end{align}
The factor of $1/2$ on the right of \eq{gQ_relation} removes the double-counting due to the complex conjugate terms: indeed, since TMDs are real, one can readily show that the expression in \eq{gQ_relation} is also real. The factor of $2 p^+$ from Eqs.~\eqref{TMD12} and \eqref{TMD_DY} is present in \eq{Qtilde} owing to the definition of the double angle brackets in \eq{double_def}.

We thus obtain
\begin{align}\label{Qg_relation}
  {\widetilde Q} (x^2_{10}, Q^2/x) = \frac{\as \, \pi^2}{2} \, \int d^2 k_\perp \, e^{i {\un k} \cdot {\un x}_{10}} \, \left[ g_{1L}^{S, \, \textrm{SIDIS}} (x, k_T^2) + g_{1L}^{S, \, \textrm{DY}} (x, k_T^2) \right]. 
\end{align}
Since helicity TMDs are PT-even, the DY and SIDIS helicity TMDs are equal, $g_{1L}^{S, \, \textrm{SIDIS}} (x, k_T^2) = g_{1L}^{S, \, \textrm{DY}} (x, k_T^2) \equiv g_{1L}^S (x, k_T^2)$, such that we can rewrite \eq{Qg_relation} as
\begin{align}\label{Qg_relation2}
 {\widetilde Q} (x^2_{10}, Q^2/x) = \as \, \pi^2 \, \int d^2 k_\perp \, e^{i {\un k} \cdot {\un x}_{10}} \, g_{1L}^S (x, k_T^2), 
\end{align}
or, equivalently, 
\begin{align}\label{Qg_relation3}
g_{1L}^S (x, k_T^2) = \frac{1}{4 \pi^4 \as} \, \int d^2 x_{10} \, e^{- i {\un k} \cdot {\un x}_{10}} \ {\widetilde Q} (x^2_{10}, Q^2/x). 
\end{align}
Equations \eqref{Qg_relation2} and \eqref{Qg_relation3} provide a relation between the quark helicity TMDs and the new object $\widetilde Q$ we have introduced in this work. 

We can also connect $\widetilde Q$ to the quark helicity PDFs by writing the flavor-singlet quark helicity PDF as an integral of the corresponding TMD over $k_T$,
\begin{align}\label{DSigma_Qtilde}
\Delta \Sigma (x, Q^2) = \sum_f \int\limits^{Q^2} d^2 k_\perp \, g_{1L}^S (x, k_T^2) = \frac{N_f}{\as \, \pi^2} \, {\widetilde Q} (x^2_{10}, Q^2/x) = \frac{N_f}{\as \, \pi^2} \ {\widetilde Q} \left( x^2_{10} = \frac{1}{Q^2} , s = \frac{Q^2}{x} \right).
\end{align}
Here, again, for simplicity we assume that all quark flavors contribute equally.

Using \eq{eq:Qevol} we write
\begin{align}\label{DSigma}
\Delta \Sigma (x, Q^2) &  = \frac{N_f}{\as \, \pi^2} \, {\widetilde Q}^{(0)} \left( x^2_{10} = \frac{1}{Q^2} , s = \frac{Q^2}{x} \right) - \frac{N_c \, N_f}{2 \, \pi^3} \, \int\limits_\frac{\Lambda^2}{s}^{1} \frac{d z}{z} \, \int\limits_{\max \left[ \frac{1}{Q^2}, \frac{x}{z \, Q^2} \right]}^{\min \left[ \frac{1}{z \, Q^2}, \frac{1}{\Lambda^2} \right]} \frac{d x^2_{10}}{x^2_{10}} \, \left[ Q (x_{10}^2, z s) + 2 \, G_2 (x_{10}^2, z s) \right] \\
& =  \frac{N_f}{\as \, \pi^2} \, {\widetilde Q}^{(0)} \left( x^2_{10} = \frac{1}{Q^2} , s = \frac{Q^2}{x} \right) - \frac{N_c \, N_f}{2 \, \pi^3} \, \int\limits_{\frac{1}{Q^2}}^{\frac{1}{\Lambda^2}} \frac{d x^2_{10}}{x^2_{10}} \, \int\limits_\frac{1}{s \, x_{10}^2}^{\frac{1}{Q^2 \, x_{10}^2}} \frac{d z}{z} \,  \left[ Q (x_{10}^2, z s) + 2 \, G_2 (x_{10}^2, z s) \right] . \notag
\end{align}
Comparing \eq{DeltaSigma} to \eq{DSigma}, we see that the latter has the extra inhomogeneous term ${\widetilde Q}^{(0)}$ which was neglected in \eq{DeltaSigma} (and in \cite{Kovchegov:2015pbl, Kovchegov:2016zex, Kovchegov:2016weo, Kovchegov:2018znm, Cougoulic:2022gbk, Adamiak:2023yhz}) as not evolving at small $x$. In addition, the lower limit of the $x_{10}$ integral is now different in the first line of \eq{DSigma} as compared to \eq{DeltaSigma}.

We conclude that equations~\eqref{DSigma_Qtilde} and \eqref{DSigma} result in an expression for $\Delta \Sigma$ at small $x$ with a slight modification as compared to \eq{DeltaSigma} used earlier in the literature. As we will see below, the present expression for $\Delta\Sigma$ appears to be close to the $\overline{\text{MS}}$ scheme for calculating the polarized DGLAP splitting functions. (The previous definition \eqref{DeltaSigma} can be thought of corresponding to the ``polarized DIS scheme" \cite{Adamiak:2023okq}.)


\subsection{Contribution to the evolution for the dipole amplitude $Q$}
\label{sec:Q_evol}

For completeness, let us determine the contribution of the transition operators \eqref{Ws2} to the DLA evolution of the dipole amplitude $Q$ defined in \eq{Qdef} above. (The dipole amplitude $G_2$ is defined by a purely gluonic operator: therefore, its evolution does not contain quarks and received no contribution from the transition operators.)

\begin{figure}[ht]
\centering
\includegraphics[width= 0.95 \textwidth]{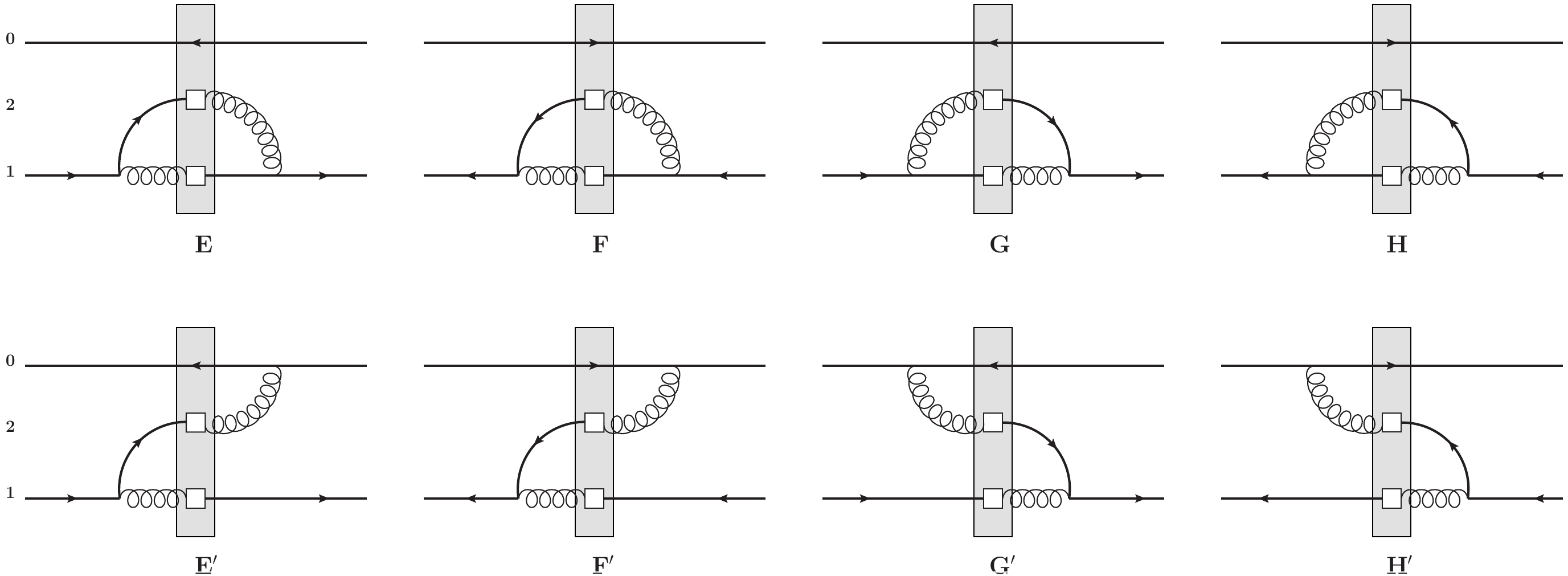}
\caption{The diagrams for the evolution of $Q_{10}$ due to the $q/{\bar q} \to G$ and $G \to q/{\bar q}$ shock-wave transition operators calculated here. The shaded rectangle denotes the shock wave, while the white square denotes the non-eikonal interaction with the shock wave mediated by the transition operators from \fig{fig:newoperators}.}
\label{FIG:Q_evol}
\end{figure}

The relevant diagrams containing contributions of the transition operators to the evolution of the dipole amplitude $Q$ are shown in \fig{FIG:Q_evol}. Their calculation parallels that in Sec.~\ref{sec:ABCD} and employs the background field propagators \eqref{psibar_a+}, \eqref{psi_a+}, \eqref{psi_a+_2}, and \eqref{psibar_a+_2}. One obtains
\begin{subequations}
\begin{align}
    & E + E' = - \frac{\as}{4 \pi^2 N_c} \, \int\limits_0^{p_2^-} \frac{d k^-}{k^-} \, \int d^2 x_2 \,\llangle \tr \Bigg[ t^a \, V_{\un 0}^\dagger \, t^b \, \bigg\{ - \frac{{\un x}_{20}}{x_{20}^2}  \cdot \frac{{\un x}_{21}}{x_{21}^2} \, \bigg[  \left( W_{\un 1}^{a \, [2]} [- \infty, \infty] \right)^\dagger \, W_{\un 2}^{b \, [1]} [\infty, -\infty] \\ 
    & + \left( W_{\un 1}^{a \, [1]} [- \infty, \infty] \right)^\dagger \, W_{\un 2}^{b \, [2]} [\infty, -\infty] \bigg] + i \, \frac{{\un x}_{20}}{x_{20}^2}  \times \frac{{\un x}_{21}}{x_{21}^2} \, \, \bigg[  \left( W_{\un 1}^{a \, [1]} [- \infty, \infty] \right)^\dagger \, W_{\un 2}^{b \, [1]} [\infty, -\infty] \notag \\ 
    & + \left( W_{\un 1}^{a \, [2]} [- \infty, \infty] \right)^\dagger \, W_{\un 2}^{b \, [2]} [\infty, -\infty] \bigg] \bigg\} \Bigg] \rrangle - ({\un x}_{20} \to {\un x}_{21}) , \notag \\
    & F + F' = (E + E')^* , \\
    & G + G' = - \frac{\as}{4 \pi^2 N_c} \, \int\limits_0^{p_2^-} \frac{d k^-}{k^-} \, \int d^2 x_2 \,\llangle \tr \Bigg[ t^a \, V_{\un 0}^\dagger \, t^b \, \bigg\{ - \frac{{\un x}_{20}}{x_{20}^2}  \cdot \frac{{\un x}_{21}}{x_{21}^2} \, \bigg[  \left( W_{\un 2}^{a \, [1]} [- \infty, \infty] \right)^\dagger \, W_{\un 1}^{b \, [2]} [\infty, -\infty] \\ 
    & + \left( W_{\un 2}^{a \, [2]} [- \infty, \infty] \right)^\dagger \, W_{\un 1}^{b \, [1]} [\infty, -\infty] \bigg] - i \, \frac{{\un x}_{20}}{x_{20}^2}  \times \frac{{\un x}_{21}}{x_{21}^2} \, \, \bigg[  \left( W_{\un 2}^{a \, [1]} [- \infty, \infty] \right)^\dagger \, W_{\un 1}^{b \, [1]} [\infty, -\infty] \notag \\ 
    & + \left( W_{\un 2}^{d \, [2]} [- \infty, \infty] \right)^\dagger \, W_{\un 1}^{b \, [2]} [\infty, -\infty] \bigg] \bigg\} \Bigg] \rrangle - ({\un x}_{20} \to {\un x}_{21}) , \notag \\
    & H + H' = (G + G')^* .
\end{align}    
\end{subequations}
It is worth noting that the diagrams in Fig.~\ref{FIG:Q_evol} were calculated in \cite{Chirilli:2021lif}. (See Eq.(4.28), Eq.(6.81) and Eq.(6.98) in \cite{Chirilli:2021lif}.) 

We are particularly interested in the DLA limit of these diagrams. Adding up the contributions and neglecting the cross-product terms in the DLA yields
\begin{align}\label{EFGH1}
    & E + E' + F + F' + G + G' + H + H' = - \frac{\as}{4 \pi N_c} \int\limits_0^{p_2^-} \frac{d k^-}{k^-} \int\limits^{x_{10}^2} \frac{d x^2_{21}}{x^2_{21}} \, \llangle \tr \Bigg[ t^a \, V_{\un 0}^\dagger \, t^b \\ 
    & \times \, \bigg[  \left( W_{\un 1}^{a \, [2]} [- \infty, \infty] \right)^\dagger \, W_{\un 2}^{b \, [1]} [\infty, -\infty] + \left( W_{\un 1}^{a \, [1]} [- \infty, \infty] \right)^\dagger \, W_{\un 2}^{b \, [2]} [\infty, -\infty] \notag \\ 
    & + \left( W_{\un 2}^{a \, [1]} [- \infty, \infty] \right)^\dagger \, W_{\un 1}^{b \, [2]} [\infty, -\infty] + \left( W_{\un 2}^{a \, [2]} [- \infty, \infty] \right)^\dagger \, W_{\un 1}^{b \, [1]} [\infty, -\infty]\bigg] \Bigg] \rrangle + \mbox{c.c.} . \notag 
\end{align}
To simplify \eq{EFGH1} we first note that
\begin{align}\label{W4}
    & \tr \Bigg[ t^a \, V_{\un 0}^\dagger \, t^b \, \, \bigg[  \left( W_{\un 1}^{a \, [2]} [- \infty, \infty] \right)^\dagger \, W_{\un 2}^{b \, [1]} [\infty, -\infty] 
    + \left( W_{\un 1}^{a \, [1]} [- \infty, \infty] \right)^\dagger \, W_{\un 2}^{b \, [2]} [\infty, -\infty] \bigg] \Bigg] \\
    & = \frac{g^2}{4 \sqrt{p_2^- \, k^-}} \, \int\limits_{-\infty}^\infty d y^- \, \int\limits_{-\infty}^\infty d z^- \, U_{\un 1}^{ca} [z^-, - \infty] \, U_{\un 2}^{bd} [\infty, y^-] \, {\bar \psi} (y^-, {\un x}_2) \, \gamma^+ \, \gamma^5 \, t^d \, V_{\un 2} [y^-, - \infty] \, t^a \, V_{\un 0}^\dagger \, t^b \, V_{\un 1} [\infty, z^-] \, t^c \, \psi (z^-, {\un x}_1) , \notag
\end{align}
which allows us to rewrite \eq{EFGH1} as
\begin{align}\label{EFGH2}
    & E + E' + F + F' + G + G' + H + H' = - \frac{\as}{4 \pi N_c} \int\limits_0^{p_2^-} \frac{d k^-}{k^-} \int\limits^{x_{10}^2} \frac{d x^2_{21}}{x^2_{21}} \, \int\limits_{-\infty}^\infty d y^- \, \int\limits_{-\infty}^\infty d z^- \, \llangle  \frac{g^2}{4 \sqrt{p_2^- \, k^-}} \, U_{\un 1}^{ca} [z^-, - \infty] \, U_{\un 2}^{bd} [\infty, y^-] \notag \\ 
    & \times \, {\bar \psi} (y^-, {\un x}_2) \, \gamma^+ \, \gamma^5 \, t^d \, V_{\un 2} [y^-, - \infty] \, t^a \, V_{\un 0}^\dagger \, t^b \, V_{\un 1} [\infty, z^-] \, t^c \, \psi (z^-, {\un x}_1) + ({\un 1} \leftrightarrow {\un 2})  \rrangle + \mbox{c.c.} .  
\end{align}
Employing \eq{Uab} and the Fierz identity one can recast \eq{W4} as
\begin{align}\label{W42}
    & \llangle \tr \Bigg[ t^a \, V_{\un 0}^\dagger \, t^b \, \, \bigg[  \left( W_{\un 1}^{a \, [2]} [- \infty, \infty] \right)^\dagger \, W_{\un 2}^{b \, [1]} [\infty, -\infty] 
    + \left( W_{\un 1}^{a \, [1]} [- \infty, \infty] \right)^\dagger \, W_{\un 2}^{b \, [2]} [\infty, -\infty] \bigg] \Bigg] \rrangle \\
    & = \frac{g^2}{16 \sqrt{p_2^- \, k^-}} \, \int\limits_{-\infty}^\infty d y^- \, \int\limits_{-\infty}^\infty d z^- \,\Bigg\{ \llangle {\bar \psi} (y^-, {\un x}_2) \, \gamma^+ \, \gamma^5 \, V_{\un 2} [y^-,\infty] \, V_{\un 1} \, V_{\un 0}^\dagger \, V_{\un 2} \, V_{\un 1} [-\infty, z^-] \, \psi (z^-, {\un x}_1) \rrangle \notag \\
    & - S_{10} \, \llangle {\bar \psi} (y^-, {\un x}_2) \, \gamma^+ \, \gamma^5 \, V_{\un 2} [y^-,-\infty] \, V_{\un 1} [-\infty, z^-] \, \psi (z^-, {\un x}_1) \rrangle - S_{20} \, \llangle {\bar \psi} (y^-, {\un x}_2) \, \gamma^+ \, \gamma^5 \, V_{\un 2} [y^-,\infty] \, V_{\un 1} [\infty, z^-] \, \psi (z^-, {\un x}_1) \rrangle \notag \\
    & + \frac{1}{N_c^2} \, \llangle {\bar \psi} (y^-, {\un x}_2) \, \gamma^+ \, \gamma^5 \, V_{\un 2} [y^-,-\infty] \, V_{\un 0}^\dagger \, V_{\un 1} [\infty, z^-] \, \psi (z^-, {\un x}_1) \rrangle  \Bigg\} . \notag 
\end{align}
In the second to last line of \eq{W42} we have employed the large-$N_c$ limit to factor out the unpolarized $S$-matrices $S_{10}$ and $S_{20}$. 

Using \eq{W42} in \eq{EFGH2} we conclude that the contribution of the diagrams in \fig{FIG:Q_evol} is sub-leading at large $N_c$ (and at large-$N_c \& N_f$): see the explicit $1/N_c$ overall factor in \eq{EFGH2}. We conclude that these diagrams do not contribute to the large-$N_c \& N_f$ (or large-$N_c$) evolution of the dipole amplitude $Q$ in the DLA. Therefore, we neglect the contribution of the diagrams in \fig{FIG:Q_evol} below.


\subsection{A new version of the large-$N_c \& N_f$ evolution equations}
\label{sec:evolution}

Let us now summarize the modified version of the large-$N_c \& N_f$ evolution equations (155) from \cite{Cougoulic:2022gbk}. Combining our results in \eq{total5} and \eq{eq:Qevol}, along with incorporating \eq{total5} into the evolution for the neighbor dipole amplitude $\widetilde \Gamma$ from \cite{Cougoulic:2022gbk}, we arrive at
\begin{subequations}\label{eq_LargeNcNf}
\begin{tcolorbox}[ams align]
& Q(x^2_{10},zs) = Q^{(0)}(x^2_{10},zs) + \frac{\alpha_sN_c}{2\pi} \int_{1/s x^2_{10}}^{z} \frac{dz'}{z'}   \int_{1/z's}^{x^2_{10}}  \frac{dx^2_{21}}{x_{21}^2}    \left[ 2 \, {\widetilde G}(x^2_{21},z's) + 2 \, {\widetilde \Gamma}(x^2_{10},x^2_{21},z's) \right. \\
&\hspace*{5cm}\left.+ \; Q(x^2_{21},z's) -  \overline{\Gamma}(x^2_{10},x^2_{21},z's) + 2 \, \Gamma_2(x^2_{10},x^2_{21},z's) + 2 \, G_2(x^2_{21},z's)   \right] \notag \\
&\hspace*{3cm}+ \frac{\alpha_sN_c}{4\pi} \int_{\Lambda^2/s}^{z} \frac{dz'}{z'}   \int_{1/z's}^{\min \{ x^2_{10}z/z', 1/\Lambda^2 \}}  \frac{dx^2_{21}}{x_{21}^2} \left[Q(x^2_{21},z's) + 2 \, G_2(x^2_{21},z's) \right] ,  \notag  \\
&\overline{\Gamma}(x^2_{10},x^2_{21},z's) = Q^{(0)}(x^2_{10},z's) + \frac{\alpha_sN_c}{2\pi} \int_{1/s x^2_{10}}^{z'} \frac{dz''}{z''}   \int_{1/z''s}^{\min\{x^2_{10}, x^2_{21}z'/z''\}}  \frac{dx^2_{32}}{x_{32}^2}    \left[ 2\, {\widetilde G} (x^2_{32},z''s)  \right. \\
&\hspace*{2.5cm}\left.+ \; 2\, {\widetilde \Gamma} (x^2_{10},x^2_{32},z''s) +  Q(x^2_{32},z''s) -  \overline{\Gamma}(x^2_{10},x^2_{32},z''s) + 2 \, \Gamma_2(x^2_{10},x^2_{32},z''s) + 2 \, G_2(x^2_{32},z''s) \right] \notag \\
&\hspace*{3cm}+ \frac{\alpha_sN_c}{4\pi} \int_{\Lambda^2/s}^{z'} \frac{dz''}{z''}   \int_{1/z''s}^{\min \{ x^2_{21}z'/z'', 1/\Lambda^2 \}}  \frac{dx^2_{32}}{x_{32}^2} \left[Q(x^2_{32},z''s) + 2 \, G_2(x^2_{32},z''s) \right] , \notag \\
& {\widetilde G}(x^2_{10},zs) = {\widetilde G}^{(0)}(x^2_{10},zs) + \frac{\alpha_s N_c}{2\pi}\int_{1/s x^2_{10}}^z\frac{dz'}{z'}\int_{1/z's}^{x^2_{10}} \frac{dx^2_{21}}{x^2_{21}} \left[3 \, {\widetilde G}(x^2_{21},z's) + {\widetilde \Gamma}(x^2_{10},x^2_{21},z's) \right. \\
&\hspace*{3cm}\left.  + \; 2\,G_2(x^2_{21},z's)  +  \left(2 - \frac{N_f}{2N_c}\right) \Gamma_2(x^2_{10},x^2_{21},z's) - \frac{N_f}{4N_c}\,\overline{\Gamma}(x^2_{10},x^2_{21},z's) \color{blue}{ - \frac{N_f}{2 N_c} \, {\widetilde Q}(x^2_{21},z's) }  \right] \notag \\
&\hspace*{3cm}- \frac{\alpha_sN_f}{8\pi}  \int_{\Lambda^2/s}^z \frac{dz'}{z'}\int_{\max\{x^2_{10},\,1/z's\}}^{\min \{ x^2_{10}z/z', 1/\Lambda^2 \}} \frac{dx^2_{21}}{x^2_{21}}  \left[   Q(x^2_{21},z's) +     2 \, G_2(x^2_{21},z's)  \right] , \notag \\
& {\widetilde \Gamma} (x^2_{10},x^2_{21},z's) = {\widetilde G}^{(0)}(x^2_{10},z's) + \frac{\alpha_s N_c}{2\pi}\int_{1/s x^2_{10}}^{z'}\frac{dz''}{z''}\int_{1/z''s}^{\min\{x^2_{10},x^2_{21}z'/z''\}} \frac{dx^2_{32}}{x^2_{32}} \left[3 \, {\widetilde G} (x^2_{32},z''s) \right. \\
&\hspace*{.1cm}\left. + \; {\widetilde \Gamma}(x^2_{10},x^2_{32},z''s) + 2 \, G_2(x^2_{32},z''s)  +  \left(2 - \frac{N_f}{2N_c}\right) \Gamma_2(x^2_{10},x^2_{32},z''s) - \frac{N_f}{4N_c} \,\overline{\Gamma}(x^2_{10},x^2_{32},z''s)  \color{blue}{ - \frac{N_f}{2 N_c} \, {\widetilde Q}(x^2_{32},z''s) } \right] \notag \\
&\hspace*{3cm}- \frac{\alpha_sN_f}{8\pi}  \int_{\Lambda^2/s}^{z'x^2_{21}/x^2_{10}} \frac{dz''}{z''}\int_{\max\{x^2_{10},\,1/z''s\}}^{\min \{ x^2_{21}z'/z'', 1/\Lambda^2 \} } \frac{dx^2_{32}}{x^2_{32}}  \left[   Q(x^2_{32},z''s) +  2  \,  G_2(x^2_{32},z''s)  \right] , \notag \\
& G_2(x_{10}^2, z s)  =  G_2^{(0)} (x_{10}^2, z s) + \frac{\as N_c}{\pi} \, \int\limits_{\frac{\Lambda^2}{s}}^z \frac{d z'}{z'} \, \int\limits_{\max \left[ x_{10}^2 , \frac{1}{z' s} \right]}^{\min \{\frac{z}{z'} x_{10}^2, 1/\Lambda^2 \}} \frac{d x^2_{21}}{x_{21}^2} \left[ {\widetilde G} (x^2_{21} , z' s) + 2 \, G_2 (x_{21}^2, z' s)  \right] , \\
& \Gamma_2 (x_{10}^2, x_{21}^2, z' s)  =  G_2^{(0)} (x_{10}^2, z' s) + \frac{\as N_c}{\pi}  \int\limits_{\frac{\Lambda^2}{s}}^{z' \frac{x_{21}^2}{x_{10}^2}} \frac{d z''}{z''}  \int\limits_{\max \left[ x_{10}^2 , \frac{1}{z'' s} \right]}^{\min \{ \frac{z'}{z''} x_{21}^2, 1/\Lambda^2 \}} \frac{d x^2_{32}}{x_{32}^2} \left[ {\widetilde G} (x^2_{32} , z'' s) + 2 \, G_2(x_{32}^2, z'' s)  \right] , \\
& \color{blue}{ {\widetilde Q} (x^2_{10}, z s) =  {\widetilde Q}^{(0)} (x^2_{10}, z s) - \frac{\as \, N_c}{2 \, \pi} \, \int\limits_\frac{\Lambda^2}{s}^{z} \frac{d z'}{z'} \, \int\limits_{\max \left[ x_{10}^2, \frac{1}{z' s} \right]}^{\min \{\frac{z}{z'} x_{10}^2, 1/\Lambda^2 \}} \frac{d x^2_{21}}{x^2_{21}} \, \left[ Q (x_{21}^2, z' s) + 2 \, G_2 (x_{21}^2, z' s) \right] }.  
\end{tcolorbox}
\end{subequations}
The new terms and the new equation are shown in blue. The equations \eqref{eq_LargeNcNf} are written in the form treating $\Lambda$ as the infrared cutoff (and not as a semi-perturbative scale characterizing the target, like the target saturation scale, see \cite{Kovchegov:2020hgb, Cougoulic:2022gbk} for a discussion of this issue). The equations \eqref{eq_LargeNcNf} are re-derived in Appendix~\ref{B} using LCPT.

For completeness, let us re-state the expressions for helicity PDFs at small $x$: 
\begin{subequations}\label{eq:DeltaGSigma_G2_Qtilde}
\begin{align}
& \Delta G (x, Q^2) = \frac{2 N_c}{\as \, \pi^2} \, G_2 \left(  x_{10}^2 = \frac{1}{Q^2} ,  s = \frac{Q^2}{x} \right) , \\
& \Delta \Sigma (x, Q^2) = \frac{N_f}{\as \, \pi^2} \, {\widetilde Q} \left( x^2_{10} = \frac{1}{Q^2} , s = \frac{Q^2}{x} \right) .
\end{align}
\end{subequations}
While the expression for $\Delta G$ was obtained earlier \cite{Kovchegov:2017lsr}, the result for $\Delta \Sigma$ is new, slightly correcting the older result \cite{Kovchegov:2015pbl, Kovchegov:2016zex, Kovchegov:2018znm, Cougoulic:2022gbk}.

\section{Iterative Solution}
\label{sec:iterative}

In this Section, we solve Eqs.~\eqref{eq_LargeNcNf} iteratively, similar to what was done in \cite{Cougoulic:2022gbk, Adamiak:2023okq}. We assume non-vanishing constant (independent of $x$ and $Q^2$) initial conditions $\Delta \Sigma^{(0)}$ and $\Delta G^{(0)}$ for the respective flavor-singlet quark and gluon helicity PDFs.  The (also constant) initial conditions for the polarized dipole amplitudes are 
\begin{equation}\label{init_cond}
G_2^{(0)} = \frac{\alpha_s \pi^2}{2N_c} \Delta G^{(0)},\quad
\widetilde{Q}^{(0)} = \frac{\alpha_s \pi^2}{N_f} \Delta \Sigma^{(0)}, \quad Q^{(0)} = -\frac{1}{2} \widetilde{Q}^{(0)}, \quad \widetilde{G}^{(0)} = \frac{N_f}{4N_c} \widetilde{Q}^{(0)}. 
\end{equation}
The first two conditions follow from Eqs.~\eqref{eq:DeltaGSigma_G2_Qtilde}. The last two conditions can be derived from the original definitions of $Q(x_{10}^2, s)$ in Eq.~\eqref{Qdef}, $\widetilde{G}(x_{10}^2, s)$ in Eq.~\eqref{G_dfn_NcNf} and $\widetilde{Q}(x_{10}^2, s)$ in Eq.~\eqref{Qtilde} by taking the zero dipole size limit, $x_{10}^2 \rightarrow 0$. Indeed, we first notice that only $V_{\un x}^{\textrm{q} [1]}$ in $V_{\un x}^{\textrm{pol} [1]}$ contributes to $Q(x_{10}^2, s)$ in this limit, and only the quark-fields part of $W_{{\un x}}^{\textrm{pol} [1] }$ in \eq{Wpol_dfn} contributes to $\widetilde{G}(x_{10}^2, s)$. Performing a little algebra and clarifying that the zero dipole size limit corresponds to $x_{10}^2 \to 1/s$, with the center-of-mass energy squared $s$ being the highest momentum scale in the problem, we arrive at the following relations between the (impact-parameter integrated) polarized dipole amplitudes:
\begin{subequations}\label{relationsGQQ}
    \begin{align}
    & \widetilde{G} \left(x_{10}^2 = \frac{1}{s}, s \right) = - \frac{N_f}{4 C_F} \, Q \left(x_{10}^2 = \frac{1}{s}, s \right) \approx - \frac{N_f}{2 N_c} \, Q \left(x_{10}^2 = \frac{1}{s}, s \right), \label{GtildeQ} \\
    & \widetilde{G} \left(x_{10}^2 = \frac{1}{s}, s \right) = \frac{N_f}{4 N_c} \, {\widetilde Q} \left(x_{10}^2 = \frac{1}{s}, s \right). \label{GtildeQtilde}
\end{align}
\end{subequations}
Here we have summed over flavors in $\widetilde G$, assuming that all flavors contribute equally. (Otherwise, one would have to replace $N_f \, Q \to \sum_f Q_f$ in \eq{GtildeQ} and $N_f \, {\widetilde Q} \to \sum_f {\widetilde Q}_f$ in \eq{GtildeQtilde}, with the flavor-dependent dipole amplitudes $Q_f$ and ${\widetilde Q}_f$.) The last transition in \eq{GtildeQ} is valid in the large-$N_c$ and large-$N_c \& N_f$ limits. Equations \eqref{relationsGQQ} are otherwise exact. Since all the integral kernels in Eqs.~\eqref{eq_LargeNcNf} vanish in the $x_{10}^2 \to 1/s$ limit (which, in those equations, corresponds to the $x_{10}^2 \to 1/(z s)$ limit), we see that the conditions \eqref{relationsGQQ} need to be satisfied by the inhomogeneous terms (the initial conditions) in Eqs.~\eqref{eq_LargeNcNf}, after which they will be always satisfied by the dipole amplitudes solving those evolution equations. 

Equations~\eqref{relationsGQQ} allowed us to fix the initial amplitudes $Q^{(0)}$ and $\widetilde{G}^{(0)}$ in \eq{init_cond}.

In what follows, we will solve Eqs.~\eqref{eq_LargeNcNf} iteratively, using Eqs.~\eqref{eq:DeltaGSigma_G2_Qtilde} to obtain order-by-order in $\as$ expansion for hPDFs,
\begin{subequations}\label{pert_exp}
    \begin{align}
        & \Delta \Sigma (x, Q^2) = \Delta\Sigma^{(0)} + \Delta\Sigma^{(1)}(x,Q^2) + \Delta\Sigma^{(2)}(x,Q^2) + \ldots , \\
        & \Delta G (x, Q^2) = \Delta G^{(0)} + \Delta G^{(1)}(x,Q^2) + \Delta G^{(2)}(x,Q^2) + \ldots , 
    \end{align}
\end{subequations}
where the index in the superscript parenthesis matches the power of $\as$ correction to $\Delta\Sigma^{(0)}$ and $\Delta G^{(0)}$. The iterative expansion of the dipole amplitudes and DGLAP splitting functions is labeled in the same way.

\subsection{Step 1}
After one step of evolution, using the dipole amplitudes from \eq{init_cond} in  Eqs.~\eqref{eq_LargeNcNf}, one obtains
\begin{subequations}\label{eq:G_2_(1)}
    \begin{align}
        G_2^{(1)}(x_{10}^2, zs) =&\frac{\alpha_s N_c}{\pi}\ln (zs x_{10}^2)\ln \left(\frac{1}{\Lambda^2 x_{10}^2}\right) \left[\frac{N_f}{4N_c} \widetilde{Q}^{(0)} + 2G_2^{(0)}\right], \\
\widetilde{Q}^{(1)}(x_{10}^2, zs) =&- \frac{\alpha_s N_c}{2\pi}\ln (zs x_{10}^2)\ln \left(\frac{1}{\Lambda^2 x_{10}^2}\right) \left[-\frac{1}{2}\widetilde{Q}^{(0)} + 2G_2^{(0)}\right]. 
    \end{align}
\end{subequations}

The hPDFs follow from Eq.~\eqref{eq:DeltaGSigma_G2_Qtilde}
\begin{subequations}
    \begin{align}
        &\Delta G^{(1)}(x, Q^2) =   \frac{\alpha_s N_c}{4\pi}\ln\frac{1}{x}\ln \frac{Q^2}{\Lambda^2}  \left[2 \Delta \Sigma^{(0)} + 8 \Delta G^{(0)}\right],\\
&\Delta \Sigma^{(1)}(x, Q^2) =  \frac{\alpha_s N_c}{4\pi }\ln\frac{1}{x}\ln \frac{Q^2}{\Lambda^2}  \left[ \Delta \Sigma^{(0)} - 2\frac{N_f}{N_c} \Delta G^{(0)}\right]. 
    \end{align}
\end{subequations}

Having these two expressions for hPDFs, one can employ the spin-dependent DGLAP evolution equations
\begin{align}\label{DGLAP_diff}
\frac{\pd}{\pd \ln Q^2}
\begin{pmatrix}
 \Delta {\Sigma} (x, Q^2) \\
 \Delta {G} (x, Q^2)
\end{pmatrix}
= 
\int\limits_x^1 \frac{dz}{z} \, 
\begin{pmatrix}
\Delta P_{qq} (z) & \Delta P_{qG} (z) \\
\Delta P_{Gq} (z)  & \Delta P_{GG} (z) 
\end{pmatrix}
\,
\begin{pmatrix}
 \Delta {\Sigma} \left( \frac{x}{z} , Q^2 \right) \\
 \Delta {G} \left( \frac{x}{z} , Q^2 \right)
\end{pmatrix}
\equiv 
\left[\Delta\mathbf{P} \otimes
\begin{pmatrix} 
\Delta {\Sigma}   \\
\Delta {G} 
\end{pmatrix} \right] (x, Q^2 )
\end{align}
to readily read out the polarized small-$x$ large-$N_c \& N_f$ splitting functions at the one-loop order 
\begin{equation}\label{eq:DeltaP_(0)}
\Delta \mathbf{P}^{(0)} (x) = \frac{\alpha_s N_c}{4\pi} \begin{pmatrix}
    1 & -2\frac{N_f}{N_c}\\[10pt]
    2 & 8 \\
\end{pmatrix}.
\end{equation}

We also calculated other polarized dipole amplitudes and the neighbor dipole amplitudes after one step of the evolution, obtaining
\begin{subequations}\label{eq:Q(1)}
    \begin{align} 
    & Q^{(1)}(x_{10}^2, zs) 
= \frac{\alpha_s N_c}{4\pi}\ln^2(zsx_{10}^2) \left[\left(\frac{N_f}{N_c} - \frac{1}{4}\right)\widetilde{Q}^{(0)} + 5 G_2^{(0)}\right] +  \frac{\alpha_s N_c}{4\pi} \ln(zsx_{10}^2) \ln\frac{1}{\Lambda^2x_{10}^2}\left[-\frac{1}{2} \widetilde{Q}^{(0)} + 2G_2^{(0)}\right], \\   & \widetilde{G}^{(1)}(x_{10}^2, zs) = \frac{\alpha_s N_c}{4\pi} \ln^2(zsx_{10}^2) \left[\frac{5N_f}{8N_c} \widetilde{Q}^{(0)} + 4 \left( 1-\frac{N_f}{8N_c} \right) G_2^{(0)}\right] \label{eq:Gtilde(1)}  \\ 
& \hspace*{2.5cm} - \frac{\alpha_s N_c}{4\pi} \ln(zsx_{10}^2)\ln\left(\frac{1}{\Lambda^2x_{10}^2}\right) \frac{N_f}{2N_c}\left[-\frac{1}{2} \widetilde{Q}^{(0)} + 2G_2^{(0)}\right] , \notag \\
& \bar{\Gamma}^{(1)}(x_{10}^2, x_{21}^2, z's) 
=\frac{\alpha_s N_c}{4\pi}  \ln^2(z'sx_{21}^2) \left[\left(\frac{N_f}{N_c} - \frac{1}{4}\right) \widetilde{Q}^{(0)}  + 5 G_2^{(0)}\right] \\ 
& \hspace*{.5cm} + \frac{\alpha_s N_c}{4\pi}  \ln(z'sx_{21}^2)\ln\frac{1}{\Lambda^2 x_{21}^2} \left[\left( \frac{2N_f}{N_c}-\frac{1}{2}\right) \widetilde{Q}^{(0)} + 10 G_2^{(0)}\right] -\frac{\alpha_s N_c}{4\pi}  \ln(z'sx_{21}^2)\ln\frac{1}{\Lambda^2 x_{10}^2} \left[\frac{2N_f}{N_c}\widetilde{Q}^{(0)} + 8 G_2^{(0)}\right] , \notag \\
& \widetilde{\Gamma}^{(1)}(x_{10}^2, x_{21}^2, z's) 
= \frac{\alpha_s N_c}{4\pi}  \ln^2(z'sx_{21}^2)  \left[\frac{5N_f}{8N_c} \widetilde{Q}^{(0)} + \left( 4-\frac{N_f}{2N_c} \right) G_2^{(0)}\right] \\
& \hspace*{.5cm} + \frac{\alpha_s N_c}{4\pi}  \ln(z'sx_{21}^2) \ln\frac{1}{\Lambda^2 x_{21}^2}   \left[\frac{5N_f}{4N_c} \widetilde{Q}^{(0)} + \left( 8-\frac{N_f}{N_c} \right) G_2^{(0)}\right]-\frac{\alpha_s N_c}{4\pi} \ln(z's x_{21}^2) \ln\frac{1}{\Lambda^2x_{10}^2}\left[\frac{N_f}{N_c} \, \widetilde{Q}^{(0)} + 8 G_2^{(0)}\right] , \notag \\ 
& \Gamma_2^{(1)}(x_{10}^2, x_{21}^2, z's) =4\frac{\alpha_s N_c}{4\pi}  \ln(z's x_{21}^2) \ln \left(\frac{1}{\Lambda^2x_{10}^2}\right)\left[\frac{N_f}{4N_c} \, \widetilde{Q}^{(0)} + 2G_2^{(0)}\right].
    \end{align}
\end{subequations}

\subsection{Step 2}

After two steps of the evolution, one gets
\begin{subequations}
    \begin{align}
        G_2^{(2)}(x_{10}^2, zs) 
=&\left(\frac{\alpha_s N_c}{4\pi} \right)^2 \frac{1}{3}\ln^3(zsx_{10}^2)\ln\left(\frac{1}{\Lambda^2x_{10}^2}\right)  \left[\frac{5N_f}{2N_c}  \widetilde{Q}^{(0)} + 2 \left( 8-\frac{N_f}{N_c} \right) G_2^{(0)}\right]\\
&+\left(\frac{\alpha_s N_c}{4\pi} \right)^2 \ln^2(zsx_{10}^2) \ln^2\left(\frac{1}{\Lambda^2x_{10}^2}\right)\left[\frac{9N_f}{4N_c} \widetilde{Q}^{(0)} + \left( 16- \frac{N_f}{N_c} \right) G_2^{(0)})\right], \notag \\
\widetilde{Q}^{(2)}(x_{10}^2, zs)
=&-\left(\frac{\alpha_s N_c}{4\pi} \right)^2 \frac{2}{3} \ln^3(zsx_{10}^2) \ln \frac{1}{\Lambda^2x_{10}^2}\left[ \left( \frac{N_f}{N_c}-\frac{1}{4} \right) \widetilde{Q}^{(0)} + 5 G_2^{(0)}\right]\\
&-\left(\frac{\alpha_s N_c}{4\pi} \right)^2  \ln^2(zsx_{10}^2)\ln^2\left(\frac{1}{\Lambda^2x_{10}^2}\right)\left[\left(-\frac{1}{4} + \frac{N_f}{N_c}\right)\widetilde{Q}^{(0)} + 9 G_2^{(0)}\right]. \notag
    \end{align}
\end{subequations}
The hPDFs from the second iteration are
\begin{subequations}
    \begin{align}
        \Delta G^{(2)} (x,Q^2) 
=&\left(\frac{\alpha_s N_c}{4\pi} \right)^2 \frac{1}{3}\ln^3\frac{1}{x}\ln\left(\frac{Q^2}{\Lambda^2}\right)  \left[5 \Delta \Sigma^{(0)} + 2 \left( 8-\frac{N_f}{N_c} \right) \Delta G^{(0)}\right]\\
&+\left(\frac{\alpha_s N_c}{4\pi} \right)^2 \ln^2\frac{1}{x} \ln^2\left(\frac{Q^2}{\Lambda^2}\right)\left[\frac{9}{2} \Delta\Sigma^{(0)} + \left(16- \frac{N_f}{N_c} \right)\Delta G^{(0)}\right], \notag \\
\Delta \Sigma^{(2)}(x, Q^2) 
=&\left(\frac{\alpha_s N_c}{4\pi} \right)^2 \frac{1}{3}\ln^3\frac{1}{x} \ln\frac{Q^2}{\Lambda^2}\left[ \left( \frac{1}{2} - \frac{2N_f}{N_c} \right) \Delta \Sigma^{(0)} - \frac{5N_f}{N_c} \Delta G^{(0)}\right]\\
&+\left(\frac{\alpha_s N_c}{4\pi} \right)^2\ln^2\frac{1}{x} \ln^2\frac{Q^2}{\Lambda^2}  \left[ \left( \frac{1}{4}-\frac{N_f}{N_c} \right) \Delta \Sigma^{(0)} -\frac{9N_f}{2N_c} \Delta G^{(0)}\right]. \notag
    \end{align}
\end{subequations}
From the terms containing $\ln^3\frac{1}{x} \ln\frac{Q^2}{\Lambda^2}$, one can read out the polarized small-$x$ large-$N_c \& N_f$ splitting functions at the two-loop order
\begin{equation}\label{eq:DeltaP_(1)}
    \Delta \mathbf{P}^{(1)} (x) = \left(\frac{\alpha_s N_c}{4\pi}\right)^2\ln^2\frac{1}{x}\begin{pmatrix}
        \frac{1}{2} - \frac{2N_f}{N_c} & -\frac{5N_f}{N_c}\\[10pt]
        5 & 16-\frac{2N_f}{N_c} \\
    \end{pmatrix}.
\end{equation}
Other polarized dipole amplitudes after two steps of the evolution that we will need for the next iteration of the solution are
\begin{subequations}
    \begin{align}
        Q^{(2)}(x_{10}^2, zs) = & \ \frac{1}{6}\left(\frac{\alpha_s N_c}{4\pi} \right)^2\ln^4(zsx_{10}^2)\left[\left(\frac{5N_f}{N_c} -\frac{1}{4}\right)\widetilde{Q}^{(0)} + \left( 35-\frac{4N_f}{N_c} \right) G_2^{(0)}\right]\\
&+\frac{1}{3} \left(\frac{\alpha_s N_c}{4\pi} \right)^2\ln^3(zsx_{10}^2) \ln \frac{1}{\Lambda^2x_{10}^2} \left[\left(\frac{7N_f}{N_c}-\frac{1}{2}\right)\widetilde{Q}^{(0)}+ \left(46-\frac{4N_f}{N_c}\right) G_2^{(0)}\right] \notag \\
&+\frac{1}{4}\left(\frac{\alpha_s N_c}{4\pi} \right)^2\ln^2(zsx_{10}^2)\ln^2\left(\frac{1}{\Lambda^2x_{10}^2}\right)\left[\left(\frac{2N_f}{N_c}-\frac{1}{2}\right)\widetilde{Q}^{(0)} + 18G_2^{(0)} \right], \notag \\
\widetilde{G}^{(2)}(x_{10}^2, zs)=& \ \frac{1}{6}\left(\frac{\alpha_s N_c}{4\pi} \right)^2 \ln^4(zsx_{10}^2) \left[\frac{N_f}{N_c} \left( -\frac{N_f}{2N_c} + \frac{35}{8} \right) \, \widetilde{Q}^{(0)} + \left( 28 - \frac{11N_f}{2N_c} \right) \, G_2^{(0)}\right] \label{eq:Gtilde_(2)_final} \\
&+\frac{1}{6}\left(\frac{\alpha_s N_c}{4\pi} \right)^2 \ln^3(zsx_{10}^2)\ln\frac{1}{\Lambda^2x_{10}^2} \left[\frac{N_f}{N_c} \left( \frac{19}{2} -\frac{2N_f}{N_c} \right) \widetilde{Q}^{(0)} + \left( 64-\frac{18N_f}{N_c} \right) \, G_2^{(0)}\right] \notag \\
&-\frac{1}{8}\left(\frac{\alpha_s N_c}{4\pi} \right)^2 \ln^2(zsx_{10}^2)\ln^2\frac{1}{\Lambda^2x_{10}^2}\left[\frac{N_f}{N_c} \left( \frac{2N_f}{N_c} - \frac{1}{2} \right) \, \widetilde{Q}^{(0)} + \frac{18N_f}{N_c}\,  G_2^{(0)}\right]. \notag
    \end{align}
\end{subequations}

\subsection{Step 3}
After three steps of the evolution, the polarized dipole amplitudes that we need to find hPDFs become
\begin{subequations}
    \begin{align}
        G^{(3)}_2(x_{10}^2, zs)
=&\frac{1}{30} \ \left(\frac{\alpha_s N_c}{4\pi} \right)^3 \ln^5(zsx_{10}^2)\ln\frac{1}{\Lambda^2x_{10}^2} \left[\frac{N_f}{N_c} \left(-\frac{2N_f}{N_c} + \frac{35}{2} \right) \widetilde{Q}^{(0)} + \left( 112 - \frac{22N_f}{N_c} \right) G_2^{(0)}\right]\\
&+\frac{1}{24}\left(\frac{\alpha_s N_c}{4\pi} \right)^3 \ln^4(zsx_{10}^2) \ln^2\frac{1}{\Lambda^2 x_{10}^2} \left[\frac{N_f}{N_c} \left( 39 -\frac{4N_f}{N_c} \right) \widetilde{Q}^{(0)} + 2 \left( 128-\frac{26N_f}{N_c} \right) G_2^{(0)}\right] \notag \\
&+\frac{1}{36} \left(\frac{\alpha_s N_c}{4\pi} \right)^3\ln^3(zsx_{10}^2) \ln^3\frac{1}{\Lambda^2 x_{10}^2}\left[\frac{N_f}{N_c} 
\left(73-\frac{4N_f}{N_c} \right) \widetilde{Q}^{(0)} + 2 \left( 256-\frac{34N_f}{N_c} \right) G_2^{(0)}\right], \notag \\
\widetilde{Q}^{(3)}(x_{10}^2, zs) 
= &-\frac{1}{30}\left(\frac{\alpha_s N_c}{4\pi} \right)^3\ln^5(zsx_{10}^2)\ln\frac{1}{\Lambda^2x_{10}^2}  \left[ \left( \frac{10N_f}{N_c}-\frac{1}{2} \right)\widetilde{Q}^{(0)} + 2 \left( 35-\frac{4N_f}{N_c} \right) G_2^{(0)}\right]\\
&-\frac{1}{24}\left(\frac{\alpha_s N_c}{4\pi} \right)^3\ln^4(zsx_{10}^2)\ln^2\frac{1}{\Lambda^2x_{10}^2} \left[ \left( \frac{24N_f}{N_c} -1 \right ) \widetilde{Q}^{(0)} + 2 \left( 78 - \frac{8N_f}{N_c} \right) G_2^{(0)}\right] \notag \\
&-\frac{1}{36}\left(\frac{\alpha_s N_c}{4\pi} \right)^3 \ln^3(zsx_{10}^2)\ln^3\frac{1}{\Lambda^2x_{10}^2} \left[ \left( \frac{40N_f}{N_c}-1 \right) \widetilde{Q}^{(0)} + 2 \left( 146-\frac{8N_f}{N_c} \right) G_2^{(0)}\right]. \notag 
    \end{align}
\end{subequations}
The corresponding hPDFs are 
\begin{subequations}
    \begin{align}
\Delta G^{(3)}(x, Q^2) 
=&\frac{1}{30}\left(\frac{\alpha_s N_c}{4\pi} \right)^3 \ln^5\frac{1}{x}\ln\frac{Q^2}{\Lambda^2}  \left[ \left( 35-\frac{4N_f}{N_c} \right) \Delta \Sigma^{(0)} + 2 \left( 56 - \frac{11N_f}{N_c} \right ) \Delta G^{(0)}\right]\\
&+\frac{1}{24}\left(\frac{\alpha_s N_c}{4\pi} \right)^3 \ln^4\frac{1}{x} \ln^2\frac{Q^2}{\Lambda^2} \left[2 \left( 39 -\frac{4N_f}{N_c} \right) \Delta\Sigma^{(0)} + 4 \left( 64-\frac{13N_f}{N_c} \right) \Delta G^{(0)}\right] \notag \\
&+ \frac{1}{36}\left(\frac{\alpha_s N_c}{4\pi} \right)^3\ln^3\frac{1}{x} \ln^3\frac{Q^2}{\Lambda^2 }\left[2 \left( 73-\frac{4N_f}{N_c} \right) \Delta \Sigma ^{(0)} + 4 \left(128-\frac{17N_f}{N_c} \right) \Delta G^{(0)}\right], \notag \\ 
\Delta \Sigma^{(3)}(x, Q^2) 
=&-\frac{1}{30}\left(\frac{\alpha_s N_c}{4\pi} \right)^3\ln^5\frac{1}{x}\ln\frac{Q^2}{\Lambda^2} \left[\left( \frac{10N_f}{N_c}-\frac{1}{2} \right) \Delta \Sigma^{(0)} + \frac{N_f}{N_c}\left( 35-\frac{4N_f}{N_c} \right) \Delta G^{(0)}\right]\\
&-\frac{1}{24}\left(\frac{\alpha_s N_c}{4\pi} \right)^3\ln^4\frac{1}{x}\ln^2\frac{Q^2}{\Lambda^2} \left[\left( \frac{24N_f}{N_c} -1 \right) \Delta\Sigma^{(0)} + \frac{N_f}{N_c} \left(78 - \frac{8N_f}{N_c} \right) \Delta G^{(0)}\right] \notag \\
&-\frac{1}{36}\left(\frac{\alpha_s N_c}{4\pi} \right)^3 \ln^3\frac{1}{x}\ln^3\frac{Q^2}{\Lambda^2} \left[\left( \frac{40N_f}{N_c}-1 \right) \Delta \Sigma^{(0)} + \frac{N_f}{N_c} \left( 146-\frac{8N_f}{N_c} \right) \Delta G^{(0)}\right]. \notag 
    \end{align}
\end{subequations}
Again, from the terms containing $\ln^5\frac{1}{x} \ln \frac{Q^2}{\Lambda^2}$, one reads out the polarized small-$x$ large-$N_c \& N_f$ splitting functions at the three-loop order 
\begin{equation}\label{eq:DeltaP_(2)}
    \Delta \mathbf{P}^{(2)} (x) = \frac{1}{6}\left(\frac{\alpha_s N_c}{4\pi}\right)^3 \ln^4\frac{1}{x}\begin{pmatrix}
    (\frac{1}{2} - \frac{10N_f}{N_c})\, & \, - \frac{N_f}{N_c}(35-\frac{4N_f}{N_c})\\[10pt]
    35-\frac{4N_f}{N_c}\, &\, 2(56-\frac{11N_f}{N_c}) \\
    \end{pmatrix}.
\end{equation}

\subsection{Summary of our results up to three loops}

To summarize the above results, we note that we obtained the splitting functions up to three loops by iteratively solving the small-$x$ helicity evolution equations. Collecting Eqs.~\eqref{eq:DeltaP_(0)},~\eqref{eq:DeltaP_(1)} and~\eqref{eq:DeltaP_(2)}, we have the following expressions
\begin{subequations}\label{eq:DeltaP_BER}
\begin{align}
& \Delta P_{qq} (x) = \frac{\as N_c}{4 \pi} +\left( \frac{\as N_c}{4 \pi} \right)^2 \left(\frac{1}{2} - 2 \, \frac{N_f}{N_c} \right)\ln^2 \frac{1}{x} \, +\left( \frac{\as N_c}{4 \pi} \right)^3 \frac{1}{12} \left(1-20\frac{N_f}{N_c}\right)\ln^4 \frac{1}{x}\,+ {\cal O} (\as^4) , \\
& \Delta P_{qG} (x) = - \left(\frac{\as N_c}{4 \pi}\right) \frac{2N_f}{N_c}-\left( \frac{\as N_c}{4 \pi} \right)^2 \, \frac{5N_f}{N_c}\ln^2 \frac{1}{x} - \left( \frac{\as N_c}{4 \pi} \right)^3\frac{1}{6}\frac{N_f}{N_c}\left( 35 - 4  \, \frac{N_f}{N_c} \right)\, \ln^4 \frac{1}{x} \, + {\cal O} (\as^4)  , \label{eq:P'_qG} \\
& \Delta P_{Gq} (x) = 2\left(\frac{\as N_c}{4 \pi}\right) + 5 \, \left( \frac{\as N_c}{4\pi} \right)^2 \, \ln^2 \frac{1}{x}+\left( \frac{\as N_c}{4 \pi} \right)^3 \, \frac{1}{6}\left( 35 - 4  \, \frac{N_f}{N_c} \right)\, \ln^4 \frac{1}{x} + {\cal O} (\as^4)  , \label{eq:P'_Gq}\\
& \Delta P_{GG} (x) = 8 \, \left(\frac{\as N_c}{4\pi}\right) + \left( \frac{\as N_c}{4\pi} \right)^2 \, \left( 16 - 2\frac{N_f}{ \, N_c} \right) \, \ln^2 \frac{1}{x} + \left( \frac{\as N_c}{4\pi} \right)^3 \, \frac{1}{3} \,  \left( 56 - 11 \, \frac{N_f}{N_c} \right) \, \ln^4 \frac{1}{x} + {\cal O} (\as^4) .
\end{align}
\end{subequations}
These splitting functions completely agree with the results of BER IREE expansion (obtained by applying the large-$N_c \& N_f$ limit to Eq.~(14) in \cite{Blumlein:1996hb}).

For comparison, in the $\overline{\text{MS}}$ scheme, the small-$x$ limits of the polarized splitting functions at large-$N_c \& N_f$ are \cite{Altarelli:1977zs,Dokshitzer:1977sg,Mertig:1995ny,Moch:2014sna},
\begin{subequations}\label{eq:DeltaP_MSbar}
\begin{align}
&\Delta \overline{P}_{qq}(x) = \left(\frac{\alpha_sN_c}{4\pi}\right) + \left(\frac{\alpha_s N_c}{4\pi}\right)^2 \left( \frac{1}{2}-2\frac{N_f}{N_c} \right)\ln^2\frac{1}{x} + \left(\frac{\alpha_sN_c}{4\pi}\right)^3\frac{1}{12} \left( 1-20\frac{N_f}{N_c} \right) \ln^4\frac{1}{x} + {\cal O} (\alpha_s^4) \,,     \label{Pqq} \\
&\Delta \overline{P}_{qG}(x) =  - \left(\frac{\alpha_sN_c}{4\pi}\right)\frac{2N_f}{N_c} - \left(\frac{\alpha_sN_c}{4\pi}\right)^2 5\frac{N_f}{N_c}\ln^2\frac{1}{x} - \left(\frac{\alpha_sN_c}{4\pi}\right)^3\frac{1}{6}\frac{N_f}{N_c} \left( 34-4\frac{N_f}{N_c} \right) \ln^4\frac{1}{x} + {\cal O} (\alpha_s^4) \,,     \label{eq:PqG} \\
&\Delta \overline{P}_{Gq}(x) =   2\left(\frac{\alpha_sN_c}{4\pi}\right) + 5\left(\frac{\alpha_s N_c}{4\pi}\right)^2 \ln^2\frac{1}{x} + \left(\frac{\alpha_s N_c}{4\pi}\right)^3\frac{1}{6} \left( 36-4\frac{N_f}{N_c} \right) \ln^4\frac{1}{x} + {\cal O} (\alpha_s^4)  \,,   \label{eq:PGq} \\
&\Delta \overline{P}_{GG}(x) =   8 \left(\frac{\alpha_sN_c}{4\pi}\right) + \left(\frac{\alpha_s N_c}{4\pi}\right)^2 \left( 16-2\frac{N_f}{N_c} \right) \ln^2\frac{1}{x} + \left(\frac{\alpha_s N_c}{4\pi}\right)^3\frac{1}{3} \left( 56-11\frac{N_f}{N_c} \right) \ln^4\frac{1}{x} + {\cal O} (\alpha_s^4) \,.   \label{PGG} 
\end{align}
\end{subequations}
We see that our (and BER) polarized splitting functions $\Delta P_{qq}$ and $\Delta P_{GG}$ agree with the $\overline{\text{MS}}$ ones, while $\Delta P_{qG}$ and $\Delta P_{Gq}$ appear to disagree slightly at the order-$\as^3$ with their $\overline{\text{MS}}$ counterparts. The same disagreement between the BER 3-loop results and $\overline{\text{MS}}$ was attributed to scheme dependence of splitting functions in \cite{Moch:2014sna}: in Appendix~\ref{B} we construct the scheme transformation relating Eqs.~\eqref{eq:DeltaP_BER} and \eqref{eq:DeltaP_MSbar} explicitly. Here we note that the three-loop calculation \cite{Moch:2014sna} was carried out in the Larin scheme \cite{Larin:1991tj, Larin:1993tq}, and the results were then transformed into the $\overline{\text{MS}}$ scheme (see \cite{Blumlein:2024euz} for a recent comparison of the two schemes).

Let us also note that it was pointed out in \cite{Blumlein:1996hb} that the splitting functions obtained from solving BER IREE evolution equations have the property that 
\begin{equation}\label{PqGGq}
    \Delta P_{qG}(x) = - \frac{N_f}{N_c} \Delta P_{Gq}(x)
\end{equation}
to all orders in $\alpha_s$. The same property seem to be true for our splitting functions \eqref{eq:DeltaP_BER} to three loops and, as we will see below, at four loops as well. However, the relation \eqref{PqGGq} is not obeyed by the polarized splitting functions \eqref{eq:DeltaP_MSbar} in the $\overline{\text{MS}}$ scheme. It is violated at the order-$\alpha_s^3$ as can be seen from Eqs.~\eqref{eq:PqG} and \eqref{eq:PGq}.


\subsection{Step 4}

One can continue iteratively solving the evolution equations (for simplicity further iteration was done while setting $\Delta\Sigma^{(0)} = 0$). The resulting four-loop splitting functions we obtain this way are
\begin{subequations}\label{splittingfunctions_4loops}
\begin{align}
    \label{Pqq_4loops}
    &\Delta P_{qq}^{(3)}(x) = \left(\frac{\alpha_s N_c}{4\pi}\right)^4 \frac{1}{720}\left(5 - 748\frac{N_f}{N_c} + 80\frac{N_f^2}{N_c^2} \right)\ln^6\frac{1}{x}, \\
    \label{PqG_4loops}
    &\Delta P_{qG}^{(3)}(x) = -\left(\frac{\alpha_s N_c}{4\pi}\right)^4 \frac{1}{360}\frac{N_f}{N_c}\left(1213 - 224\frac{N_f}{N_c}\right)\ln^6\frac{1}{x}, \\
    \label{PGq_4loops}
    &\Delta P_{Gq}^{(3)}(x) = \left(\frac{\alpha_s N_c}{4\pi}\right)^4 \frac{1}{360}\left(1213 - 224\frac{N_f}{N_c}\right)\ln^6\frac{1}{x}, \\
     \label{PGG_4loops}
    &\Delta P_{GG}^{(3)}(x) = \left(\frac{\alpha_s N_c}{4\pi}\right)^4 \frac{1}{180}\left(1984 - 549\frac{N_f}{N_c} + 20\frac{N_f^2}{N_c^2} \right)\ln^6\frac{1}{x}.
\end{align}
\end{subequations}
This is our prediction for the small-$x$ large-$N_c \& N_f$ polarized splitting functions at four loops. It obeys the property \eqref{PqGGq} stated above. We hope to be able to compare it to the appropriate limit of the finite-order four-loop result, when the latter is calculated. 

The four-loop splitting functions \eqref{splittingfunctions_4loops} are to be compared with the four loop predictions from the BER IREE expansion. Taking Eq.~(15) in \cite{Blumlein:1996hb} and applying the large-$N_c\&N_f$ limit, these are 
\begin{subequations}\label{splittingfunctions_4loops_BER}
\begin{align}
    \label{Pqq_4loops_BER}
    &\Delta P_{qq}^{(3)\,\text{(BER)}}(x) = \left(\frac{\alpha_s N_c}{4\pi}\right)^4 \frac{1}{720}\left(5 - 764\frac{N_f}{N_c} + 80\frac{N_f^2}{N_c^2} \right)\ln^6\frac{1}{x}, \\
    \label{PqG_4loops_BER}
    &\Delta P_{qG}^{(3)\,\text{(BER)}}(x) = -\left(\frac{\alpha_s N_c}{4\pi}\right)^4 \frac{1}{360}\frac{N_f}{N_c}\left(1229 - 224\frac{N_f}{N_c}\right)\ln^6\frac{1}{x}, \\
    \label{PGq_4loops_BER}
    &\Delta P_{Gq}^{(3)\,\text{(BER)}}(x) = \left(\frac{\alpha_s N_c}{4\pi}\right)^4 \frac{1}{360}\left(1229 - 224\frac{N_f}{N_c}\right)\ln^6\frac{1}{x}, \\
     \label{PGG_4loops_BER}
    &\Delta P_{GG}^{(3)\,\text{(BER)}}(x) = \left(\frac{\alpha_s N_c}{4\pi}\right)^4 \frac{1}{180}\left(2016 - 557\frac{N_f}{N_c} + 20\frac{N_f^2}{N_c^2} \right)\ln^6\frac{1}{x}.
\end{align}
\end{subequations}
By comparing Eqs.~\eqref{splittingfunctions_4loops} to Eqs.~\eqref{splittingfunctions_4loops_BER} we see that evidently the polarized splitting functions we obtain exhibit a very minor disagreement with those from the BER IREE expansion beginning at four loops. This is consistent with \cite{Borden:2023ugd}, where an analytic all-orders in $\as$ prediction for the polarized small-$x$ GG anomalous dimension at large-$N_c$ was derived from the large-$N_c$ version of this small-$x$ helicity evolution and also disagreed with that derived from the BER IREE beginning at four loops. The difference between Eqs.~\eqref{splittingfunctions_4loops} and \eqref{splittingfunctions_4loops_BER} cannot be accounted for by a scheme transformation since it persists even in the gluons-only large-$N_c$ limit, where there is only one splitting function, $\Delta P_{GG}$ \cite{Borden:2023ugd}.


\section{Conclusions and Outlook}
\label{sec:conclusions}

In this paper we have calculated the contribution of the $q/{\bar q} \to G$ and $G \to q/{\bar q}$ shock-wave transition operators to the small-$x$ evolution of the flavor-singlet quark and gluon helicity PDFs. These operators were previously considered in \cite{Chirilli:2021lif} in the framework of helicity evolution: however, a direct comparison with \cite{Chirilli:2021lif} appears to be complicated due to the difference in our decompositions into flavor-singlet and non-singlet components. While the transition operators do not contribute to helicity evolution in the large-$N_c$ limit which contains no quarks, they do contribute in the large-$N_c \& N_f$ limit. The main result of our work is the modified large-$N_c \& N_f$ helicity evolution equations \eqref{eq_LargeNcNf}. We showed that this evolution agrees with the (small-$x$ and large-$N_c \& N_f$ limit of the) existing polarized DGLAP splitting functions up to the three known loops, indicating that the equations  \eqref{eq_LargeNcNf} are likely to include all the contributions needed for small-$x$ helicity evolution. In the future, an analytic solution of Eqs.~\eqref{eq_LargeNcNf} can be found following the approach of \cite{Borden:2023ugd}.

Our calculations can also be used to further correct the helicity version of the Jalilian-Marian--Iancu--McLerran--Weigert--Leonidov--Kovner
(JIMWLK)
\cite{Jalilian-Marian:1997dw,Jalilian-Marian:1997gr,Weigert:2000gi,Iancu:2001ad,Iancu:2000hn,Ferreiro:2001qy} evolution equation, which was constructed in \cite{Cougoulic:2019aja} using the older version of helicity evolution, not containing the $\cev{D}^i \, D^i$ operator employed in \cite{Cougoulic:2022gbk} and the transition operators calculated here. Performing these corrections for the helicity version of JIMWLK evolution is left for the future work. 

The new corrections in equations \eqref{eq_LargeNcNf} imply that the results of the helicity phenomenology project \cite{Adamiak:2023yhz} need to be updated as well. Introducing the new operator $\widetilde Q$ will lead to more parameters describing the initial conditions (inhomogeneous terms) for Eqs.~\eqref{eq_LargeNcNf} than in the version of these equations used in \cite{Adamiak:2023yhz}. On the other hand, the new relations \eqref{relationsGQQ} may help one constrain some of these parameters, by relating the parameters to each other. The simplicity of Eqs.~\eqref{eq:DeltaGSigma_G2_Qtilde} may also help facilitate the future small-$x$ helicity phenomenology.


\section*{Acknowledgments}

One of the authors (YK) is grateful to Frank Krauss and IPPP Durham for their hospitality during the final stages of this work. 

This material is based upon work supported by
the U.S. Department of Energy, Office of Science, Office of Nuclear
Physics under Award Number DE-SC0004286 and within the framework of the Saturated Glue (SURGE) Topical Theory Collaboration.

The Feynman diagrams in this work were drawn using JaxoDraw \cite{BINOSI200476}. \\


\appendix

 \section{Derivation of the large-$N_c \& N_f$ Evolution Equations using Light-Cone Perturbation Theory}
 \renewcommand{\theequation}{A\arabic{equation}}
   \setcounter{equation}{0}
 \label{A}

The contribution of the shock-wave transition operators to the evolution of $G^{\text{adj}}_{10}$ and the evolution of the new structure $\widetilde{Q}_{10}$ were derived in Sections~\ref{sec:Gadj_evol} and \ref{sec:Qtilde_evol} using the LCOT approach \cite{Kovchegov:2017lsr,Kovchegov:2018znm,Cougoulic:2022gbk,Kovchegov:2021iyc}, resulting in Eqs.~\eqref{eq_LargeNcNf}. The goal of this Appendix is to show that the same evolution equations can be obtained using the tools of light-cone perturbation theory (LCPT), similarly to that done in \cite{Kovchegov:2015pbl} and in Sec.~4.2 of \cite{Tawabutr:2022gei}. As in the main text, we still work in the $A^- = 0$ gauge with the gluon polarization vector still given by the expression in the main text following Eqs.~\eqref{Ws}. The spirit of the method used in this Appendix is to treat the interactions with the shock wave in terms of operator insertions (this is the same as in LCOT) but to treat everything outside the shock wave in terms of the light-cone wavefunctions of LCPT. To that end we will need expressions for these wavefunctions. To keep this Appendix self-contained, we first reproduce here the necessary elements from Section~III of \cite{Kovchegov:2015pbl}.


\subsection{Light-cone Wavefunctions}

\begin{figure}[ht]
    \centering
    \includegraphics[scale=0.9]{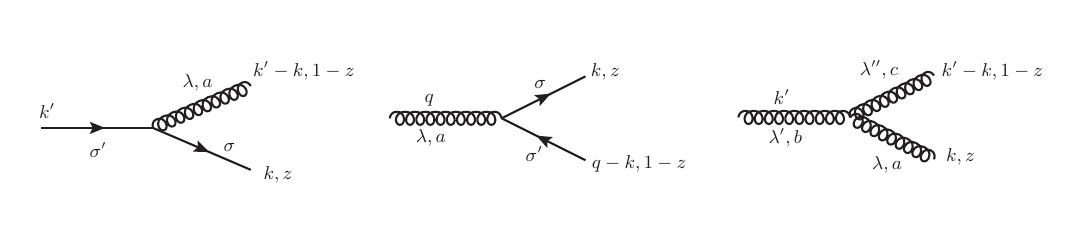}
    \caption{Leading order contribution to the $q\rightarrow qG$, $G\rightarrow q\bar{q}$, and $G\rightarrow GG$ light-cone wavefunctions.}
    \label{fig:appA_lcwfs}
\end{figure}

The leading order in $\alpha_s$ momentum-space expression for the $q\rightarrow qG$ wavefunction shown in \fig{fig:appA_lcwfs} is
\begin{align}\label{appA_q_to_qG_wf}
    \psi^{q\rightarrow qG} = -g t^a \delta_{\sigma\sigma'} \sqrt{z}\,\frac{\underline{\epsilon}_{\lambda}^* \cdot \left(\underline{k}-z\underline{k}'\right)}{|\underline{k}-z\underline{k}'|^2} \left[1+z+\sigma\lambda\left(1-z\right)\right].
\end{align}
In particular we will need the limits where one of the outgoing particles is soft. These limits, Fourier-transformed to transverse coordinate space, are
\begin{subequations}
    \begin{align}\label{appA_q_to_qG_wf_softq}
    & \psi^{q\rightarrow qG}|_{z\rightarrow 0} \approx -\frac{igt^a}{2\pi}\delta_{\sigma\sigma'}\sqrt{z}\, \frac{\underline{\epsilon}_\lambda^*\cdot \underline{x}}{\underline{x}^2}\left[1+\sigma\lambda\right], \quad \text{(soft quark)}, \\
\label{appA_q_to_qG_wf_softG}
    & \psi^{q\rightarrow qG}|_{z\rightarrow 1} \approx \frac{igt^a}{2\pi}\delta_{\sigma\sigma'} \frac{\underline{\epsilon}_\lambda^*\cdot \underline{x}}{\underline{x}^2}\left[2+\sigma\lambda\left(1-z\right)\right], \quad \text{(soft gluon)}.
\end{align}
\end{subequations}
In the above expressions, $\underline{x}$ is the transverse separation between the soft outgoing particle and the incoming particle. In addition, \eq{appA_q_to_qG_wf_softG} omits some polarization-independent terms that do not contribute to the helicity evolution at hand. Note that we can also use these same expressions, Eqs.~\eqref{appA_q_to_qG_wf_softq} and \eqref{appA_q_to_qG_wf_softG}, for $\bar{q}\rightarrow \bar{q}G$ splitting in the diagrams to follow, with the caveat that we need an additional overall minus sign to properly account for the Wick contractions associated with the antiquark fields. 

The leading order in $\alpha_s$ momentum-space expression for the $G\rightarrow q\bar{q}$ wavefunction shown in \fig{fig:appA_lcwfs} is
\begin{align}\label{appA_G_to_qq_wf}
    \psi^{G\rightarrow q\bar{q}} = gt^a \sqrt{z\left(1-z\right)}\, \frac{\underline{\epsilon}_\lambda\cdot\left(\underline{k}-z\underline{q}\right)}{|\underline{k}-z\underline{q}|^2} \delta_{\sigma,-\sigma'}\left[1-2z-\sigma\lambda\right].
\end{align}
The transverse coordinate space expressions in the limits where an outgoing particle is soft are
\begin{subequations}
    \begin{align}\label{appA_G_to_qq_wf_softq}
    & \psi^{G\rightarrow q\bar{q}}|_{z\rightarrow 0} \approx \frac{igt^a}{2\pi} \sqrt{z}\, \frac{\underline{\epsilon}_\lambda\cdot\underline{x}}{\underline{x}^2} \delta_{\sigma,-\sigma'}\left[1-\sigma\lambda\right], \quad \text{(soft quark)}, \\
\label{appA_G_to_qq_wf_softqbar}
    & \psi^{G\rightarrow q\bar{q}}|_{z\rightarrow 1} \approx \frac{igt^a}{2\pi} \sqrt{1-z}\, \frac{\underline{\epsilon}_\lambda\cdot\underline{x}}{\underline{x}^2} \delta_{\sigma,-\sigma'}\left[1-\sigma'\lambda\right], \quad \text{(soft antiquark)}.
\end{align}
\end{subequations}

Finally, the leading order in $\alpha_s$ momentum-space expression for the $G\rightarrow GG$ wavefunction shown in Fig.~\ref{fig:appA_lcwfs} is 
\begin{align}\label{appA_G_to_GG_wf}
    \psi^{G\rightarrow GG} = 2igf^{abc}\, \frac{z\left(1-z\right)}{|\underline{k}-z\underline{k}'|^2} \left[\frac{1}{1-z} \underline{\epsilon}_{\lambda''}^*\cdot\left(\underline{k}-z\underline{k}' \right)\underline{\epsilon}_\lambda^*\cdot \underline{\epsilon}_{\lambda'} + \frac{1}{z}\underline{\epsilon}_\lambda^*\cdot \left( \underline{k}-z\underline{k}'\right) \underline{\epsilon}_{\lambda''}^*\cdot \underline{\epsilon}_{\lambda'} - \underline{\epsilon}_{\lambda'}\cdot \left( \underline{k}-z\underline{k}'\right) \underline{\epsilon}_{\lambda''}^*\cdot \underline{\epsilon}_{\lambda'}^*  \right].
\end{align}
The resulting transverse coordinate space expression for the soft outgoing gluon limit is 
\begin{align}\label{appA_G_to_GG_wf_softG}
    \psi^{G\rightarrow GG}|_{z\rightarrow 0} \approx \frac{-g f^{abc}}{\pi}\frac{\underline{\epsilon}_\lambda^*\cdot \underline{x}}{\underline{x}^2}\delta_{\lambda''\lambda'}, \quad \text{(soft gluon)}.
\end{align}
Here again we have kept only the leading sub-eikonal, polarization-dependent terms.


\subsection{Contribution to the evolution of the adjoint dipole amplitude of the first type}
\label{sec:adj_dip}

In this Subsection we derive the contribution of the shock-wave transition operators to the evolution of $G^{\text{adj}}_{10}$, in order to ultimately reproduce the results of Sec.~\ref{sec:Gadj_evol}. We begin with diagrams A and A' from \fig{FIG:Gadj_evol}, reproduced here with a few modifications to the labels.

\begin{figure}[ht]
    \centering
    \includegraphics{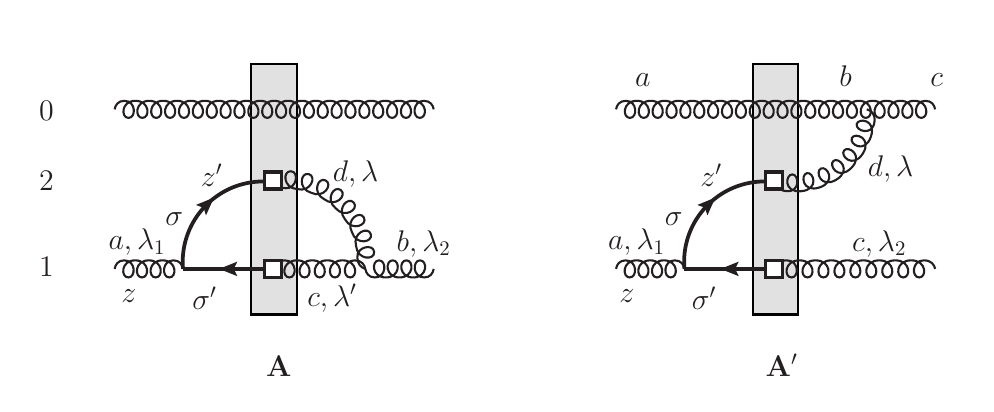}
    \caption{Diagrams A and A', originally considered in \fig{FIG:Gadj_evol}.}
    \label{fig:appA_diagramsAA'}
\end{figure}

We will still treat the interactions with the shock wave in terms of operators, specifically those in Eqs.~\eqref{Ws}. For everything outside the shock wave, we employ the light cone wavefunctions from \fig{fig:appA_lcwfs}. We write for the diagram A:
\begin{align}\label{appA_diagramA1}
    &A = \frac{1}{2\left(N_c^2 - 1\right)}\left(\frac{1}{2}\sum_{\lambda_1,\lambda_2}\lambda_1\right)\sum_f\sum_{\sigma,\sigma',\lambda,\lambda'} \int\frac{\mathrm{d}z'}{z'}\int \frac{\mathrm{d}^2\underline{x}_2}{4\pi}\theta\left(x_{10}^2z - x_{21}^2z'\right)\\
    &\hspace{1cm}\times\bigg\langle U_{\underline{0}}^{ba}\left[\psi^{G\rightarrow GG}_{dbc;\lambda,\lambda_2,\lambda'}\left(\underline{x}_{21}, \frac{z'}{z}\right)\bigg|_{\tfrac{z'}{z}\rightarrow 0}\right]^*\left(W_{\underline{2}}^{Gq}\left[\infty,-\infty\right]\right)^{d,\lambda;i,\sigma} \left[\left(\psi^{G\rightarrow q\bar{q}}_{a,\lambda_1;\sigma,\sigma'}\right)^{ij}\left(\underline{x}_{21},\frac{z'}{z}\right)\bigg|_{\tfrac{z'}{z}\rightarrow 0}  \right] \notag \\
    &\hspace{2cm}\times\left(W_{\underline{1}}^{G\bar{q}}\left[\infty,-\infty\right]\right)^{c,\lambda'; j,\sigma'} \bigg\rangle \notag.
\end{align}
The factor of $1/2$ in front averages over incoming helicities. We also weigh our sum by the incoming helicity $\lambda_1$ in order to pick out the helicity-dependent term of the sub-eikonal gluon S-matrix, since this is the one that contributes to $G^{\text{adj}}_{10}$ (see Eqs.~\eqref{Uxy_sub-eikonal}, \eqref{G_adj_def}, and \eqref{Upol1_def}). 
Here we have included both adjoint and fundamental color indices as well as polarization arguments for clarity. Note also that we treat both the shock-wave transition operators as being part of the (forward) amplitude rather than one being in the amplitude and another one in the complex conjugate amplitude. Hence we use $W^{Gq}$ and $W^{G\bar{q}}$, both with endpoints running from $-\infty$ to $\infty$. In addition, we have a theta function in the first line of \eq{appA_diagramA1} to enforce light-cone lifetime ordering \cite{Kovchegov:2015pbl,Kovchegov:2018znm,Cougoulic:2022gbk}.

Now we substitute our splitting wavefunctions and shock-wave transition operators into \eq{appA_diagramA1}. Here we opt to employ the explicit expressions in Eqs.~\eqref{Ws} for the transition operators from the outset, obtaining
\begin{align}\label{appA_diagramA2}
    &A = \frac{1}{2\left(N_c^2 - 1\right)}\left(\frac{1}{2}\sum_{\lambda_1,\lambda_2}\lambda_1\right)\sum_f\sum_{\sigma,\sigma',\lambda,\lambda'} \int\frac{\mathrm{d}z'}{z'}\int \frac{\mathrm{d}^2\underline{x}_2}{4\pi}\theta\left(x_{10}^2z - x_{21}^2z'\right)\\
    &\hspace{0.25cm}\times\Bigg\langle U_{\underline{0}}^{ba}\left[\frac{-g}{\pi} f^{dbc} \delta_{\lambda_2,\lambda'} \frac{\underline{\epsilon}_{\lambda}^*\cdot\underline{x}_{21}}{x_{21}^2}\right]^* \notag \\
    &\hspace{0.75cm}\left(\int\limits_{-\infty}^{\infty}\mathrm{d}w_2^- \,U_{\underline{2}}^{de}\left[\infty,w_2^-\right] \frac{-ig}{2\sqrt{\sqrt{2}k^-}}\bar{\psi}_\alpha\left(w_2^-,\underline{x}_2\right) t^{e} \delta_{\sigma,\lambda} \left[\sigma\left(\rho_\alpha(-)-\rho_\alpha(+) \right) + \left(\rho_\alpha(-)+\rho_\alpha(+) \right)   \right] V_{\underline{2}}\left[w_2^-,-\infty \right]\right) \notag \\
    &\hspace{0.75cm} \left[\frac{ig}{2\pi}t^a\delta_{\sigma,-\sigma'}\sqrt{\frac{z'}{z}}\left(1-\sigma\lambda_1 \right)\frac{\underline{\epsilon}_{\lambda_1}\cdot\underline{x}_{21}}{x_{21}^2}   \right] \notag \\
    &\hspace{0.75cm}\left(\int\limits_{-\infty}^{\infty}\mathrm{d}w_1^- \,U_{\underline{1}}^{cf}\left[\infty,w_1^-\right] \frac{-ig}{2\sqrt{\sqrt{2}p_2^-}}    V_{\underline{1}}^\dagger\left[w_1^-,-\infty \right] \delta_{\sigma',\lambda'} \left[ \sigma'\left(\bar{\rho}_\beta(-)-\bar{\rho}_\beta(+) \right) - \left(\bar{\rho}_\beta(-)+\bar{\rho}_\beta(+) \right)  \right]  t^f \psi_\beta\left(w_1^-,\underline{x}_1 \right)\right)\bigg\rangle \notag.
\end{align}
We have now suppressed the fundamental color indices but explicitly included the spinor indices $\alpha,\beta$. We also suppose that the soft quark entering the shock wave at transverse position 2 comes in with momentum $k$, while the momentum of the parent dipole is $p_2$. Both of these momenta are assumed to have large minus components. With these explicit factors of momentum, and with the factor $\tfrac{1}{\sqrt{z}}$ from the $G\rightarrow q\bar{q}$ light-cone wave function, we can write
\begin{align}\label{appA_diagramA3}
    \frac{1}{\sqrt{k^-p_2^-}}\frac{1}{\sqrt{z}} \approx \frac{1}{\sqrt{\left(z'P^- \right)\left(zP^-\right) } } \frac{1}{\sqrt{z}} = \frac{1}{\sqrt{z'}}\frac{2p_1^+}{zs}
\end{align}
where we have taken the projectile to have a large minus component of momentum $P^-$ and the target proton to have a large plus component of momentum $p_1^+$ with $s=2p_1^+P^-$ the center of mass energy squared between them. (The adjoint parent dipole in \fig{fig:appA_diagramsAA'} need not be the original projectile and could have instead been produced by previous steps of evolution). Employing this fact and carrying out some algebra we arrive at 
\begin{align}\label{appA_diagramA4}
    &A = \frac{1}{z}\frac{1}{2\left(N_c^2 - 1\right)}4i \frac{\alpha_s N_f}{\pi^2}\int\frac{\mathrm{d}z'}{z'}\sqrt{z'}\int \frac{\mathrm{d}^2\underline{x}_2}{x_{21}^2}\theta\left(x_{10}^2z - x_{21}^2z'\right) \\
    &\hspace{2cm}\times\bigg\langle \frac{g^2p_1^+}{8\sqrt{z'}s}\int\limits_{-\infty}^{\infty}\mathrm{d}w_1^-\int\limits_{-\infty}^{\infty}\mathrm{d}w_2^- \,U_{\underline{0}}^{ba} f^{dbc} U_{\underline{2}}^{de}\left[\infty, w_2^-\right] \bar{\psi}_\alpha\left(w_2^-,\underline{x}_2\right) t^e V_{\underline{2}}\left[w_2^-,-\infty\right] t^a U_{\underline{1}}^{cf}\left[\infty,w_1^-\right] \notag \\
    &\hspace{3cm} \times V_{\underline{1}}^\dagger\left[w_1^-,-\infty\right] t^f \left(\frac{1}{2}\gamma^+\gamma^5 \right)_{\alpha\beta}\psi_{\beta}\left(w_1^-,\underline{x}_1\right)\bigg\rangle . \notag
\end{align}
We have replaced $\sum_f$ with $N_f$ for simplicity. Employing the double angle brackets defined in \eq{double_def}, we obtain
\begin{align}\label{appA_diagramA5}
    &A = \frac{1}{zs}\frac{1}{2\left(N_c^2 - 1\right)}4i \frac{\alpha_s N_f}{\pi^2}\int\frac{\mathrm{d}z'}{z'}\int \frac{\mathrm{d}^2\underline{x}_2}{x_{21}^2}\theta\left(x_{10}^2z - x_{21}^2z'\right) \\
    &\hspace{2cm}\times\llangle \frac{g^2p_1^+}{8\sqrt{z'}s}\int\limits_{-\infty}^{\infty}\mathrm{d}w_1^-\int\limits_{-\infty}^{\infty}\mathrm{d}w_2^- \,U_{\underline{0}}^{ba} f^{dbc} U_{\underline{2}}^{de}\left[\infty, w_2^-\right] \bar{\psi}_\alpha\left(w_2^-,\underline{x}_2\right) t^e V_{\underline{2}}\left[w_2^-,-\infty\right] t^a U_{\underline{1}}^{cf}\left[\infty,w_1^-\right] \notag \\
    &\hspace{3cm} \times V_{\underline{1}}^\dagger\left[w_1^-,-\infty\right] t^f \left(\frac{1}{2}\gamma^+\gamma^5 \right)_{\alpha\beta}\psi_{\beta}\left(w_1^-,\underline{x}_1\right)\rrangle \notag.
\end{align}
Diagram A' is similar. We write
\begin{align}\label{appA_diagramA'1}
    &A' = \frac{1}{2\left(N_c^2 - 1\right)}\left(\frac{1}{2}\sum_{\lambda_1,\lambda_2}\lambda_1\right)\sum_f\sum_{\sigma,\sigma',\lambda} \int\frac{\mathrm{d}z'}{z'}\int \frac{\mathrm{d}^2\underline{x}_2}{4\pi}\theta\left(x_{10}^2z - \text{max}\{x_{20}^2,x_{21}^2\}z'\right)\\
    &\hspace{1cm}\times\bigg\langle U_{\underline{0}}^{ba}\left[\psi^{G\rightarrow GG}_{dcb;\lambda}\left(\underline{x}_{20}, \frac{z'}{z}\right)\bigg|_{\tfrac{z'}{z}\rightarrow 0}\right]^*\left(W_{\underline{2}}^{Gq}\left[\infty,-\infty\right]\right)^{d,\lambda;i,\sigma} \left[\left(\psi^{G\rightarrow q\bar{q}}_{a,\lambda_1;\sigma,\sigma'}\right)^{ij}\left(\underline{x}_{21},\frac{z'}{z}\right)\bigg|_{\tfrac{z'}{z}\rightarrow 0}  \right] \notag \\
    &\hspace{2cm}\times\left(W_{\underline{1}}^{G\bar{q}}\left[\infty,-\infty\right]\right)^{c,\lambda_2; j,\sigma'} \bigg\rangle \notag.
\end{align}
Note that the light-cone lifetime-ordering theta function has been modified to account for both resulting (daughter) dipole sizes \cite{Kovchegov:2015pbl,Kovchegov:2018znm,Cougoulic:2022gbk}. Following the same steps as with diagram A, we obtain the following result for diagram A',
\begin{align}\label{appA_diagramA'3}
    &A' = \frac{1}{zs}\frac{1}{2\left(N_c^2-1\right)}4i \frac{\alpha_s N_f}{\pi^2}\int\frac{\mathrm{d}z'}{z'}\int \mathrm{d}^2\underline{x}_2 \theta\left(x_{10}^2z - \text{max}\{x_{20}^2,x_{21}^2\}z'\right)\frac{\underline{x}_{20}\cdot\underline{x}_{21}}{x_{20}^2x_{21}^2} \\ &\hspace{1cm}\times\llangle \frac{g^2p_1^+}{8\sqrt{z'}s} \int\limits_{-\infty}^{\infty}\mathrm{d}w_1^-\int\limits_{-\infty}^{\infty}\mathrm{d}w_2^-\, U_{\underline{0}}^{ba}f^{dcb}U_{\underline{2}}^{de}\left[\infty,w_2^-\right] \bar{\psi}_{\alpha}\left(w_2^-,\underline{x}_2\right)  t^e V_{\underline{2}}\left[w_2^-,-\infty\right] t^a U_{\underline{1}}^{cf}\left[\infty,w_1^-\right] V_{\underline{1}}^\dagger\left[w_1^-,-\infty \right] \notag \\
    &\hspace{5cm}\times t^f \left(\frac{1}{2}\gamma^+\gamma^5\right)_{\alpha\beta}\psi_{\beta}\left(w_1^-,\underline{x}_1 \right)\rrangle \notag.
\end{align}
We have neglected a term proportional to $\underline{x}_{20} \times \underline{x}_{21}$ which does not contribute in DLA. 

For diagrams B and B' from \fig{FIG:Gadj_evol} it remains true as in Sec. \ref{sec:Gadj_evol} that 
\begin{align}\label{appA_diagramBB'1}
    \left(B+B'\right) = \left(A+A'\right)^*.
\end{align}
A direct calculation within the framework of this Appendix bears this out, but for brevity we do not show it here. 

Next we consider diagrams C and C' from \fig{FIG:Gadj_evol}. For diagram C we can write
\begin{align}\label{appA_diagramC1}
    &C = \frac{1}{2\left(N_c^2 - 1\right)}\left(\frac{1}{2}\sum_{\lambda_1,\lambda_2}\lambda_1\right)\sum_f\sum_{\sigma,\sigma',\lambda,\lambda'} \int\frac{\mathrm{d}z'}{z'}\int \frac{\mathrm{d}^2\underline{x}_2}{4\pi}\theta\left(x_{10}^2z - x_{21}^2z'\right)\\
    &\hspace{1cm}\times\bigg\langle U_{\underline{0}}^{ba}\left[\psi^{G\rightarrow GG}_{dac;\lambda,\lambda_1,\lambda'}\left(\underline{x}_{21}, \frac{z'}{z}\right)\bigg|_{\tfrac{z'}{z}\rightarrow 0}\right]\left(W_{\underline{1}}^{\bar{q}G}\left[\infty,-\infty\right]\right)^{j,\sigma';c,\lambda'} \left[\left(\psi^{G\rightarrow q\bar{q}}_{b,\lambda_2;\sigma,\sigma'}\right)^{ji}\left(\underline{x}_{21},\frac{z'}{z}\right)\bigg|_{\tfrac{z'}{z}\rightarrow 0}  \right]^* \notag \\
    &\hspace{2cm}\times\left(W_{\underline{2}}^{qG}\left[\infty,-\infty\right]\right)^{i,\sigma;d,\lambda} \bigg\rangle \notag.
\end{align}
Just as we did for diagram A, we substitute the light-cone wave functions and the explicit expressions for the shock-wave transition operators. Some simplification yields
\begin{align}\label{appA_diagramC2}
    &C = -\frac{1}{zs}\frac{1}{2\left(N_c^2-1\right)} 4i\frac{\alpha_s N_f}{\pi^2}\int\frac{\mathrm{d}z'}{z'}\int\frac{\mathrm{d}^2\underline{x}_2}{x_{21}^2}\theta\left(x_{10}^2z-x_{21}^2z'\right) \\
    &\hspace{1.5cm}\times\llangle \frac{g^2 p_1^+}{8\sqrt{z'}s} \int\limits_{-\infty}^{\infty}\mathrm{d}w_1^-\int\limits_{-\infty}^{\infty}\mathrm{d}w_2^- \,U_{\underline{0}}^{ba}f^{dac}\bar{\psi}_{\alpha}\left(w_1^-,\underline{x}_{1}\right) t^{e} V_{\underline{1}}^\dagger\left[\infty,w_1^-\right] U_{\underline{1}}^{ec}\left[w_1^-,-\infty\right] t^{b}V_{\underline{2}}\left[\infty,w_2^-\right]t^{f} \notag \\
    &\hspace{2.5cm}\times\left(\frac{1}{2}\gamma^+\gamma^5 \right)_{\alpha\beta} \psi_\beta\left(w_2^-,\underline{x}_2 \right) U_{\underline{2}}^{fd}\left[w_2^-,-\infty\right]\rrangle. \notag
\end{align}

Diagram C' from \fig{FIG:Gadj_evol} is calculated similarly. The result is 
\begin{align}\label{appA_diagramC'1}
    &C' = -\frac{1}{zs}\frac{1}{2\left(N_c^2-1\right)} 4i \frac{\alpha_s N_f}{\pi^2} \int\frac{\mathrm{d}z'}{z'}\int\mathrm{d}^2\underline{x}_2\, \theta\left(x_{10}^2z - \text{max}\{x_{20}^2,x_{21}^2\}z' \right) \frac{\underline{x}_{20}\cdot \underline{x}_{21}}{x_{20}^2x_{21}^2} \\
    &\hspace{1.5cm}\times \llangle \frac{g^2p_1^+}{8\sqrt{z'}s} \int\limits_{-\infty}^{\infty}\mathrm{d}w_1^-\int\limits_{-\infty}^{\infty}\mathrm{d}w_2^-\, U_{\underline{0}}^{ba} f^{dca} \bar{\psi}_{\alpha}\left(w_1^-,\underline{x}_1\right) t^{e} V_{\underline{1}}^{\dagger}\left[\infty,w_1^-\right] U_{\underline{1}}^{ec}\left[w_1^-,-\infty\right] t^{d} V_{\underline{2}}\left[\infty,w_2^-\right] t^{f} \notag \\
    &\hspace{2.5cm}\times\left(\frac{1}{2}\gamma^+\gamma^5 \right)_{\alpha\beta} \psi_{\beta}\left(w_2^-,\underline{x}_2 \right) U_{\underline{2}}^{fd}\left[w_2^-,-\infty\right]\rrangle \notag.
\end{align}
Again a straightforward calculation in the framework of this Appendix shows that
\begin{align}\label{appA_diagramDD'}
    (D+D') = (C+C')^*
\end{align}
Adding together all the contributions to the evolution of $G^{\text{adj}}_{10}$ we obtain
\begin{align}\label{appA_alladjdiagrams1}
    &A+A'+B+B'+C+C'+D+D' \bigg|_{\text{DLA}} = 
    \frac{4i}{2\left(N_c^2 - 1\right)} \frac{\alpha_s N_f}{\pi}\int\limits_{\Lambda^2/s}^{z}\frac{\mathrm{d}z'}{z'}\int\limits_{1/z's}^{x_{10}^2}\frac{\mathrm{d}x_{21}^2}{x_{21}^2} \\
    &\hspace{1cm}\times\llangle \frac{g^2p_1^+}{8\sqrt{z'}s}\int\limits_{-\infty}^{\infty}\mathrm{d}w_1^-\int\limits_{-\infty}^{\infty}\mathrm{d}w_2^- \,\bigg(U_{\underline{0}}^{ba} f^{dbc} U_{\underline{2}}^{de}\left[\infty, w_2^-\right] \bar{\psi}_\alpha\left(w_2^-,\underline{x}_2\right) t^e V_{\underline{2}}\left[w_2^-,-\infty\right] t^a U_{\underline{1}}^{cf}\left[\infty,w_1^-\right] \notag \\
    &\hspace{6cm} \times V_{\underline{1}}^\dagger\left[w_1^-,-\infty\right] t^f \left(\frac{1}{2}\gamma^+\gamma^5 \right)_{\alpha\beta}\psi_{\beta}\left(w_1^-,\underline{x}_1\right) \notag \\
    &\hspace{5cm} + U_{\underline{0}}^{ba}f^{dca}\bar{\psi}_{\alpha}\left(w_1^-,\underline{x}_{1}\right) t^{e} V_{\underline{1}}^\dagger\left[\infty,w_1^-\right] U_{\underline{1}}^{ec}\left[w_1^-,-\infty\right] t^{b}V_{\underline{2}}\left[\infty,w_2^-\right]t^{f} \notag \\
    &\hspace{6cm}\times\left(\frac{1}{2}\gamma^+\gamma^5 \right)_{\alpha\beta} \psi_\beta\left(w_2^-,\underline{x}_2 \right) U_{\underline{2}}^{fd}\left[w_2^-,-\infty\right] \quad +\quad \text{c.c.} \bigg) \rrangle \notag
\end{align}
where we cancelled the factor of $1/zs$ since the same factor appears on the left hand side of the evolution equation (upon applying the double angle brackets) and we also simplified our kernel in DLA, where the transverse integral is only logarithmic for $x_{21}\ll x_{10} \sim x_{20}$. Next we take the large-$N_c$ limit, using \eq{Uab}
along with the Fierz identity. We neglect all terms sub-leading in $N_c$ and set the unpolarized S-matrices we obtain to 1 (which is valid at DLA). At large-$N_c$ the structure in double angle brackets in \eq{appA_alladjdiagrams1} becomes $\widetilde{Q}_{21}$ from Sec.~\ref{sec:Gadj_evol} (see \eq{Qtilde} and simply use the relations $k^- = z'p_2^-$, $s = 2p_1^+p_2^-$ from that section to reconcile the prefactors). The result at large $N_c\&N_f$ is
\begin{align}
    A+A'+B+B'+C+C'+D+D' \bigg|_{\text{DLA}} = - \frac{\alpha_s N_f}{\pi}\int\limits_{\Lambda^2/s}^{z}\frac{\mathrm{d}z'}{z'}\int\limits_{1/z's}^{x_{10}^2}\frac{\mathrm{d}x_{21}^2}{x_{21}^2} \, \widetilde{Q}_{12}\left(z's\right).
\end{align}
Lastly, to write the evolution equations in terms of the dipole amplitude $\widetilde{G}_{10}$ instead of $G^{\text{adj}}_{10}$, we divide the contribution obtained here by $4$ (see \eq{Gadj_Gtilde2}). The result is 
\begin{align}\label{appA_alldiagrams4}
    \widetilde{G}_{10}\left(zs\right) \supset -\frac{\alpha_s N_f}{4\pi}\int\limits_{\Lambda^2/s}^{z}\frac{\mathrm{d}z'}{z'}\int\limits_{1/z's}^{x_{10}^2}\frac{\mathrm{d}x_{21}^2}{x_{21}^2} \widetilde{Q}_{12}\left(z's\right)
\end{align}
in agreement with \eq{total5}.


\subsection{\texorpdfstring{Evolution for $\widetilde{Q}$}{Evolution for Qtilde}}

\begin{figure}[ht]
    \centering
    \includegraphics[scale=0.9]{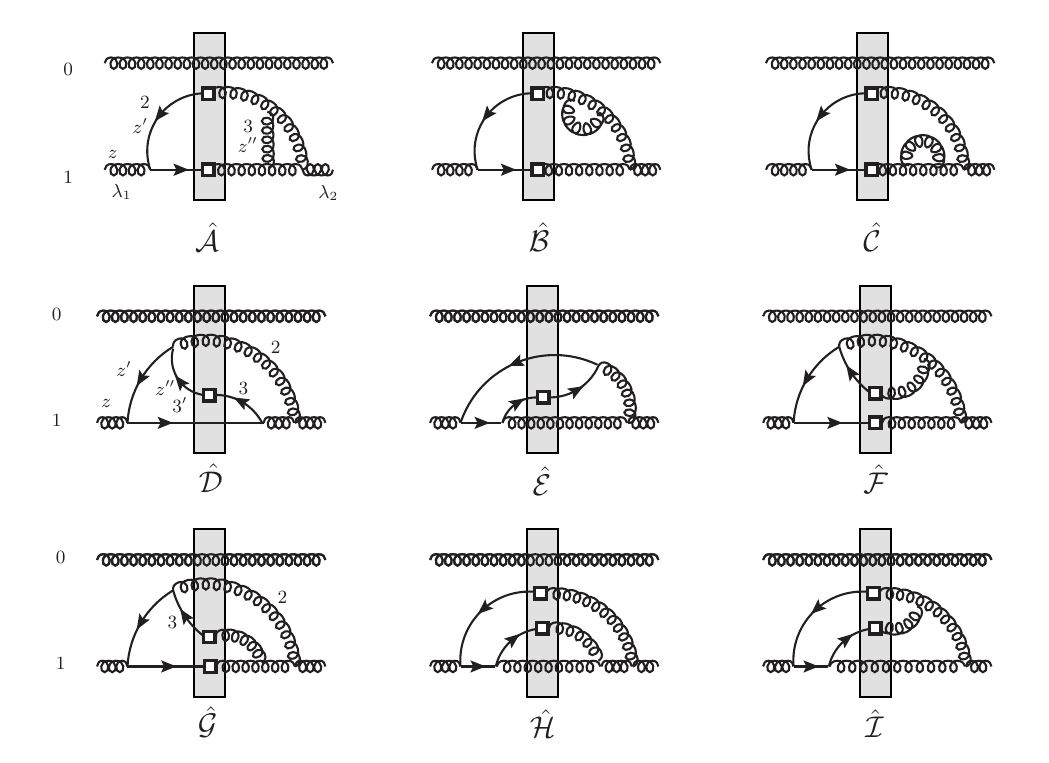}
    \caption{The diagrams containing two steps of evolution: a step of evolution for $G_{10}^\textrm{adj}$ calculated in Sec.~\ref{sec:adj_dip} followed by a step of evolution for $\widetilde{Q}_{21}$. Upon taking the large-$N_c$ limit and factoring out the evolution kernel for $G_{10}^\textrm{adj}$ found in Sec.~\ref{sec:adj_dip}, the contributions of these diagrams will reduce to those in \fig{FIG:Qtilde_evol}.}
    \label{fig:appA_Qtilde_alldiagrams}
\end{figure}

Next we derive the evolution of the new structure $\widetilde{Q}$, in an effort to reproduce the results of Sec. \ref{subsec:Qtilde_evol_I}. We need to consider the diagrams in \fig{FIG:Qtilde_evol}. However, in order to use the LCPT light-cone wavefunctions we begin without yet having taken the large-$N_c$ limit. That is, we start from the same diagrams for the evolution of the adjoint dipole we have just calculated in Sec.~\ref{sec:adj_dip} and add an additional step of evolution to those diagrams. In this procedure we will obtain two evolution kernels. We can factor out and remove the first kernel, corresponding to the first step of our evolution calculated above, and what remains will be the second step of the evolution that we need to find, the evolution of $\widetilde{Q}$. The diagrams we consider here are shown in \fig{fig:appA_Qtilde_alldiagrams}. As these diagrams are similar to but distinct from those in \fig{FIG:Qtilde_evol}, we denote them by the same calligraphic letters but with hats. Upon ultimately taking the large-$N_c$ limit we will obtain the contributions of the diagrams in \fig{FIG:Qtilde_evol}. 

We begin with the virtual corrections labeled $\hat{\cal A}$, $\hat{\cal B}$, and $\hat{\cal C}$ shown in the first row of Fig.~\ref{fig:appA_Qtilde_alldiagrams}. To calculate these diagrams we treat the extra emitted gluon (the one at transverse position $\underline{x}_3$) as real (going through the shock wave, but not interacting with it) and multiply the result of such calculation by an overall factor of $-\frac{1}{2}$, in accordance with the unitarity arguments relating the sum of real and virtual corrections \cite{Kovchegov:2012mbw} which we use to extract the virtual contribution.
For diagram $\hat{\cal A}$, which represents the sum of two time-orderings, we write
\begin{align}\label{appA_eikA}
    &\hat{\cal A} = \left(-\frac{1}{2}\right)\frac{1}{2\left(N_c^2 - 1\right)} \left(\frac{1}{2}\sum_{\lambda_1,\lambda_2}\lambda_1\right)\sum_f\sum_{\text{internal}}\int\frac{\mathrm{d}z'}{z'}\int\frac{\mathrm{d}^2\underline{x}_2}{4\pi}\theta\left(x_{10}^2z - x_{21}^2z'\right) \\
    &\hspace{.5cm}\times \int\frac{\mathrm{d}z''}{z''}\int\frac{\mathrm{d}^2\underline{x}_3}{4\pi}\theta\left(x_{21}^2z' - \text{max}\{x_{32}^2,x_{31}^2\}z''\right) \bigg\langle \left[\psi^{G\rightarrow GG}_{ebf;\lambda_3,\lambda_2,\lambda_4}\left(\underline{x}_{21}, \frac{z'}{z}\right)\bigg|_{\tfrac{z'}{z}\rightarrow 0}\right]^* \left(W^{Gq}_{\underline{1}}\left[\infty,-\infty\right] \right)^{c,\lambda';j,\sigma'} \notag \\
    &\hspace{1cm}\times \left[\left(\psi^{G\rightarrow q\bar{q}}_{a\lambda_1;\sigma',\sigma}\right)^{ji}\left(\underline{x}_{21},1-\frac{z'}{z}\right)\bigg|_{1-\tfrac{z'}{z}\rightarrow 1}\right]\left(W^{G\bar{q}}_{\underline{2}}\left[\infty,-\infty\right]\right)^{d,\lambda; i, \sigma} U_{\underline{0}}^{ba} \notag \\
    &\hspace{1cm}\times \bigg(\left[\psi^{G\rightarrow GG}_{gde; \lambda_5,\lambda,\lambda_3}\left(\underline{x}_{32},\frac{z''}{z'} \right)\bigg|_{\tfrac{z''}{z'}\rightarrow 0}\right] \left[\psi^{G\rightarrow GG}_{gfc; \lambda_5,\lambda_4,\lambda'}\left(\underline{x}_{31},\frac{z''}{z'} \right)\bigg|_{\tfrac{z''}{z'}\rightarrow 0}\right]^* \notag \\
    &\hspace{3cm} + \left[\psi^{G\rightarrow GG}_{gcf; \lambda_5,\lambda',\lambda_4}\left(\underline{x}_{31},\frac{z''}{z'} \right)\bigg|_{\tfrac{z''}{z'}\rightarrow 0}\right] \left[\psi^{G\rightarrow GG}_{ged; \lambda_5,\lambda_3,\lambda}\left(\underline{x}_{32},\frac{z''}{z'} \right)\bigg|_{\tfrac{z''}{z'}\rightarrow 0}\right]^*    \bigg) \bigg\rangle. \notag
\end{align}

Meanwhile, for diagrams $\hat{\cal B}$ and $\hat{\cal C}$ we write
\begin{align}\label{appA_eikBC}
    &\hat{\cal B} + \hat{\cal C} = \frac{1}{2}\left(-\frac{1}{2}\right)\frac{1}{2\left(N_c^2 - 1\right)} \left(\frac{1}{2}\sum_{\lambda_1,\lambda_2}\lambda_1\right)\sum_f\sum_{\text{internal}}\int\frac{\mathrm{d}z'}{z'}\int\frac{\mathrm{d}^2\underline{x}_2}{4\pi}\theta\left(x_{10}^2z - x_{21}^2z'\right) \\
    &\hspace{.5cm}\times \int\frac{\mathrm{d}z''}{z''}\int\frac{\mathrm{d}^2\underline{x}_3}{4\pi}\theta\left(x_{21}^2z' - x_{32}^2 z''\right) \bigg\langle \left[\psi^{G\rightarrow GG}_{ebf;\lambda_3,\lambda_2,\lambda_4}\left(\underline{x}_{21}, \frac{z'}{z}\right)\bigg|_{\tfrac{z'}{z}\rightarrow 0}\right]^* \left(W^{Gq}_{\underline{1}\left[\infty,-\infty\right]} \right)^{c,\lambda';j,\sigma'} \notag \\
    &\hspace{1cm}\times \left[\left(\psi^{G\rightarrow q\bar{q}}_{a\lambda_1;\sigma',\sigma}\right)^{ji}\left(\underline{x}_{21},1-\frac{z'}{z}\right)\bigg|_{1-\tfrac{z'}{z}\rightarrow 1}\right]\left(W^{G\bar{q}}_{\underline{2}}\left[\infty,-\infty\right]\right)^{d,\lambda; i, \sigma} U_{\underline{0}}^{ba} \notag \\
    &\hspace{1cm}\times \left[\psi^{G\rightarrow GG}_{gde; \lambda_5,\lambda,\lambda_4}\left(\underline{x}_{32},\frac{z''}{z'} \right)\bigg|_{\tfrac{z''}{z'}\rightarrow 0}\right] \left[\psi^{G\rightarrow GG}_{gfe; \lambda_5,\lambda_3,\lambda_4}\left(\underline{x}_{32},\frac{z''}{z'} \right)\bigg|_{\tfrac{z''}{z'}\rightarrow 0}\right]^* \bigg \rangle \quad + \quad \left(\underline{x}_{32}\rightarrow \underline{x}_{31} \right). \notag
\end{align}
We have multiplied diagrams $\hat{\cal B}$ and $\hat{\cal C}$ by an additional factor of $\tfrac{1}{2}$. When we eventually take the large-$N_c$ limit, the gluon that connects transverse positions $\underline{x}_2$ and $\underline{x}_1$ becomes a quark-antiquark pair and the virtual gluon at transverse position $\underline{x}_3$ could couple to either of these lines. But only the quark line (in this diagram) becomes a part of the large-$N_c$ structure $\widetilde{Q}_{21}$, and so we only want the contribution where the virtual gluon couples to this line. The factor of $1/2$ gives us this contribution only.

The diagrams we consider here represent an additional step of evolution done to diagram B in \fig{FIG:Gadj_evol}. In addition we need to include this same second step of evolution done to all the other diagrams in \fig{FIG:Gadj_evol}. This is straightforward as the contribution from diagram A is simply the conjugate of what we have just written, while that from diagrams C and D adds in the contribution to the past-pointing staple (see the second term in parentheses in \eq{Qtilde}) and its conjugate. Meanwhile diagrams A', B', C', and D' simply modify the DLA kernel for the first step of the evolution (the same modification occurred in arriving at \eq{appA_alladjdiagrams1}). Adding in all these contributions, carrying out the algebra, taking the large-$N_c$ limit, and multiplying by an additional factor of $\tfrac{1}{4}$ to switch from $G^{\text{adj}}_{21}$ to $\widetilde{G}_{21}$ we obtain
\begin{align}\label{appA_alleik1}
    &\hat{\cal A} + \hat{\cal B} + \hat{\cal C} = \left(-\frac{\alpha_s N_f}{4\pi} \int\limits_{\Lambda^2/s}^{z}\frac{\mathrm{d}z'}{z'}\int\limits_{1/z's}^{x_{10}^2}\frac{\mathrm{d}x_{21}^2}{x_{21}^2} \right) \left(-\frac{\alpha_s N_c}{4\pi^2} \right)\int\frac{\mathrm{d}z''}{z''}\int\mathrm{d}^2\underline{x}_3 \\
    &\hspace{2cm}\times\left[\frac{1}{x_{32}^2}\theta\left(x_{21}^2z'-x_{32}^2z''\right) + \frac{1}{x_{31}^2}\theta\left(x_{21}^2z'-x_{31}^2z''\right) - 2 \frac{\underline{x}_{32}\cdot \underline{x}_{31}}{x_{32}^2x_{31}^2}\theta\left(x_{21}^2z' - \text{max}\{x_{32}^2,x_{31}^2\}z'' \right)\right] \widetilde{Q}_{21}(z''s).\notag
\end{align}
We have cancelled a factor of $\tfrac{1}{zs}$ for the same reason stated following \eq{appA_alladjdiagrams1}. We also set unpolarized S-matrices to $1$. Now the first kernel in \eq{appA_alleik1} (the large parentheses in the first line) is simply the kernel for the first step of evolution. So we remove it, leaving just the second step of evolution which, now that we have taken the large-$N_c$ limit, corresponds to diagrams $\cal A$, $\cal B$, and $\cal C$ in \fig{FIG:Qtilde_evol}. Simplifying the kernel at DLA we obtain
\begin{align}\label{appA_alleik2}
    {\cal A}+{\cal B}+{\cal C} = -\frac{\alpha_s N_c}{4 \pi} \int\limits_{\Lambda^2/s}^{z'}\frac{\mathrm{d}z''}{z''} \left[\int\limits_{1/(z''s)}^{x_{21}^2} \frac{\mathrm{d}x_{32}^2}{x_{32}^2} +  \int\limits_{1/(z''s)}^{x_{21}^2} \frac{\mathrm{d}x_{31}^2}{x_{31}^2} \right] \widetilde{Q}_{21}(z''s),
\end{align}
which agrees with \eq{ABC} in the main text upon substituting $\underline{x}_1 \rightarrow \underline{x}_{2}$, $\underline{x}_0 \rightarrow \underline{x}_{1}$, $\underline{x}_2 \rightarrow \underline{x}_{3}$, $z' \rightarrow z''$, and $z\rightarrow z''$ in the latter. 

The remaining diagrams we need are shown in the last two rows of \fig{fig:appA_Qtilde_alldiagrams}. 
We first consider diagrams $\hat{\cal F}$ and $\hat{\cal G}$. For diagram $\hat{\cal F}$ we write 
\begin{align}\label{appA_Qtilde_2steps1}
    &\hat{\cal F} = \frac{1}{2\left(N_c^2 - 1\right)}\left(\frac{1}{2}\sum_{\lambda_1,\lambda_2}\lambda_1 \right)\sum_{f} \sum_{\text{internal}} \int\frac{\mathrm{d}z'}{z'}\int\frac{\mathrm{d}^2\underline{x}_2}{4\pi} \int\frac{\mathrm{d}z''}{z''}\int\frac{\mathrm{d}^2\underline{x}_3}{4\pi} \\
    &\times \bigg\langle \left[\psi^{G\rightarrow GG}_{fbc;\lambda_4,\lambda_2,\lambda'}\left(\underline{x}_{21}, \tfrac{z'}{z} \right)\bigg|_{\tfrac{z'}{z}\rightarrow 0} \right]^* \left[\psi^{G\rightarrow GG}_{gfd;\lambda_3,\lambda_4,\lambda}\left(\underline{x}_{32},\tfrac{z''}{z'} \right)\bigg|_{\tfrac{z''}{z'}\rightarrow 0} \right]^* U_{\underline{2}}^{de}U_{\underline{0}}^{ba} \left(W_{\underline{1}}^{Gq} \left[\infty,-\infty\right]\right)^{c,\lambda';j,\sigma_0'} \notag \\
    & \hspace{0.35cm}\times \left[\left(\psi^{G\rightarrow q\bar{q}}\right)^{ji}_{a\lambda_1;\sigma_0',\sigma_0}\left(\underline{x}_{21},1-\tfrac{z'}{z} \right)\bigg|_{1-\tfrac{z'}{z}\rightarrow 1} \right] \left[-\left(\psi^{q\rightarrow qG}\right)^{ii'}_{e,\lambda;\sigma_0,\sigma}\left(\underline{x}_{32},\tfrac{z''}{z'} \right)\bigg|_{\tfrac{z''}{z}\rightarrow 0}  \right] \left(W_{\underline{3}}^{G\bar{q}}\left[\infty,-\infty \right]\right)^{g,\lambda_3;i'\sigma}\bigg\rangle \notag .
\end{align}
Here we have used used $-\psi^{q\rightarrow qG}$ for the $\bar{q}\rightarrow \bar{q}G$ splitting, as mentioned in the paragraph just after \eq{appA_q_to_qG_wf_softG}. We have also suppressed the light-cone lifetime-ordering theta functions for brevity. Carrying out the algebra we obtain 
\begin{align}\label{appA_Qtilde_2steps3}
    &\hat{\cal F} = \frac{1}{2\left(N_c^2-1\right)}\left(-\frac{\alpha_s N_f}{\pi^2}\right) \int\frac{\mathrm{d}z'}{z'}\int\frac{\mathrm{d}^2\underline{x}_2}{x_{21}^2} \left(\frac{2\alpha_s}{\pi^2}\right)\int\frac{\mathrm{d}z''}{z''}\int\frac{\mathrm{d}^2\underline{x}_3}{x_{32}^2} \\
    &\hspace{0.5cm}\times\llangle \frac{g^2P^+}{8\sqrt{z''}s}\int\limits_{-\infty}^{\infty}\mathrm{d}w_1^-\int\limits_{-\infty}^{\infty}\mathrm{d}w_3^- f^{fbc}f^{gfd}U_{\underline{1}}^{ch}\left[\infty,w_1^-\right]\bar{\psi}_{\alpha}\left(w_1^-,\underline{x}_1\right) t^h \left(\frac{1}{2}\gamma^+\gamma^5\right)_{\alpha\beta} V_{\underline{1}}\left[w_1^-,-\infty\right] t^a t^e \notag \\
    &\hspace{4.5cm}\times U_{\underline{3}}^{gh'}\left[\infty,w_3^-\right] V_{\underline{3}}^\dagger\left[w_3^-,-\infty\right] t^{h'} \psi_\beta\left(w_3^-,\underline{x}_3\right) U_{\underline{2}}^{de}U_{\underline{0}}^{ba}\rrangle \notag,
\end{align}
where we have again cancelled a factor of $\tfrac{1}{zs}$ for the same reason as above. Again, diagram $\hat{\cal F}$ represents a second step of evolution done to diagram B in \fig{FIG:Gadj_evol}, so we need to add in this same step of evolution done to the rest of the diagrams in \fig{FIG:Gadj_evol}, just as discussed before \eq{appA_alleik1}. We again obtain the original kernel from the first step of evolution, which we remove. Doing this, then taking the large-$N_c$ limit, setting the unpolarized S-matrices that result to $1$, and including the factor of $\tfrac{1}{4}$ for the normalization of $\widetilde{G}$, we obtain the contribution of diagram ${\cal F}$. Diagram $\hat{\cal G}$ is calculated similarly. Simplifying the kernel at DLA, we have for the contributions of diagrams ${\cal F}$ and ${\cal G}$
\begin{align}\label{appA_Qtilde_2steps4}
    {\cal F} + {\cal G} = \frac{\alpha_s N_c}{4\pi}\int\limits_{\Lambda^2/s}^{z'}\frac{\mathrm{d}z''}{z''} \int\limits_{1/(z''s)}^{x_{21}^2}\frac{\mathrm{d}x_{32}^2}{x_{32}^2} \widetilde{Q}_{31}(z''s).
\end{align} 
\eq{appA_Qtilde_2steps4} agrees with \eq{FG3} upon substituting $\underline{x}_1 \rightarrow \underline{x}_{2}$, $\underline{x}_0 \rightarrow \underline{x}_{1}$, $\underline{x}_2 \rightarrow \underline{x}_{3}$, $z' \rightarrow z''$, and $z\rightarrow z'$ in the latter. Further, \eq{appA_Qtilde_2steps4} cancels the first term of \eq{appA_alleik2} with the DLA accuracy, just as in \eq{Q-Q}. Subsequently diagrams $\hat{\cal H}$ and $\hat{\cal I}$ from Fig.~ \ref{fig:appA_Qtilde_alldiagrams} yield contributions that cancel the second term of \eq{appA_alleik2} with DLA accuracy, just as in Sec.~ \ref{subsec:Qtilde_evol_I}.

What remains is to calculate the contributions from diagrams $\hat{\cal D}$ and $\hat{\cal E}$. We write
\begin{align}\label{appA_Qtilde_2steps5}
    &\hat{\cal D} + \hat{\cal E} = (-1) \frac{1}{2\left(N_c^2 - 1\right)}\left(\frac{1}{2}\sum_{\lambda_1,\lambda_2}\lambda_1 \right) \sum_{f} \sum_{\text{internal}} \int\frac{\mathrm{d}z'}{z'}\int\frac{\mathrm{d}^2\underline{x}_2}{4\pi} \int\frac{\mathrm{d}z''}{z''}\int\frac{\mathrm{d}^2\underline{x}_3\mathrm{d}^2\underline{x}_3'}{4\pi} \\
    &\hspace{0.5cm} \times \bigg\langle \left[\psi^{G\rightarrow GG}_{dbc;\lambda,\lambda_2,\lambda'}\left(\underline{x}_{21},\frac{z'}{z} \right)\bigg|_{\tfrac{z'}{z}\rightarrow 0} \right]^* \left[\left(\psi^{G\rightarrow q\bar{q}}\right)^{j'j''}_{c\lambda'; \sigma',\sigma_0'}\left(\underline{x}_{31},\frac{z''}{z}  \right)\bigg|_{\tfrac{z''}{z}\rightarrow 0}\right]^* V_{\underline{1}}^{j''j} \notag \\
    &\hspace{0.85cm}\times\left[\left(\psi^{G\rightarrow q\bar{q}}\right)^{ji}_{a\lambda_1;\sigma_0'\sigma_0}\left(\underline{x}_{21}, 1-\frac{z'}{z}\right)\bigg|_{1-\tfrac{z'}{z}\rightarrow 1 }  \right] \left[-\left(\psi^{q\rightarrow qG}\right)^{ii'}_{e\lambda; \sigma_0,\sigma}\left(\underline{x}_{3'2}, \frac{z''}{z'} \right)\bigg|_{\tfrac{z''}{z'}\rightarrow 0} \right] \notag \\
    &\hspace{1cm}\times\left(-\sigma \delta_{\sigma,\sigma'}V_{\underline{3}}^{\text{pol}[1]\,\dagger} \delta^2\left(\underline{x}_3-\underline{x}_{3'} \right) + \delta_{\sigma,\sigma'}V_{\underline{3},\underline{3'}}^{\text{pol}[2]\,\dagger}  \right)^{i'j'} U_{\underline{2}}^{de}U_{\underline{0}}^{ba}  \quad + \quad \text{c.c.}  \bigg\rangle \notag.
\end{align}
As we did for diagrams $\hat{\cal F}$ and $\hat{\cal G}$, we again use $-\psi^{q\rightarrow qG}$ for the $\bar{q}\rightarrow \bar{q}G$ splitting. In addition we have an overall minus sign for the fermion loop. This is still required even with an operator insertion mediating the interaction with the shock-wave. We have also once again omitted the lifetime-ordering theta functions. Following the same steps as in the other diagrams and taking the large-$N_c$ limit, we obtain for the contribution of diagrams ${\cal D} + {\cal E}$:
\begin{align}\label{appA_Qtilde_2steps7}
    {\cal D}+{\cal E} = \frac{\alpha_s}{4\pi^2}\int\limits_{\Lambda^2/s}^{z'}\frac{\mathrm{d}z''}{z''}\left\{-\int \mathrm{d}^2\underline{x}_3 \frac{\underline{x}_{32}\cdot \underline{x}_{31}}{x_{32}^2x_{31}^2} \llangle \text{tr}\left[V_{\underline{2}}V_{\underline{3}}^{\text{pol}[1]\dagger} \right] \rrangle + i\int\mathrm{d}^2\underline{x}_{3}\mathrm{d}^2\underline{x}_{3'} \frac{\underline{x}_{3'2}}{x_{3'2}^2}\times\frac{\underline{x}_{31}}{x_{31}^2} \llangle \text{tr}\left[V_{\underline{2}}V_{\underline{3},\underline{3'}}^{\text{pol}[2]\dagger} \right]\rrangle  + \text{c.c} \right\}.
\end{align}
\eq{appA_Qtilde_2steps7} agrees with \eq{DE_app1} upon substituting $\underline{x}_0 \rightarrow \underline{x}_{1}$, $\underline{x}_1 \rightarrow \underline{x}_{2}$, $\underline{x}_{2} \rightarrow \underline{x}_{3}$, $\underline{x}_{2'} \rightarrow \underline{x}_{3'}$, $z\rightarrow z'$, and $z' \rightarrow z''$ in the latter equation. The rest of the calculation then proceeds exactly as that following \eq{DE_app1} in the main text, resulting in 
\begin{align}\label{appA_Qtilde_2steps8}
    \int \mathrm{d}^2\left(\frac{x_1+x_2}{2}\right)\left[{\cal D} + {\cal E} \right] = -\frac{\alpha_s N_c}{2\pi}\int\limits_{\Lambda^2/s}^{z'}\frac{\mathrm{d}z''}{z''}\int\limits_{\text{max}\left[x_{21}^2,\tfrac{1}{z''s}\right]}^{\text{min}\left[\tfrac{z'}{z''}x_{21}^2, \tfrac{1}{\Lambda^2} \right]} \frac{\mathrm{d}x_{32}^2}{x_{32}^2}\left[Q(x_{32}^2,z''s) + 2G_2(x_{32}^2,z''s)  \right]
\end{align}
in agreement with \eq{DE_app2} upon making the same substitutions.

To summarize, in this Appendix we have derived the contribution of the shock-wave transition operators to the evolution of the adjoint type-1 polarized dipole amplitude and additionally derived the evolution of the new structure $\widetilde{Q}$ itself. While the results of Sec.~\ref{sec:Gadj_evol} and \ref{sec:Qtilde_evol} in the main text were derived using light-cone operator treatment, the derivation in this Appendix employed a slightly different method, making use of the light-cone wave functions of LCPT. We have found complete agreement in the resulting evolution equations.

 \section{Scheme Transformation}
 \renewcommand{\theequation}{B\arabic{equation}}
   \setcounter{equation}{0}
 \label{B}

The differences between the polarized small-$x$ large-$N_c \& N_f$ splitting functions in our/BER scheme \eqref{eq:DeltaP_BER} and in the $\overline{\text{MS}}$ scheme  \eqref{eq:DeltaP_MSbar} can be accounted for by a scheme transformation. In the following, we explicitly derive the scheme transformation matrix $Z$ that transforms $\Delta \mathbf{P}$ in our/BER scheme to $\Delta \overline{\mathbf{P}}$ in the $\overline{\text{MS}}$ scheme. In our notation, over-lined quantities are understood to be in the $\overline{\text{MS}}$ scheme. The derivation is performed in the Mellin space. In Mellin space, the matrix of anomalous dimensions is
\begin{equation}
\Delta \gamma (\omega) = \int_0^1 dx\, x^{\omega - 1} \Delta \mathbf{P} (x).
\end{equation}
We follow the method in \cite{Moch:2014sna}. 

Let $\Delta f(\omega, \mu^2) \equiv (\Delta \Sigma(\omega, \mu^2), \Delta G(\omega, \mu^2)^T$ denote the quark and gluon helicity PDFs in Mellin space.
The hPDFs in the two schemes are related by
\begin{equation}
\Delta \overline{f}(\omega, \mu^2) = Z(\omega, \mu^2) \, \Delta f(\omega, \mu^2).
\end{equation}
The polarized DGLAP equations \eqref{DGLAP_diff} in Mellin space can be formally expressed as a matrix equation
\begin{equation}
\frac{\pd \Delta \overline{f}(\omega, \mu^2)}{ \pd \ln \mu^2} = \Delta \overline{\gamma}(\omega) \, \Delta \overline{f}(\omega, \mu^2),
\end{equation} 
which can be further written as 
\begin{subequations}
\begin{align}
&\frac{\pd Z(\omega, \mu^2)}{\pd \ln\mu^2} \Delta f(\omega, \mu^2) + Z(\omega, \mu^2)\frac{\pd \Delta f(\omega, \mu^2)}{\pd \ln\mu^2} = \Delta \overline{\gamma}(\omega) Z(\omega,\mu^2) \Delta f(\omega, \mu^2),\\
\Longrightarrow &\,\,\, \frac{\pd Z(\omega, \mu^2)}{\pd \ln\mu^2} \Delta f(\omega, \mu^2) + Z(\omega, \mu^2) \Delta \gamma(\omega) \Delta f(\omega, \mu^2)= \Delta \overline{\gamma}(\omega) Z(\omega, \mu^2) \Delta f(\omega, \mu^2).
\end{align}
\end{subequations}
One therefore obtains
\begin{equation}\label{eq:scheme_tranformation_eq}
\Delta \overline{\gamma}(\omega) = Z(\omega, \mu^2)\Delta \gamma(\omega) Z^{-1}(\omega, \mu^2)  + \frac{\pd Z(\omega, \mu^2)}{\pd \ln\mu^2} \, Z^{-1}(\omega, \mu^2).
\end{equation}
This is the equation to determine the scheme transformation matrix $Z$, given $\Delta \gamma$ and $\Delta \overline{\gamma}$.  In the following, we solve this equation perturbatively order by order in $a_s = \alpha_s(\mu^2)/4\pi$.  The anomalous dimensions have the expansions
\begin{subequations}
\begin{align}
&\Delta \overline{\gamma}(\omega) = a_s \Delta \overline{\gamma}^{(0)}_{\omega} + a_s^2 \Delta \overline{\gamma}^{(1)}_{\omega} + a_s^3 \Delta \overline{\gamma}^{(2)}_{\omega} + \ldots,\\
&\Delta \gamma(\omega) = a_s \Delta \gamma^{(0)}_{\omega} + a_s^2 \Delta \gamma^{(1)}_{\omega} + a_s^3 \Delta \gamma^{(2)}_{\omega} + \ldots.
\end{align}
\end{subequations}
The scheme transformation matrix $Z$ can also be expanded as 
\begin{subequations}
\begin{align}
&Z(\omega, \mu^2) = 1 + a_s Z_{\omega}^{(1)} + a_s^2 Z_{\omega}^{(2)} + \ldots,\\
&Z^{-1}(\omega, \mu^2) = 1- a_s Z_{\omega}^{(1)} - a_s^2\left(Z_{\omega}^{(2)}- \left[Z_{\omega}^{(1)}\right]^2\right) + \mathcal{O}(a_s^3).
\end{align}
\end{subequations}
The dependence on $\mu^2$ comes solely from the running coupling constant $a_s$. 
We start with the first term on the right-hand side of Eq.~\eqref{eq:scheme_tranformation_eq}.
\begin{equation}
\begin{split}
Z(\omega, \mu^2)\Delta \gamma(\omega)Z^{-1}(\omega, \mu^2)
= & a_s \Delta \gamma^{(0)}_{\omega}+a_s^2\left(\Delta \gamma^{(1)}_{\omega}  + \left[Z^{(1)}_{\omega}, \Delta \gamma^{(0)}_{\omega}\right]\right)\\
&+a_s^3 \left(\Delta \gamma^{(2)}_{\omega} +\left[Z^{(1)}_{\omega}, \Delta \gamma^{(1)}_{\omega}\right] +  \left[Z^{(2)}_{\omega}, \Delta \gamma^{(0)}_{\omega}\right] - \left[Z^{(1)}_{\omega}, \Delta \gamma^{(0)}_{\omega}\right] Z^{(1)}_{\omega}\right) +\mathcal{O}(a_s^4).
\end{split}
\end{equation}
We then compute the second term on the right-hand side of Eq.~\eqref{eq:scheme_tranformation_eq} by first noting that 
\begin{equation}
\frac{ \pd Z(\omega, \mu^2)}{ \pd \ln\mu^2} = \frac{da_s}{d\ln\mu^2} Z^{(1)}_{\omega} + 2a_s \frac{da_s}{d\ln \mu^2} Z^{(2)}_{\omega} + \ldots .
\end{equation}
Here running coupling constant satisfies the standard renormalization-group equation
\begin{equation}
\frac{da_s}{d\ln\mu^2} = \beta(a_s) = -\sum_{l=0}^{\infty} a_s^{l+2} \beta_l
\end{equation}
with the leading coefficients in the large $N_c\& N_f$ limit being 
\begin{subequations}
\begin{align}
&\beta_0 = \frac{11}{3}N_c-\frac{2}{3}N_f, \\
&\beta_1 = \frac{34}{3}N_c^2 - \frac{13}{3}N_cN_f.
\end{align}
\end{subequations}
One then gets
\begin{equation}
\frac{\pd Z(\omega, \mu^2)}{\pd \ln\mu^2} Z^{-1}(\omega, \mu^2) = -a_s^2 \beta_0 Z^{(1)}_{\omega} -a_s^3 \left(\beta_1 Z^{(1)}_{\omega} + 2\beta_0 Z^{(2)}_{\omega} - \beta_0 \left[Z^{(1)}_{\omega}\right]^2\right) + \mathcal{O}(a_s^4).
\end{equation}
Using the above expressions, Eq.~\eqref{eq:scheme_tranformation_eq} becomes
\begin{equation}
\begin{split}
&a_s \Delta \overline{\gamma}^{(0)}_{\omega} + a_s^2 \Delta \overline{\gamma}^{(1)}_{\omega}+ a_s^3 \Delta \overline{\gamma}^{(2)}_{\omega} + \mathcal{O}(a_s^4)\\
 = & a_s \Delta \gamma^{(0)}_{\omega} +a_s^2\left(\Delta \gamma^{(1)}_{\omega}  + \left[Z^{(1)}_{\omega}, \Delta \gamma^{(0)}_{\omega}\right]- \beta_0 Z^{(1)}_{\omega}\right)+a_s^3 \Big(\Delta \gamma^{(2)}_{\omega} + \left[Z^{(1)}_{\omega}, \Delta \gamma^{(1)}_{\omega}\right] +  \left[Z^{(2)}_{\omega}, \Delta \gamma^{(0)}_{\omega}\right] \\
 &- \left[Z^{(1)}_{\omega}, \Delta \gamma^{(0)}_{\omega}\right] Z^{(1)}_{\omega}  - \left(\beta_1 Z^{(1)}_{\omega} + 2\beta_0 Z^{(2)}_{\omega} - \beta_0 [Z^{(1)}_{\omega}]^2\right)\Big) +\mathcal{O}(a_s^4).\\
\end{split}
\end{equation}
Comparing the left-hand side and the right-hand side of the equation, one arrives at 
\begin{subequations}
\begin{align}
&\Delta \overline{\gamma}^{(0)}_{\omega} = \Delta \gamma^{(0)}_{\omega}, \\
& \Delta \overline{\gamma}^{(1)}_{\omega}  = \Delta \gamma_{\omega}^{(1)}  + \left[Z^{(1)}_{\omega}, \Delta \gamma_{\omega}^{(0)}\right]-\beta_0 Z^{(1)}_{\omega},\label{eq:ObyO_relation_(1)}\\
\begin{split}
& \Delta \overline{\gamma}_{\omega}^{(2)} =  \Delta \gamma_{\omega}^{(2)} + \left[Z^{(1)}_{\omega}, \Delta \gamma_{\omega}^{(1)}\right] +  \left[Z^{(2)}_{\omega}, \Delta \gamma_{\omega}^{(0)}\right] - \left[Z_{\omega}^{(1)}, \Delta \gamma_{\omega}^{(0)}\right] Z_{\omega}^{(1)} \\
&\qquad \qquad- \left(\beta_1 Z_{\omega}^{(1)} + 2\beta_0 Z_{\omega}^{(2)} - \beta_0 [Z_{\omega}^{(1)}]^2\right).  \label{eq:ObyO_relation_(2)}
\end{split}
\end{align}
\end{subequations}
From the explicit expressions for the splitting functions in Eq.~\eqref{eq:DeltaP_BER}
and in Eq.~\eqref{eq:DeltaP_MSbar}, we have $\Delta \overline{\gamma}_{\omega}^{(0)} = \Delta \gamma_{\omega}^{(0)}$ and $\Delta \overline{\gamma}_{\omega}^{(1)} = \Delta \gamma_{\omega}^{(1)}$.  Eq.~\eqref{eq:ObyO_relation_(1)} becomes 
\begin{equation}
\left[Z_{\omega}^{(1)}, \Delta \gamma_{\omega}^{(0)}\right]-\beta_0 Z_{\omega}^{(1)}=0.
\end{equation}
One can readily see that the only solution is $Z_{\omega}^{(1)} =0$. As a result, Eq.~\eqref{eq:ObyO_relation_(2)} reduces to
\begin{equation}\label{eq:DeltaP(2)_equation}
\Delta \overline{\gamma}_{\omega}^{(2)} - \Delta \gamma_{\omega}^{(2)} =  \left[Z_{\omega}^{(2)}, \Delta \gamma_{\omega}^{(0)}\right]  - 2\beta_0 Z_{\omega}^{(2)}.
\end{equation}
Let us denote the elements of the scheme transformation matrix at order $a_s^2$ as
\begin{equation}
Z^{(2)}_{\omega} = \begin{pmatrix}
Z^{(2)}_{qq}(\omega)\, & \,Z^{(2)}_{qG}(\omega)\\[10pt]
Z^{(2)}_{Gq}(\omega)\, & \,Z^{(2)}_{GG}(\omega)\\
\end{pmatrix} .
\end{equation}
The four component equations from the matrix equation in Eq.~\eqref{eq:DeltaP(2)_equation} are
\begin{subequations}\label{eq:4components}
\begin{align}
&\Delta \overline{\gamma}^{(2)}_{qq}(\omega) - \Delta \gamma^{(2)}_{qq}(\omega) = Z^{(2)}_{qG}(\omega)\Delta \gamma^{(0)}_{Gq}(\omega) - \Delta \gamma^{(0)}_{qG}(\omega)Z^{(2)}_{Gq}(\omega) - 2\beta_0 Z^{(2)}_{qq}(\omega) =0,\\
&\Delta \overline{\gamma}^{(2)}_{GG}(\omega) - \Delta \gamma^{(2)}_{GG}(\omega) = Z^{(2)}_{Gq}(\omega)\Delta \gamma^{(0)}_{qG}(\omega) - \Delta \gamma^{(0)}_{Gq}(\omega) Z^{(2)}_{qG}(\omega) - 2\beta_0 Z^{(2)}_{GG}(\omega) =0, \\
&\Delta \overline{\gamma}^{(2)}_{qG}(\omega) - \Delta \gamma^{(2)}_{qG}(\omega) = \Delta \gamma^{(0)}_{qG}(\omega)\left(Z^{(2)}_{qq}(\omega)-Z^{(2)}_{GG}(\omega)\right) + Z^{(2)}_{qG}(\omega)\left(\Delta \gamma^{(0)}_{GG}(\omega) - \Delta \gamma^{(0)}_{qq}(\omega)\right) ,\\
&\Delta \overline{\gamma}^{(2)}_{Gq}(\omega) - \Delta \gamma^{(2)}_{Gq}(\omega) = \Delta \gamma^{(0)}_{Gq}(\omega)\left(Z^{(2)}_{GG}(\omega)- Z^{(2)}_{qq}(\omega)\right) + Z^{(2)}_{Gq}(\omega)\left(\Delta \gamma^{(0)}_{qq}(\omega) - \Delta \gamma^{(0)}_{GG}(\omega)\right).
\end{align}
\end{subequations}
The solutions are 
\begin{subequations}\label{eq:formal_solutions_Z(2)}
    \begin{align}
        &Z^{(2)}_{GG}(\omega)=-Z^{(2)}_{qq}(\omega),\\
        \begin{split}
&Z^{(2)}_{qq}(\omega) = \frac{1}{2\beta_0} \left[\frac{\Delta \gamma^{(0)}_{Gq}(\omega)}{\Delta \gamma^{(0)}_{GG}(\omega) - \Delta \gamma^{(0)}_{qq}(\omega)} (\Delta \overline{\gamma}^{(2)}_{qG}(\omega) - \Delta \gamma^{(2)}_{qG}(\omega)) \right. \\
& \hspace*{3cm} \left. + \frac{\Delta \gamma^{(0)}_{qG}(\omega)}{\Delta \gamma^{(0)}_{GG}(\omega) - \Delta \gamma^{(0)}_{qq}(\omega)} (\Delta \overline{\gamma}^{(2)}_{Gq}(\omega) - \Delta \gamma^{(2)}_{Gq}(\omega)) \right],\\
\end{split}\\
&Z^{(2)}_{qG}(\omega) = \frac{\Delta \overline{\gamma}^{(2)}_{qG}(\omega) - \Delta \gamma^{(2)}_{qG}(\omega)}{\Delta \gamma^{(0)}_{GG}(\omega) - \Delta \gamma^{(0)}_{qq}(\omega)} - \frac{2\Delta \gamma^{(0)}_{qG}(\omega)}{\Delta \gamma^{(0)}_{GG}(\omega) - \Delta \gamma^{(0)}_{qq}(\omega)} Z^{(2)}_{qq}(\omega),\\
&Z^{(2)}_{Gq}(\omega) = - \frac{\Delta \overline{\gamma}^{(2)}_{Gq}(\omega) - \Delta \gamma^{(2)}_{Gq}(\omega)}{\Delta \gamma^{(0)}_{GG}(\omega) - \Delta \gamma^{(0)}_{qq}(\omega)}- \frac{2\Delta \gamma^{(0)}_{Gq}(\omega)}{\Delta \gamma^{(0)}_{GG}(\omega) - \Delta \gamma^{(0)}_{qq}(\omega)} Z^{(2)}_{qq}(\omega).
    \end{align}
\end{subequations}
The relevant anomalous dimensions at fixed orders are 
\begin{equation}
\Delta \gamma_{qq}^{(0)}(\omega) = N_c\frac{1}{\omega}, \quad \Delta \gamma_{GG}^{(0)}(\omega) = 8N_c \frac{1}{\omega}, \quad \Delta \gamma^{(0)}_{qG}(\omega) = -2N_f\frac{1}{\omega}, \quad \Delta \gamma^{(0)}_{Gq}(\omega) = 2N_c\frac{1}{\omega}
\end{equation}
and 
\begin{subequations}
\begin{align}
&\Delta \overline{\gamma}^{(2)}_{qG}(\omega) - \Delta \gamma^{(2)}_{qG}(\omega) = 4N_c^2N_f \frac{1}{\omega^5},\\
&\Delta \overline{\gamma}^{(2)}_{Gq}(\omega)- \Delta \gamma^{(2)}_{Gq}(\omega) = 4N_c^3 \frac{1}{\omega}.
\end{align}
\end{subequations}
Substituting these expressions into Eq.~\eqref{eq:formal_solutions_Z(2)}, one obtains 
\begin{subequations}
\begin{align}
&Z^{(2)}_{qq}(\omega)=-Z^{(2)}_{GG}(\omega)=0,\\
&Z^{(2)}_{qG}(\omega) = \frac{4}{7}N_cN_f \frac{1}{\omega^4},\\
&Z^{(2)}_{Gq}(\omega) = -\frac{4}{7} N_c^2 \frac{1}{\omega^4}.
\end{align}
\end{subequations}
In the matrix form this can be written as
\begin{equation}
Z^{(2)}_{\omega} = \begin{pmatrix}
0 & \frac{4}{7} N_cN_f \frac{1}{\omega^4} \\[10pt]
-\frac{4}{7}N_c^2 \frac{1}{\omega^4} & 0 \\
\end{pmatrix}.
\end{equation}
Transforming back to the $x$-space, one gets
\begin{equation}
Z^{(2)}(x) = \begin{pmatrix}
0 & \frac{2}{21}N_cN_f \ln^3\frac{1}{x} \\[10pt]
-\frac{2}{21}N_c^2 \ln^3\frac{1}{x} & 0 \\
\end{pmatrix}.
\end{equation}

We conclude that the matrix of the scheme transformation relating the splitting functions \eqref{eq:DeltaP_MSbar} to those in   \eqref{eq:DeltaP_BER} up to three loops is given by the following matrix (with $a_s = \alpha_s(\mu^2)/4\pi$): 
\begin{equation}
Z (x) = \begin{pmatrix}
x \, \delta (1-x) & \frac{2}{21} \, a_s^2 \, N_cN_f \ln^3\frac{1}{x} \\[10pt]
-\frac{2}{21} \, a_s^2 \, N_c^2 \ln^3\frac{1}{x} & x \, \delta (1-x) \\
\end{pmatrix}.
\end{equation}



\providecommand{\href}[2]{#2}\begingroup\raggedright\endgroup

\end{document}